\documentclass[a4paper,11pt]{article}
\usepackage{jheppub} 
\usepackage[utf8]{inputenc}
\usepackage[english]{babel}
\usepackage{csquotes}
\usepackage{physics}
\usepackage[T1]{fontenc}
\usepackage{graphicx}
\usepackage{chngcntr}
\usepackage{subcaption}
\usepackage{tabularx}
\usepackage{caption}
\renewcommand{\figurename}{figure}
\usepackage{hyperref}
\usepackage{bbm}
\usepackage[colorinlistoftodos]{todonotes}
\usepackage{epigraph}
\usepackage[export]{adjustbox}
\usepackage{amsmath}
\allowdisplaybreaks
\usepackage{amsthm}
\usepackage{amssymb}
\usepackage{amscdx}
\usepackage{nccmath}
\usepackage{mathtools}
\usepackage{mathrsfs}
\usepackage{blindtext}
\usepackage{titlesec}
\usepackage{natbib}
\usepackage{comment}
\usepackage{simplewick}
\usepackage{dsfont}
\usepackage{cancel}
\usepackage{yfonts}
\usepackage{colonequals}
\usepackage{cellspace}
\usepackage{subfiles}

\newcommand{\beq}{\begin{equation}}
\newcommand{\eeq}{\end{equation}}

\renewcommand{\a}{\alpha}
\renewcommand{\b}{\beta}

\renewcommand{\d}{\delta}

\newcommand{\g}{\gamma}

\newcommand{\e}{\epsilon}
\newcommand{\z}{\zeta}

\renewcommand{\k}{\kappa}

\newcommand{\x}{\chi}

\newcommand{\s}{\sigma}
\renewcommand{\S}{\Sigma}

\newcommand{\y}{\upsilon}

\newcommand{\zb}{\bar{z}}

\DeclareFontFamily{U}{stix2bb}{}
\DeclareFontShape{U}{stix2bb}{m}{n} {<-> stix2-mathbb}{}

\NewDocumentCommand{\id}{}{\text{\usefont{U}{stix2bb}{m}{n}1}}

\newcommand{\nn}{\nonumber}

\usepackage{tikz}
\usetikzlibrary{shapes}
\usetikzlibrary{shapes.misc}

\tikzset{
  vtx/.style={
    circle,
    draw=blue,
    fill=blue,
    inner sep=1pt
  },
  wcirc/.style={
    circle,
    draw=white,
    fill=white,
    inner sep=2pt
  },
  bcirc/.style={
    circle,
    draw=black,
    fill=black,
    inner sep=1pt
  },
  dcirc/.style={
    circle,
    draw=blue,
    fill=blue,
    inner sep=1pt
  },
  rcirc/.style={
    circle,
    draw=red,
    fill=red,
    inner sep=1pt
  },
  phi/.style={
    thick
  },
  sigma/.style={
    thick,
    dashed
  },
  vl1/.style={
    thick,
    blue
  },
  vl2/.style={
    thick,
    dashed,
    blue
  },
  valign/.style={
    baseline={([yshift=-.55ex]current bounding box.center)}
  }
}

\preprint{}

\title{Surface defects in $\mathcal{N}=4$ Super Yang-Mills:\\
exact results and holographic bootstrap}

\author[a,b]{Lorenzo Bianchi,}
\author[c]{Davide Bonomi,}
\author[d]{Maddalena Ferragatta,}
\author[e]{Valentina Forini.}
 \affiliation[a]{Dipartimento di Fisica, Università di Torino,\\Via P. Giuria 1, 10125 Torino, Italy}
\affiliation[b]{INFN - Sezione di Torino,\\Via P. Giuria 1, 10125 Torino, Italy}
\affiliation[c]{Deutsches Elektronen-Synchrotron DESY,\\ Notkestr. 85, 22607 Hamburg, Germany}
\affiliation[d]{CPHT, CNRS, École Polytechnique, Institut Polytechnique de Paris, 91120 Palaiseau, France}
\affiliation[e]{Institut f\"ur Physik, Humboldt-Universit\"at zu Berlin and IRIS Adlershof,\\
Zum Großen Windkanal 2, 12489 Berlin, Germany}

\emailAdd{lorenzo.bianchi@unito.it, davide.bonomi@desy.de, maddalena.ferragatta@polytechnique.edu, valentina.forini@hu-berlin.de}

\abstract{We study two-point functions of chiral primary operators in $\mathcal{N}=4$ Super Yang-Mills in the presence of a half-BPS surface defect. Such a defect preserves only the diagonal combination of transverse rotations and of a global $U(1)$ R-symmetry, so that the correlator depends on one additional cross-ratio and splits into sectors of definite angular charge. We solve the superconformal Ward identities and construct the bulk and defect superconformal blocks in all sectors, including the charged ones, which have no analogue for defects preserving transverse rotations. The solution contains a holomorphic sector, which we fix non-perturbatively by mapping it to a chiral algebra correlator. We extend this map to defects breaking transverse rotations, where it requires a redefinition of the modes of the two-dimensional current algebra, and use it to fix the charged sectors as well. We then use the defect dispersion relation to reconstruct the full correlator of the dimension-two chiral primary at leading order at large $N$ and strong coupling. In the charged sectors long multiplets are absent and the answer is conjecturally exact in both $N$ and the coupling; expanding it at weak coupling, we find perfect agreement with known results. From these correlators we extract infinitely many new pieces of defect CFT data. Finally, we reproduce the strong-coupling answer by an independent Witten diagram bootstrap on the rotation-invariant locus.}


\begin{document}

\thispagestyle{empty} 

\begin{flushright} 
\footnotesize{DESY-26-135}\\
\footnotesize{HU-EP-26/21}
\end{flushright}

\maketitle
\flushbottom

\newpage

\section{Introduction}

A defect is an excitation of a quantum field theory, which modifies the theory along an extended locus without
decoupling it from the ambient degrees of freedom.  The resulting operator
algebra is hybrid: bulk operators may fuse among themselves or be expanded in
operators living on the defect. Conformal symmetry turns this feature into a quantitative consistency
problem.  A bulk two-point function admits both a bulk and a defect OPE
expansion, and crossing symmetry places powerful constraints on the interaction of bulk and defect degrees of freedom \cite{Billo:2016cpy}. This is the main idea of the defect bootstrap program.

In a holographic setting~\cite{Maldacena:1997re}, the
same structure becomes an interplay between closed-string propagation and brane excitations. A rich yet tractable setting to explore this interaction is provided by half-BPS defects in $\mathcal{N}=4$ Super Yang-Mills (SYM) theory. In the dual supergravity theory, perturbative results can be obtained using Witten diagrams in the presence of branes \cite{Giombi, Bianchi:2020hsz,Gimenez-Grau:2023fcy,Carmi:2026mhc,Castellani:2026wwh}, but direct computations of this kind quickly become technically demanding, both in deriving the relevant interaction vertices and in evaluating the resulting integrals. On the other hand, combining analytic bootstrap methods \cite{Lemos:2017vnx,Liendo:2019jpu,Bianchi:2022ppi,Barrat:2022psm} with superconformal Ward identities and supersymmetric localization \cite{Pestun:2007rz} has proved powerful enough to fix strong-coupling results for the half-BPS Wilson line \cite{Barrat:2021yvp,Bianchi:2022ppi,Barrat:2022psm,Artico:2024wnt} as well as for surface defects in six dimensions \cite{Meneghelli:2022gps,Chen:2023yvw,Chen:2024orp}.\footnote{Similar techniques can be used to bootstrap statistical impurities in the $\epsilon$-expansion \cite{Gimenez-Grau:2022ebb,Bianchi:2022sbz,Bianchi:2023gkk}.}

Surface defects in $\mathcal{N}=4$ SYM are, in many respects, the natural next case. This example, however, is not just a straightforward extension of those mentioned above. What makes surface defects qualitatively different from lines and boundaries is the fate of the transverse rotation group. A half-BPS surface defect in $\mathcal{N}=4$ SYM preserves only the diagonal combination of transverse rotations and of a global $U(1)$ R-symmetry. As a consequence, one-point functions acquire a dependence on an angle, and two-point functions depend on an additional cross-ratio with respect to the more familiar case of defects that preserve the two $U(1)$'s separately. Invariance under transverse rotations has always been assumed in previous bootstrap analyses and the lack of it has important consequences for the construction of the conformal blocks and the crossing equation. This is the main novelty of the present analysis.

The first explicit realization of half-BPS surface defects in $\mathcal{N}=4$ SYM is due to Gukov and Witten \cite{Gukov:2006jk,Gukov:2008sn}. They introduced codimension-two disorder operators, defined by prescribing a singular behavior of the fields along a two-dimensional surface; these defects are labeled by a Levi subgroup and by a set of continuous parameters, which are exactly marginal defect couplings. Equivalent order-type realizations exist in which two-dimensional degrees of freedom are coupled to the bulk \cite{Gaiotto:2009fs,Gaiotto:2013sma,Gomis:2014eya}; we refer to \cite{Gukov:2014gja} for a review of the two constructions.

What makes this family of defects particularly tractable is that several complementary non-perturbative methods apply to it. Supersymmetric localization reduces a large class of defect observables to finite-dimensional matrix integrals, or to two-dimensional Yang-Mills theory on the sphere \cite{Pestun:2007rz,Pestun:2009nn}. For Gukov-Witten defects it determines the expectation value of the surface operator and the one-point functions of chiral primary operators exactly in $N$ and in the 't~Hooft coupling \cite{Drukker:2008wr,Choi:2024ktc} and, as we show here, it also fixes a topological component of the two-point function. A second handle is the chiral algebra construction of \cite{Beem:2013sza}, which maps the Schur sector of a four-dimensional $\mathcal{N}=2$ theory to a two-dimensional chiral algebra. A superconformal surface defect becomes a module over that algebra \cite{Cordova:2017mhb,Pan:2017zie,Bianchi:2019sxz}, so that an infinite amount of protected defect data is computed by purely two-dimensional means. At the level of the Schur index, surface defects of Gukov-Witten type are conjectured to correspond to non-vacuum modules of the associated vertex operator algebra, whose characters are the defect indices \cite{Pan:2021ulr,Zheng:2022zkm}. Finally, the integrability techniques developed for one-point functions in defect CFT \cite{deLeeuw:2015hxa,Cavaglia:2021bnz} have recently been extended to surface operators, where integrable loci in the space of Gukov-Witten parameters have been identified \cite{Chalabi:2025nbg,Holguin:2025bfe}. The first two of these methods play a central role in this work, both as input for the bootstrap and as independent checks on the final answer.

At large $N$ and strong 't~Hooft coupling, the Gukov-Witten defect admits a dual description in terms of a probe D3-brane wrapping $\mathrm{AdS}_3 \times S^1 \subset \mathrm{AdS}_5 \times S^5$, with its embedding moduli and worldvolume holonomies encoding the defect parameters \cite{Constable:2002xt,Drukker:2008wr}. This description holds when the number of defect branes is kept fixed as $N\to\infty$; when it is instead comparable to $N$, the backreaction cannot be neglected and the defect is described by the bubbling supergravity solutions of \cite{Gomis:2007fi}. We work throughout in the probe regime, which we review in section \ref{sec:holo}. The chiral primary operators $\mathcal{O}_p$ are dual to Kaluza-Klein modes of type IIB supergravity, and their two-point functions in the presence of the defect are therefore represented by AdS bulk-exchange, D3-worldvolume-exchange, and brane-localized contact Witten diagrams~\cite{Gimenez-Grau:2023fcy}.

In this work we study the two-point function of chiral primary operators in the presence of a half-BPS
surface defect.  Building on \cite{Holguin:2025bfe}, we provide a full derivation of bulk and defect superconformal blocks in all allowed charged sectors.  In particular, the
charged bulk blocks are new and they can be mapped to bulk blocks for
non-identical external scalars, with the difference of external dimensions
identified with the charge.  We also analyze the protected subsector using the chiral algebra construction, which fixes the
holomorphic part of the correlator. For the Gukov-Witten defects, we also perform a second topological twist, and we relate the resulting constant to the localization result in two-dimensional Yang-Mills.

Our main dynamical tool is the defect conformal dispersion relation of
\cite{Bianchi:2022ppi,Barrat:2022psm}.  It reconstructs the correlator from its discontinuities at the two lightcone cuts.  At the leading
non-trivial order in $1/N$ at strong
coupling, only the bulk identity and the stress-tensor multiplet contribute to
these discontinuities.  The remaining low-transverse-spin ambiguity is fixed
by the Ward identities, consistency with the chiral algebra, crossing and
analyticity. Remarkably, in the
charged sectors long multiplets are absent altogether, and we find a result that we conjecture to be exact at finite $N$ and arbitrary
coupling.  Its agreement with the weak-coupling calculation of
\cite{Holguin:2025bfe} provides a strong non-trivial check.

Finally, we give an independent holographic derivation on the
rotation-invariant locus.  Rather than extracting all interaction vertices
from type IIB supergravity and the D3-brane action, we bootstrap the most
general combination of bulk-exchange, defect-exchange, and contact Witten
diagrams compatible with the supergravity and brane spectra.  Supersymmetry
fixes their relative coefficients, and one protected bulk-OPE coefficient
fixes the overall normalization.  The resulting position-space correlator
agrees exactly with the dispersion-relation answer.

The present analysis can be readily extended to two-point functions of arbitrary chiral primary operators $\mathcal{O}_p$, as well as to more general half-BPS surface defects, since it relies solely on symmetry and general analytic properties.

\subsection*{Summary of results}\label{sec:intro-summary}

Since the analysis is somewhat technical, we collect here the main results, referring to the body of the paper for the derivations and for all definitions. Throughout, $\mathcal{O}_2$ is the dimension-two superprimary of the stress-tensor multiplet, transforming in the $[0,2,0]$ representation of $SU(4)_R$, and $\Sigma$ denotes the surface defect. The general form of the two-point function is given in \eqref{2pt}. Its dependence on the fifth cross-ratio $\vartheta$, which is present because the defect preserves only a combination of transverse rotations and transverse R-symmetry, is resolved by the Fourier decomposition \eqref{Fourier}. For $\mathcal{O}_2$ only the charges $q=0,\pm2,\pm4$ occur, so that
\begin{equation}\label{intro:fourier}
 \langle \mathcal{O}_2(P_1,Y_1)\,\mathcal{O}_2(P_2,Y_2)\rangle_\Sigma
 =\frac{(Y_1\!\circ\! Y_1)(Y_2\!\circ\! Y_2)}{(P_1\!\circ\! P_1)(P_2\!\circ\! P_2)}
 \sum_{q=0,\pm2,\pm4}e^{iq\frac{\vartheta}{2}}\,
 \mathcal{F}^{(q)}(z,\bar z,\alpha,\bar\alpha)\, .
\end{equation}
The superconformal Ward identities for this two-point function were derived in  \cite{Holguin:2025bfe} and we report them in \eqref{WI}.
We solve them separately in each sector and we write the general solution as \begin{eqnarray}\label{intro:wardsolution}
&&\mathcal{F}^{(q)}(z,\bar z,\alpha,\bar\alpha) = \mathcal{C}^{(q)} + \frac{(z-\alpha)(\bar\alpha z-1)(\alpha\bar z-1)(\bar z-\bar\alpha)}{\alpha \bar \alpha \, z \bar z} F^{(q)}(z,\bar z,\alpha,\bar\alpha)\\
&&+\frac{ (\bar\alpha z-1)(\alpha\bar z-1)
\left[f^{(q)}(z,\alpha)+f^{(q)}(\bar z,\bar\alpha)\right] }{ (\alpha\bar\alpha-1)(z\bar z-1) }-\frac{ (z-\alpha)(\bar z-\bar\alpha) \left[f^{(q)}\!\left(z,\frac{1}{\bar\alpha}\right) +f^{(q)}\!\left(\bar{z},\frac{1}{\alpha}\right)\right] }{ (\alpha\bar\alpha-1)(z\bar z-1)}\,, \nonumber
\end{eqnarray}
where $\mathcal{C}^{(q)}$ are constants and $f^{(q)}(z,\alpha)$ are holomorphic functions that are related to the chiral algebra construction of \cite{Beem:2013sza}. We also construct the bulk and defect superconformal blocks in all sectors, see tables \ref{tab:blocks}, \ref{tab:blocksq} and \ref{tab:defblocks}. The charged blocks are not simply related to the neutral ones: the bulk blocks coincide with blocks for external operators of unequal dimensions, with the difference of dimensions set by $q$, while the defect blocks depend on the angular cross-ratio $\omega$ of \eqref{udef} through the bulk-to-defect couplings.

Our results are expressed in terms of protected defect data: the Fourier modes $a^{(q)}_{\mathcal{O}_2}$ of the one-point function \eqref{1ptfourier}, the shorthand $a\lambda^{(q)}_{\mathcal{B}_{[0,2,0]}}=\lambda_{\mathcal{B}_{[0,2,0]}}\,a^{(q)}_{\mathcal{O}_2}$ of \eqref{alambda} associated with the exchange of the stress-tensor multiplet in the bulk channel, and the topological constant $\mathcal{C}^{(0)}$. For a Gukov-Witten defect all of them are known in closed form: the $a^{(q)}_{\mathcal{O}_2}$ are given by localization in \eqref{a2broken}, the bulk OPE coefficient $\lambda_{\mathcal{B}_{[0,2,0]}}=\lambda_{\mathcal{O}_2\mathcal{O}_2\mathcal{O}_2}$ in \eqref{lambdaO2}, and we obtain
\begin{eqnarray}\label{intro:Cloc}
\mathcal{C}^{(0)}&=&-\left[a^{(0)}_{\mathcal{O}_2}\right]^2-2a^{(+2)}_{\mathcal{O}_2} a^{(-2)}_{\mathcal{O}_2} -\frac14+ \frac12\, \lambda_{\mathcal{B}_{[0,2,0]}}\,a^{(0)}_{\mathcal{O}_2}-\frac14\, \sum_{l=1}^M \frac{N_l(N-N_l)}{N^2}\,,\nonumber\\
\mathcal{C}^{(\pm 2)}&=&-2a^{(0)}_{\mathcal{O}_2} a^{(\pm 2)}_{\mathcal{O}_2} + \frac12 \lambda_{\mathcal{B}_{[0,2,0]}}\,a^{(\pm2)}_{\mathcal{O}_2}\,,\\
\mathcal{C}^{(\pm4)}&=& -\left[a^{(\pm 2)}_{\mathcal{O}_2}\right]^2\,,\nonumber
\end{eqnarray}
by extending the localized two-dimensional Yang-Mills computation to the bulk two-point function, see \eqref{Cloc}. Here $M$ is the number of blocks of the Levi subgroup \eqref{levi} and $N_l$ are the ranks of its factors.

\paragraph{Exact holomorphic correlators.} The functions $f^{(q)}$ are fixed by protected data alone, to all orders in $1/N$ and in the 't~Hooft coupling. In the neutral sector we find \eqref{smallf0},
\begin{equation}\label{intro:smallf0}
\begin{aligned}
f^{(0)}(z,\alpha)=\ &-\frac14-2a^{(+2)}_{\mathcal{O}_2} a^{(-2)}_{\mathcal{O}_2}-\mathcal{C}^{(0)}
+a^{(+2)}_{\mathcal{O}_2} a^{(-2)}_{\mathcal{O}_2}\left(\frac{\alpha}{z}+\frac{z}{\alpha}\right)\\
&+\frac{z(1-\alpha)^2}{4\alpha(1-z)^2}
-a\lambda_{\mathcal{B}_{[0,2,0]}}^{(0)}\frac{(\alpha+1)(z-\alpha)}{4\alpha(1-z)}\, ,
\end{aligned}
\end{equation}
while in the charged sectors \eqref{smallf2},
\begin{eqnarray}\label{intro:smallf2}
f^{(\pm2)}(z,\alpha)&=&a^{(0)}_{\mathcal{O}_2} a^{(\pm 2)}_{\mathcal{O}_2}
\left(\sqrt{\frac{z}{\alpha}}+\sqrt{\frac{\alpha}{z}}\ \right)
-\frac{a\lambda_{\mathcal{B}_{[0,2,0]}}^{(\pm 2)}}{2}\frac{\sqrt{z}\,(1-\alpha)}{(1-z)\sqrt{\alpha}}\,,\nonumber\\
f^{(\pm4)}(z,\alpha)&=&\left[a^{(\pm2)}_{\mathcal{O}_2}\right]^2 .
\end{eqnarray}

\paragraph{Chiral algebra and spectral flow.} These holomorphic results have a purely two-dimensional origin through the chiral algebra construction of \cite{Beem:2013sza}. The Schur sector of the theory is governed by an affine $\mathfrak{su}(2)$ current algebra at level $2 k$, and the defect defines a module over it. We explore how the breaking of $\mathcal{R}_+$ affects this module and we find that the defect state is not an affine primary of the original algebra, but it becomes one after a spectral flow by one unit \eqref{spectralflow},
\begin{equation}\label{intro:spectralflow}
	\widetilde J^{+}_{n}=J^{+}_{n+1}\, ,
	\qquad
	\widetilde J^{3}_{n}=J^{3}_{n}+k\,\delta_{n,0}\, ,
	\qquad
	\widetilde J^{-}_{n}=J^{-}_{n-1}\, .
\end{equation}
In the flowed frame all the components of the current acquire a standard simple-pole one-point function, whose residue \eqref{flowedonept} encodes the charged coefficients $a^{(\pm2)}_{\mathcal{O}_2}$. 

\paragraph{Charged sectors.} No long supermultiplet can contribute when $q\neq0$. The charged correlators are therefore completely determined by the holomorphic data above, and we obtain \eqref{chargedres},
\begin{equation}\label{intro:chargedres}
\begin{aligned}
\mathcal{F}^{(\pm2)}(z,\bar z,\alpha,\bar \alpha)&= a^{(0)}_{\mathcal{O}_2} a^{(\pm 2)}_{\mathcal{O}_2} \left(\sqrt{\frac{\bar{\alpha} z}{\alpha \bar z}}+\sqrt{\frac{\alpha \bar z}{\bar{\alpha} z}}\,\right)-\frac{a\lambda_{\mathcal{B}_{[0,2,0]}}^{(\pm 2)}}{2}\frac{\sqrt{z \bar{z}}(1-\alpha) (1-\bar{\alpha}) }{(1-z) (1-\bar{z}) \sqrt{\alpha  \bar{\alpha}}}\,, \\
\mathcal{F}^{(\pm4)}(z,\bar z,\alpha,\bar \alpha)&=\left[a^{(\pm 2)}_{\mathcal{O}_2}\right]^2\, .
\end{aligned}
\end{equation}
We conjecture that \eqref{intro:chargedres} is exact in both $N$ and the coupling. Its agreement with the weak-coupling computation of \cite{Holguin:2025bfe}, which we check in appendix \ref{app:weak}, is a strong test of the conjecture. In the maximally charged sector the correlator reduces to the product of two one-point functions.

\paragraph{Neutral sector at strong coupling.} In the sector $q=0$ long multiplets do contribute, and we reconstruct the correlator from the dispersion relation of \cite{Bianchi:2022ppi,Barrat:2022psm}. At the leading non-trivial order at large $N$ and strong coupling only the bulk identity and the stress-tensor multiplet contribute to the discontinuity. To fix the residual low-spin ambiguity we need to make some further assumptions, which are detailed in section \ref{sec:bootstrapO2}. Under these assumptions, the result \eqref{q0result} is
\begin{align}\label{intro:q0result}
\mathcal F^{(0)}(z,\bar z,\alpha,\bar\alpha)
= &
\frac{z\bar z(1-\alpha)^2(1-\bar\alpha)^2}
{4\alpha\bar\alpha(1-z)^2(1-\bar z)^2}
+\bigl[a_{\mathcal O_2}^{(0)}\bigr]^2
+a_{\mathcal O_2}^{(+2)}a_{\mathcal O_2}^{(-2)}
\left(\frac{\bar z\alpha}{z\bar\alpha}
     +\frac{z\bar\alpha}{\bar z\alpha}\right)\nonumber\\
    &-\frac{a\lambda_{\mathcal B_{[0,2,0]}}^{(0)}}{4}
\Biggl[
-1+\frac{(1-\alpha)(1-\bar\alpha)}
{(1-z)(1-\bar z)}
\left(\frac{z}{\alpha}+\frac{\bar z}{\bar\alpha}\right)\nonumber\\
&+\frac{(z-\alpha)(\bar z-\bar\alpha)
(1-\bar z\alpha)(1-z\bar\alpha)}
{\alpha\bar\alpha(1-z)(1-\bar z)}
\frac{1-z^2\bar z^2+2z\bar z\log(z\bar z)}
{(1-z\bar z)^3}
\Biggr]
\\
&+\left(
\frac14 +\mathcal C^{(0)}
+\bigl[a_{\mathcal O_2}^{(0)}\bigr]^2
+2a_{\mathcal O_2}^{(+2)}a_{\mathcal O_2}^{(-2)}
- \frac{a\lambda_{\mathcal B_{[0,2,0]}}^{(0)}}{4}
\right)
\nonumber \\
&\times\Biggl[
\frac{(z-\alpha)(\bar z-\bar\alpha)
(1-\bar z\alpha)(1-z\bar\alpha)}
{\alpha\bar\alpha}\frac{
1+10z\bar z+z^2\bar z^2}
{(1-z\bar z)^4}\nonumber \\
&+\frac{(z-\alpha)(\bar z-\bar\alpha)
(1-\bar z\alpha)(1-z\bar\alpha)}
{\alpha\bar\alpha}\frac{6z\bar z(1+z\bar z)\log(z\bar z)}
{(1-z\bar z)^5}
-1\Biggr]+\mathcal{O}(1/N^2)\, .\nonumber
\end{align}
The first line collects the bulk identity and the factorized one-point functions, the term proportional to $a\lambda^{(0)}_{\mathcal{B}_{[0,2,0]}}$ is reconstructed from the protected $\mathcal{B}_{[0,2,0]}$ exchange, and the last three lines are the unique allowed low-spin ambiguity. The combination multiplying them starts at order $1/N$, because the terms that would dominate cancel against the disconnected part of $\mathcal{C}^{(0)}$ in \eqref{intro:Cloc}. Expanding the correlators in either channel produces infinite families of new defect CFT data, for both protected and unprotected operators. These results are collected in tables~\ref{tab:bulkOPEdata}--\ref{tab:defOPEdata2}. 

\paragraph{Holographic check.} On the rotation-invariant locus $\beta_l=\gamma_l=0$ we reproduce the neutral correlator by a position-space Witten diagram bootstrap. The allowed exchanges are dictated by the type IIB Kaluza-Klein spectrum and by the spectrum of D3-brane fluctuations, and the superconformal Ward identities fix all relative coefficients \eqref{coeffans}. The single remaining normalization is determined by the bulk OPE limit \eqref{normans},
\begin{equation}\label{intro:normans}
    A_s= - \frac12 a \lambda_{\mathcal{B}_{[0,2,0]}}^{(0)}= - \frac12 \lambda_{\mathcal{O}_2 \mathcal{O}_2 \mathcal{O}_2} a_{\mathcal{O}_2}^{(0)} = -\frac{1}{N} +\mathcal{O}(1/N^2)\, ,
\end{equation}
in the single probe brane limit. The resulting sum of Witten diagrams agrees with \eqref{intro:q0result}, providing an independent strong-coupling derivation and making explicit the relation between the defect dispersion relation and Polyakov-Regge blocks.

\vspace{0.5cm}

The remainder of the paper is organized as follows. In section \ref{sec:setup1pt}, we review the setup and the relevant properties of half-BPS surface defects in $\mathcal{N}=4$ SYM. We also discuss one-point functions of chiral primary operators $\mathcal{O}_p$ in the presence of the defect. In section \ref{sec:2ptkine}, we analyze the kinematics of two-point functions of $\mathcal{O}_p$. In particular, we solve the superconformal Ward identities and discuss the bulk and defect superconformal block expansions. Moreover, we determine the protected sector of the correlator from the chiral algebra and supersymmetric localization. In section \ref{sec:bootstrapO2}, we apply analytic bootstrap techniques to reconstruct the two-point function of $\mathcal{O}_2$ at large $N$ and strong coupling and extract the corresponding CFT data. We also discuss exact results and the strategy to fix the low-spin ambiguities that affect the dispersion relation. 
Finally, in section \ref{sec:holo-witten}, we reproduce the rotation-invariant correlator with a holographic Witten diagram bootstrap. In section \ref{sec:concl}, we present our conclusions and outline possible future directions. Conventions and additional technical details are collected in the appendices.

\section{Setup and one-point functions of bulk operators}\label{sec:setup1pt}
\subsection{Half-BPS surface defects in \texorpdfstring{$\mathcal{N}=4$}{N=4} SYM}
We consider $\mathcal{N}=4$ Super Yang-Mills (SYM) theory with gauge group $U(N)$ in the presence of a surface defect $\mathcal{O}_\Sigma$ supported on  a two-dimensional plane $\Sigma$. We focus on $1/2$ BPS surface defects, which preserve a subalgebra of the bulk superconformal symmetry
\begin{equation}\label{symbreak}
\mathfrak{psu}(1,1|2)\oplus\mathfrak{psu}(1,1|2)\oplus \mathfrak{u}(1)_- \, ,
\end{equation}
where the two $\mathfrak{psu}(1,1|2)$ factors form the global part of the two-dimensional $\mathcal{N}=(4,4)$ superconformal algebra, centrally extended by $\mathfrak{u}(1)_{-}$ (see appendix \ref{app:conv} for details on the preserved superconformal algebra). At the level of the bosonic subgroup, the four-dimensional conformal group $SO(2,4)$ is broken to $SO(2,2)\times SO(2)_{\mathcal{M}_{\perp}}$, i.e. the product of the two-dimensional conformal group and orthogonal rotations. Similarly, the R-symmetry group $SU(4)_R \cong SO(6)_R$ is broken to $SO(4)_\mathcal{R}\times SO(2)_{\mathcal{R}_\perp}$. Computing the commutators of the preserved generators, one finds that the preserved superalgebra (see \eqref{QS}) contains a central extension 
\begin{equation}
\mathcal{R}_-=\mathcal{R}_{\perp}-\mathcal{M}_{\perp}\,.
\end{equation}
The opposite combination $\mathcal{R}_+=\mathcal{R}_{\perp}+\mathcal{M}_{\perp}$ acts as an outer automorphism on the superalgebra (see \eqref{automorph}) and it may or may not be preserved by the defect. In this work we will consider the general case when the additional $SO(2)$ generated by $\mathcal{R}_+$ is not preserved, i.e. the defect does not preserve transverse rotations and ``transverse'' R-symmetry independently, but only a diagonal subgroup.

\subsubsection{Gukov-Witten realization}
A prominent realization of such defects is provided by Gukov-Witten surface operators \cite{Gukov:2006jk}. These are \emph{disorder} operators: rather than being built out of the fields of the theory, they are defined by declaring that the path integral is performed over field configurations with a prescribed singularity along the two-dimensional surface $\Sigma$.

To identify the admissible singularities, we split the four-dimensional spacetime as $\Sigma\times\mathbb{R}^2_\perp$ and we introduce a complex coordinate $\zeta=\rho\,e^{i\theta}$ on the transverse plane. Of the six real scalars of $\mathcal{N}=4$ SYM it is convenient to single out the complex combination $Z = \tfrac{\Phi_1 + i \Phi_2}{\sqrt{2}}$, whose phase rotation generates the $SO(2)_{\mathcal{R}_\perp}$ that will be tied to transverse rotations\footnote{$Z$ carries charge $-1$ under the generator $\mathcal{R}_\perp$, i.e. $[\mathcal{R}_\perp,Z]=-Z$.}. Requiring the configuration to preserve half of the supercharges, and to be independent of the coordinates along $\Sigma$, reduces the BPS equations to a Hitchin system on the transverse plane, whose singular solution is, in a gauge in which the radial component of the gauge field vanishes \cite{Gukov:2006jk},
\begin{equation}\label{classprof}
Z = \frac{\beta+i\gamma}{\sqrt{2} \zeta}\,, \qquad A = \alpha\, d\theta \,. 
\end{equation}
The parameters $\alpha$, $\beta$, and $\gamma$ are mutually commuting matrices valued in the Lie algebra of $U(N)$, and can therefore be simultaneously brought to block diagonal form
\begin{eqnarray}
    \alpha&=& \begin{pmatrix}
\alpha_1\otimes   \mathds{1}_{N_1}& 0 &\ldots &0\\
0 & \alpha_2\otimes   \mathds{1}_{N_2}&  \ldots & 0\\
\vdots& \vdots & \ddots & \vdots\\
0&0&\ldots& \alpha_M\otimes   \mathds{1}_{N_M}
\end{pmatrix}\,, \\
\beta+i\gamma&=&\begin{pmatrix}
(\beta_1+i \gamma_1)\otimes   \mathds{1}_{N_1}& 0 &\ldots &0\\
0 & (\beta_2+i\gamma_2)\otimes   \mathds{1}_{N_2}&  \ldots & 0\\
\vdots& \vdots & \ddots & \vdots\\
0&0&\ldots& (\beta_M+i \gamma_M)\otimes   \mathds{1}_{N_M}
\end{pmatrix}\,. \label{alphabetagammadef}
\end{eqnarray}
The three parameters play rather different roles. The only gauge-invariant datum carried by $\alpha$ is the monodromy $\exp(2\pi i \alpha)$ of the gauge field around the defect, so that $\alpha$ is defined modulo the identification $\alpha_l\simeq \alpha_l+1$. In contrast, $\beta$ and $\gamma$ take values in a Cartan subalgebra of $U(N)$ and their eigenvalues $\beta_l$, $\gamma_l$ are noncompact. They control the scalar profile and, as we will see, they are the parameters that will be relevant in all our final formulas. One may also introduce an additional parameter $\eta$,\footnote{The Gukov-Witten parameters $\alpha$ and $\eta$ that appear in this section should not be confused with the cross-ratios defined in \eqref{lightcone} and \eqref{etatheta} which are used throughout the rest of the paper.} which couples to a theta term localized on the
defect \cite{Gukov:2006jk}. This parameter will not be relevant for the analysis presented in the remainder of this work. All four parameters are dimensionless and, modulo the action of the Weyl group, they define a moduli space for the surface operator \cite{Gukov:2006jk}. As we will see, this is consistent with the existence of SUSY protected marginal operators in the defect spectrum.

The singular profile \eqref{classprof} breaks the gauge symmetry group $U(N)$ to the Levi subgroup that commutes with $\alpha$, $\beta$ and $\gamma$,
\begin{equation}\label{levi}
    \mathbb L=\prod_{l=1}^M U(N_l)\,, \quad \text{with } \sum_{l=1}^M N_l=N\,,
\end{equation}
the pattern of symmetry breaking being controlled by the coincidences among the eigenvalues. The maximal case $M=N$ corresponds to the full breaking to the maximal torus, while $M=1$ corresponds to $\mathbb{L}=U(N)$, for which the gauge symmetry is unbroken.

The residual spacetime and R-symmetry invariance of \eqref{classprof} is precisely the one advertised in \eqref{symbreak}. The profile is invariant under translations and conformal transformations along $\Sigma$. A transverse rotation, instead, is a symmetry only if accompanied by a compensating R-symmetry rotation, and the unbroken generator is the diagonal combination $\mathcal{R}_-=\mathcal{R}_\perp-\mathcal{M}_\perp$.  When $\beta=\gamma=0$ the scalar profile is absent and both $SO(2)$'s are preserved separately. At the quantum level, the Gukov-Witten surface defect $\mathcal{O}_\Sigma$ is defined by restricting the path integral to field configurations that satisfy the singular boundary conditions \eqref{classprof} along the support of the defect $\Sigma$.

The Gukov-Witten construction is of \emph{disorder} type: the defect is specified by a
singularity of the four-dimensional fields, and no new degrees of freedom are introduced on
$\Sigma$. This is not the only possibility. Surface defects can equally well be defined as
\emph{order} operators, by placing a two-dimensional field theory with flavor symmetry $G$ on
$\Sigma$ and gauging that symmetry through the coupling to the four-dimensional gauge fields;
we refer to \cite{Gukov:2014gja} for a review of the two constructions. In
$\mathcal{N}=2$ theories, coupled two-dimensional/four-dimensional systems of this type have been
studied systematically \cite{Gaiotto:2009fs,Gaiotto:2013sma,Gomis:2014eya}. Within
$\mathcal{N}=4$ SYM, an explicit Lagrangian realization of a half-BPS surface defect preserving a
two-dimensional $(4,4)$ superconformal algebra --- that is, precisely the symmetry
\eqref{symbreak} --- is provided by two stacks of D3-branes intersecting along two common
directions, which localizes hypermultiplets on $\Sigma$ \cite{Constable:2002xt}. We stress,
however, that order-type constructions need not preserve \eqref{symbreak}: brane intersections
realizing chiral surface defects, such as the $(0,8)$ defect of \cite{Harvey:2008zz}, preserve a
different superconformal algebra and fall outside the scope of the present analysis.

\subsubsection{Holographic picture}\label{sec:holo}
The Gukov-Witten surface defect admits a holographic realization, valid at large $N$ and strong 't Hooft coupling $\lambda$,  in terms of D3-branes  embedded in the $AdS_5\times S^5$ background of type IIB string theory \cite{Gukov:2006jk,Drukker:2008wr}. This system can be analyzed in two complementary regimes. When the number of
D3-branes is small, they can be treated as probes in a fixed $AdS_5\times S^5$ background. Their worldvolumes reach the conformal boundary along the support $\Sigma$ of the defect \cite{Drukker:2008wr}. 
When the number of defect branes is large and comparable to $N$,  their backreaction must be taken into account and the appropriate description is in terms of certain type IIB supergravity solutions known as bubbling geometries \cite{Gomis:2007fi}.

In this work we are interested in the probe approximation.
The integers $N_l$ in the Levi subgroup \eqref{levi} specify the ranks of the unbroken gauge-theory factors,
with $\sum_{l=1}^{M}N_l=N$.
For the $U(1)\times U(N-1)$ configuration considered below,
$N_1=1$ and $N_2=N-1$ with the rank-one factor representing a single probe D3-brane. 
The dynamics of the probe is governed by the Dirac-Born-Infeld action coupled to the background Ramond-Ramond flux. This description provides a direct geometric realization of the defect degrees of freedom. The probe D3-brane worldvolume is $AdS_3\times S^1\subset AdS_5\times S^5$~\footnote{This embedding was first studied in the holographic description of intersecting D3-branes \cite{Constable:2002xt}.},
so that  a convenient parametrization of the geometry is obtained by slicing $AdS_5$ by $AdS_3\times S^1$ and $S^5$ by $S^3\times S^1$ as in
\begin{equation}
ds^2= \cosh^2 u\, ds_{AdS_3}^2+ du^2+ \sinh^2 u\, d\theta^2 + \cos^2 \chi\, ds^2_{S^3} + d\chi^2+ \sin^2 \chi\, d\theta_R^2\,,
\end{equation}
where the common radius of $AdS_5$ and $S^5$ is set to one.
The preserved symmetries are explicit and the angles are named in analogy with the gauge theory description: $\theta$ and $\theta_R$ span the two circles associated to transverse rotations and $SO(2)_{\mathcal{R}_\perp}$. The surface operator is supported at the conformal boundary of $AdS_3$.\footnote{The choice of global
coordinates on $AdS_3$ corresponds to a spherical defect $\Sigma=S^2$, while Poincar\'e coordinates describe a planar defect $\Sigma=\mathbb R^2$~\cite{Drukker:2008wr}.}
In this coordinate system the brane sits at a fixed $u=u_0$ and $\chi=\pi/2$, it wraps an $AdS_3$ slice and extends along the circle parametrized by \cite{Constable:2002xt}
\begin{equation}\label{BPSbulk}
\theta_R=\eta_0-\theta\,,\qquad\text{i.e.}\qquad
\eta\equiv\theta+\theta_R=\eta_0\,,
\end{equation}
for a constant $\eta_0$, where $\eta$ is the mixed angle of \eqref{etatheta}.
By locking the $S^1$ inside $AdS_5$ to the circle in $S^5$, this condition is the geometric origin of the symmetry breaking pattern discussed above. In the dual gauge theory, it corresponds to the fact that the defect preserves only a diagonal combination of transverse rotations and R-symmetry. 

The parameters of the surface defect are encoded in the moduli of the
probe-brane embedding and in the worldvolume gauge field.  For a single
probe D3-brane, the holographic dictionary of \cite{Drukker:2008wr} reads
\begin{equation}\label{u0def}
  \beta+i\gamma
  =
  \frac{\sqrt{\lambda}}{2\pi}\,
  \sinh u_0\, e^{i\eta_0}\,.
\end{equation}
The modulus and the phase of the complex combination in \eqref{u0def} thus have a direct geometric meaning: $|\beta+i\gamma|\propto\sinh u_0$ sets the
radial position of the brane, with $\sinh u_0$ being the proper radius of the $S^1\subset AdS_5$ it wraps, while $\arg(\beta+i\gamma)=\eta_0$ is the
constant value of the mixed angle $\eta$ along the singular profile
\eqref{classprof}, which the brane condition \eqref{BPSbulk} reproduces.
For this single-probe configuration, the parameters in \eqref{u0def} are those of the rank-one block, $\beta=\beta_1$ and $\gamma=\gamma_1$.
The remaining probe parameters, $\alpha_1$ and $\eta_1$, are encoded in the holonomies of its worldvolume gauge field and of its dual along the $S^1$.
Following \cite{Drukker:2008wr}, we set $\beta_2=\gamma_2=0$ for the rank-$(N-1)$ block and accordingly drop the label $l = 1$ of the probe block. 

The special locus $\beta=\gamma=0$, or equivalently $u_0=0$, is the one relevant for the Witten diagram bootstrap discussed in section \ref{sec:holo-witten}.  In this case the $S^1$ inside $AdS_5$ collapses and the brane becomes $AdS_3\subset AdS_5$ times an
internal circle in $S^5$.  This is the rotation-invariant configuration in which $SO(2)_{\mathcal M_\perp}$ and $SO(2)_{\mathcal R_\perp}$ are preserved separately.

Let us finally spell out the holographic meaning of the observables studied below.  Chiral primary operators (CPOs) are dual to Kaluza-Klein (KK) modes of type IIB supergravity on $S^5$ \cite{Witten:1998qj,Arutyunov:1999en}.\footnote{More precisely, KK modes are dual to a combination of a CPO and non-chiral composite operators, whose contributions are subleading at large $N$ \cite{Arutyunov:1999en}.} 
These correspond to fluctuations of the metric and Ramond-Ramond fields, decomposed into representations of the $SO(6)_R$ R-symmetry. A local operator insertion in $\mathcal{N}=4$ SYM is thus described by turning on the corresponding supergravity mode in $AdS$. 
In the presence of a surface defect, realized by a D3-brane, the correlator with a local operator is obtained by evaluating how this supergravity field couples to the brane. A one-point function is computed by the linear coupling of the
corresponding supergravity fluctuation in the expansion of the DBI plus Wess--Zumino
action.  Similarly, connected two-point functions of bulk operators are represented, at
leading order, by Witten diagrams with one brane insertion: contact interactions localized
on the D3-brane, bulk-exchange diagrams ending on a brane tadpole, and defect-exchange
diagrams mediated by D3-brane worldvolume fluctuations.  This geometric picture will be
used later to formulate the holographic Witten diagram bootstrap for the correlator of two
stress-tensor multiplet superprimaries. Beyond the probe regime, the same observables are described by studying fluctuations of fully backreacted supergravity solutions.

\subsection{One-point functions of protected operators}\label{sec:onept}
In this paper we probe $1/2$ BPS surface operators by studying correlation functions of local operators in the presence of the defect. Such correlators are strongly constrained by superconformal invariance. To make these constraints manifest, we parametrize the positions of the operators using the embedding space formalism \cite{Costa:2011mg} in the defect setting \cite{Billo:2016cpy} (see appendix \ref{app:conv} for a summary of our conventions).
We focus on correlators of single-trace
chiral primary operators $\mathcal{O}_p$, defined as
\begin{equation}\label{CPO}
  \mathcal{O}_p (P, Y)\colonequals \frac{(8\pi^2)^{p / 2}}{\lambda^{p/2}\sqrt{p}} \operatorname{Tr}\left[(Y \cdot \Phi(P))^p\right]\,,
\end{equation}
where $\Phi^M$ are the six adjoint real scalar fields appearing in the $\mathcal{N}=4$ SYM action, the trace is taken over the gauge indices, and $\lambda = g_{\mathrm{YM}}^2 N$ is the 't Hooft coupling. The prefactor is chosen such that the two-point function of the operator is normalized to one at large $N$. The operators \eqref{CPO} are half-BPS superprimaries of the short multiplets $\mathcal{B}_{[0,p,0]}$ of the $\mathfrak{psu}(2,2|4)$ superalgebra \cite{Dolan:2002zh}. They transform in the symmetric traceless representation $[0,p,0]$ of the $SO(6)_R$ R-symmetry group and their indices are contracted with a complex six-dimensional polarization vector $Y^M$, satisfying
\begin{equation}
 Y\cdot Y = 0\,.
\end{equation}
As usual, one can open the indices using a Todorov differential operator \cite{Dobrev:1975ru}. 

In the absence of a defect, their two- and three-point functions at large $N$ are 
\begin{eqnarray}\label{2pt3ptbulk}
    \langle \mathcal O_p(P_1,Y_1)\,\mathcal O_q(P_2,Y_2)\rangle &=& \delta_{pq}
\frac{(Y_1\!\cdot Y_2)^p}{(-2 P_{1} \cdot P_2)^p}\,, \\
\langle \mathcal O_{p_1}(P_1,Y_1)\, \mathcal O_{p_2}(P_2,Y_2)\, \mathcal O_{p_3}(P_3,Y_3)\rangle &=& 
\frac{
\lambda_{\mathcal{O}_{p_1}\mathcal{O}_{p_2}\mathcal{O}_{p_3}}(Y_1\cdot Y_2)^{\alpha_{3}}
(Y_2\cdot Y_3)^{\alpha_{1}}
(Y_3\cdot Y_1)^{\alpha_{2}}
}{
(-2P_{1} \cdot P_2)^{\alpha_{3}}
(-2P_2 \cdot P_3)^{\alpha_{1}}
(-2P_3 \cdot P_1)^{\alpha_{2}}
}\,, \nonumber
\end{eqnarray}
where\footnote{These exponents $\alpha_i$ should not be confused with the eigenvalues $\alpha_l$ of the Gukov-Witten parameter $\alpha$.} $\alpha_i=\tfrac{1}{2}\left(p_j+p_l-p_i\right)$ with $j \neq l \neq i$ and \cite{Lee:1998bxa}
\begin{equation}\label{lambdaO2}
    \lambda_{\mathcal{O}_{p_1} \mathcal{O}_{p_2} \mathcal{O}_{p_3}}= \frac{\sqrt{p_1 p_2 p_3}}{N}\,.
\end{equation}
A particularly important example is the $\mathcal{O}_2$ operator, which transforms in the $[0,2,0]$ (i.e.\ $\mathbf{20'}$) representation of $SU(4)_R$ and corresponds to the superprimary of the stress-tensor supermultiplet. This multiplet contains, among other operators, the stress-energy tensor, the $SU(4)_R$ currents, and the supercurrents. As a consequence, correlation functions involving $\mathcal{O}_2$ play a distinguished role in the study of the dynamical properties of the theory. In the case of $\mathcal{O}_2$, both \eqref{2pt3ptbulk} and \eqref{lambdaO2} are exact in $N$ and in the coupling.

The one-point function of CPOs in the presence of a $1/2$ BPS surface defect takes the form \cite{Holguin:2025bfe}~\footnote{Throughout this paper, all correlators in the presence of the defect are normalized by the expectation value of the defect.} 
\begin{equation}\label{1ptbroken}
\langle \mathcal O_p(P,Y)\rangle_{\Sigma} \colonequals \frac{\langle \mathcal{O}_\Sigma \ \mathcal{O}_p (P,Y)\rangle}{\langle \mathcal{O}_\Sigma\rangle} = \frac{\left(Y\!\circ\! Y\right)^{p/2}}{\left(P\!\circ\! P\right)^{p/2}}\, A_{\mathcal{O}_p}(\eta)\,,
\end{equation}
where the contraction $Y\circ Y$ accounts for the R-symmetry breaking $SO(6)_R\to SO(4)_{\mathcal{R}}\times SO(2)_{\mathcal{R}_{\perp}}$ in a way that is completely analogous to the spacetime symmetry breaking $SO(2,4)\to SO(2,2)\times SO(2)_{\mathcal{M}_{\perp}}$, encoded in the product $P\circ P$.  The angle $\eta\colonequals\theta+\theta_R$\footnote{Our definition of $\eta$ differs from that of \cite{Holguin:2025bfe}, where $\eta_{\text{there}}=\theta-\theta_R$, in order to be consistent with \eqref{classprof}.}, where $\theta$ and $\theta_R$ denote the angular coordinates in the transverse spacetime and R-symmetry planes respectively (see \eqref{etatheta}), is associated with the breaking of the generator $\mathcal{R}_+$. This is in contrast with the more familiar case of defects that preserve transverse rotations, where symmetry forbids any angular dependence and the one-point function depends only on the distance from the defect. We can decompose the one-point function in Fourier modes as
\begin{equation}\label{1ptfourier}
A_{\mathcal{O}_p}(\eta) = \sum_q a^{(q)}_{\mathcal{O}_p}\, e^{i q\eta}\,, \quad q=-p, -p+2, \ldots, p\, .
\end{equation}
The sum is truncated because the l.h.s. of \eqref{1ptbroken} is a polynomial in the $Y$'s and $\eta$ depends on them through the angle $\theta_R$. In other words, only values of charge appearing in the decomposition of $[0,p,0]$ under the symmetry breaking pattern $SO(6)_R \rightarrow SO(4)_\mathcal{R} \times SO(2)_{\mathcal{R}_{\perp}}$ are allowed. As an example, which will be relevant for the rest of this work, the one-point function of the $\mathcal{O}_2$ operator takes the form 
\begin{equation}\label{1ptO2}
    A_{\mathcal{O}_2}\!\left(\eta\right)= a^{(0)}_{\mathcal {O}_2}+ a^{(2)}_{\mathcal {O}_2}\,e^{2i\eta}+ a^{(-2)}_{\mathcal {O}_2}e^{-2i\eta}\, .
\end{equation}
The coefficients $a^{(q)}_{\mathcal{O}_p}$  depend on the parameters specifying the surface operator. 

In the case of a Gukov-Witten defect, these coefficients can be computed exactly using supersymmetric localization \cite{Choi:2024ktc}.
Remarkably, the result is a finite polynomial in the inverse ’t Hooft coupling $1/\lambda$. Moreover, in the large-$N$ limit and for arbitrary values of $\lambda$, it agrees exactly with the supergravity computation \cite{Drukker:2008wr}. For $\mathcal{O}_2$ the result reads\footnote{In the notation of \cite{Choi:2024ktc}, $A_{\mathcal{O}_2}\propto \frac12 \left(\sqrt{3} \, C_{2,0}+ C_{2,-2}\,e^{2i\eta}+ C_{2,2}\, e^{-2i\eta}\right)$. The prefactor of the rotation-invariant contribution originates from the decomposition of the $SO(4)_\mathcal{R}$-singlet component of $\mathcal{O}_2$ in terms of $\mathcal{O}_{2,q}$, see equations (2.7) and (4.56) of \cite{Choi:2024ktc}. There are relative sign differences in the one-point functions since we are using the Hermitian convention of \cite{Drukker:2008wr}, as opposed to the anti-Hermitian convention used in \cite{Choi:2024ktc}.} \cite{Drukker:2008wr, Choi:2024ktc}
\begin{align}\label{a2broken}
    a^{(0)}_{\mathcal {O}_2} =\frac{2 \pi^2}{\sqrt{2} \lambda}\left[\sum_{l=1}^M N_l\!\left(2\,(\beta^2_l+\gamma^2_l)+\frac{\lambda}{4\pi^2} \frac{N-N_l}{N}\right)\!\right]\,, \quad
      a^{(\pm2)}_{\mathcal {O}_2} =\frac{2 \pi^2}{\sqrt{2} \lambda}\left[\sum_{l=1}^M N_l(\beta_l\mp i \gamma_l)^2\!\right].
      \end{align}

\section{Two-point functions of bulk operators}\label{sec:2ptkine}
\subsection{Kinematics}
We now turn to the main observable studied in this work, the two-point correlation function of bulk CPOs $\mathcal{O}_p$ in the presence of a $1/2$ BPS surface defect. The most general form for this correlator is \cite{Holguin:2025bfe}
\begin{equation}\label{2pt}
    \langle \mathcal{O}_{p} \left(P_1,Y_1 \right) \,\mathcal{O}_{p} \left(P_2,Y_2\right)\rangle_\Sigma = \frac{\left(Y_1 \circ Y_1\right)^{p / 2}\left(Y_2 \circ Y_2\right)^{p / 2}}{\left(P_1 \circ P_1\right)^{p / 2}\left(P_2 \circ P_2\right)^{p / 2}} \ \mathcal{F}(\xi, \phi, \xi_R, \phi_R,\vartheta)\,,
\end{equation}
where $\{\xi, \phi, \xi_R, \phi_R\}$ are four conformally invariant cross-ratios, whose explicit expression is in \eqref{crossratio}. 
The breaking of the $\mathcal{R}_+$ generator results in an additional cross-ratio \cite{Holguin:2025bfe}
\begin{equation} \label{theta}
\vartheta=\eta_1+\eta_2\,,
\end{equation}
with the $\eta_i$ defined in \eqref{etatheta} for each external operator. This combination is neutral under $\mathcal{R}_-$, but not under $\mathcal{R}_+$. This cross-ratio is then absent when $\mathcal{M}_{\perp}$ and $\mathcal{R}_{\perp}$ are both preserved. Its geometrical interpretation is detailed in appendix \ref{app:cross}.

In the following it will often be convenient to express the correlator
in terms of the lightcone cross-ratios
$\{z,\bar z,\alpha,\bar\alpha\}$, defined in \eqref{lightcone}.
Furthermore, it is natural
to expand the function $\mathcal{F}$ in Fourier modes with respect to $\vartheta$
\begin{equation}\label{Fourier}
\mathcal{F}(z,\bar z,\alpha,\bar\alpha,\vartheta) = \sum_{q} e^{iq\frac{\vartheta}{2}}\,
\mathcal{F}^{(q)}(z,\bar z,\alpha,\bar\alpha)\,.
\end{equation}
Each function $\mathcal{F}^{(q)}$ is associated to a different charged sector of the correlator. When decomposing the tensor product $[0,p,0]\times[0,p,0]$ under $ SO(6)_R \rightarrow SO(4)_{\mathcal R} \times SO(2)_{\mathcal{R}_{\perp}}$ one does not need to project on the $\mathcal{R}_\perp$-neutral sector because only the combination $\mathcal{R}_-$ needs to be preserved. Therefore, the sum over $q$ runs over the allowed charges in this decomposition. For instance, for the two-point function of $\mathcal{O}_2$ the allowed charges are $q= 0,\pm 2,\pm 4$. The neutral sector $q=0$ is invariant under $\mathcal{M}_\perp$ and $\mathcal{R}_\perp$ separately, whereas the sectors with $q\neq 0$ are invariant only under $\mathcal{R}_-$.

The correlator is invariant under transformations that leave the cross-ratios \eqref{crossratio} unchanged. In terms of the lightcone variables, this implies the symmetry relations
\begin{eqnarray}\label{symF}
\mathcal{F}(z,\bar z,\alpha,\bar\alpha,\vartheta)
&=&
\mathcal{F}(\bar z, z, \bar\alpha,\alpha,\vartheta)\,,\nonumber \\
\mathcal{F}(z,\bar z,\alpha,\bar\alpha,\vartheta)
=\mathcal{F}(\tfrac{1}{\bar z},\tfrac{1}{z},\alpha,\bar \alpha,\vartheta)&=&\mathcal{F}(z,\bar z,\tfrac{1}{\bar\alpha},\tfrac{1}{\alpha},\vartheta)=\mathcal{F}(\tfrac{1}{\bar z},\tfrac{1}{z},\tfrac{1}{\bar\alpha},\tfrac{1}{\alpha},\vartheta)\,.
\end{eqnarray}
While the second line of \eqref{symF} is just a redundancy in the definition of the lightcone cross-ratios, the first one leaves $\cos \phi$ and $\cos \phi_R$ invariant, but it sends $\phi\leftrightarrow-\phi$ and $\phi_R\leftrightarrow-\phi_R$. For a codimension-two defect a dependence on $\sin\phi$ and $\sin\phi_R$ (see \eqref{sinphi}), i.e. parity-odd cross structures, is allowed and it is actually required by supersymmetry.\footnote{For the same reason parity-odd structures are necessary to solve superconformal Ward identities for the one-point function of the stress tensor multiplet \cite{Bianchi:2020hsz}.} However, since we are considering identical operators, the effect of the transformation $P_1\leftrightarrow P_2$ and $Y_1\leftrightarrow Y_2$ is precisely changing the sign of $\phi$ and $\phi_R$ and therefore the first line of \eqref{symF} is enforced by Bose symmetry. The immediate consequence is that parity-odd structures must appear in the combination $\sin \phi \sin\phi_R$. In terms of the radial cross-ratios \eqref{radialcoord} the first transformation in \eqref{symF} corresponds to
$w\leftrightarrow1/w$ and $w_R\leftrightarrow1/w_R$, while the second transformations act as
$r\leftrightarrow1/r$ or $r_R\leftrightarrow1/r_R$ independently.

\subsection{Superconformal Ward identities}
The two-point function of $\mathcal{O}_p$ is further constrained by the residual superconformal symmetry \eqref{symbreak}. The associated superconformal Ward identities, derived in \cite{Holguin:2025bfe}, take the form
\begin{eqnarray}\label{WI}
\left. \left(\partial_z+\partial_\alpha\right) \mathcal{F}(z,\bar z,\alpha,\bar\alpha,\vartheta) \right|_{\alpha\to z} &=& 0 \,, \nonumber\\
\left. \left(\partial_{\bar z}+\partial_{\bar\alpha}\right) \mathcal{F}(z,\bar z,\alpha,\bar\alpha,\vartheta) \right|_{\bar\alpha\to \bar z} &=& 0\,. 
\end{eqnarray}
A crucial observation is that these equations do not involve the
variable $\vartheta$. As a consequence, they hold independently in each charged sector of the Fourier expansion of the correlator \eqref{Fourier}. 
Remarkably, these equations are identical in form to the Ward identities satisfied by four-point functions of chiral operators in the absence of a defect \cite{Nirschl:2004pa,Dolan:2004mu}. The only difference lies in the symmetry relations \eqref{symF}, which are not the same as those of the four-point function, and we need to take this into account when writing the solution.

The starting point of our analysis is the observation that the correlator \eqref{2pt} must be a polynomial in the $Y$ variables, which immediately translates into the fact that the function $\mathcal{F}(\xi,\phi,\xi_R,\phi_R,\vartheta)$ must be a polynomial in the cross-ratios $\xi_R$ and $\cos \phi_R$ (or $\sin\phi_R$) in \eqref{crossratio}. Equivalently, the dependence of the correlator on the R-symmetry variables is entirely dictated by representation theory and it contains finitely many terms. Therefore, we can write each Fourier mode $\mathcal{F}^{(q)}(z,\bar z,\alpha,\bar\alpha)$ in the general form
\begin{equation}\label{Rexpansion}
\mathcal{F}^{(q)}(z,\bar z,\alpha,\bar\alpha)=\sum_{-p\leq m,n\leq p} \mathcal F^{(q)}_{m \, n}(z,\bar z) \left(\frac{\alpha \bar \alpha}{z \bar z}\right)^{ m/2}\left(\frac{\alpha \bar z}{\bar \alpha z}\right)^{ n/2}\,,
\end{equation}
for some coefficients $\mathcal F^{(q)}_{m\, n}(z,\bar z)$ that are arbitrary functions of the spacetime coordinates. The rational factors in \eqref{Rexpansion} are annihilated by the Ward identities \eqref{WI}. Therefore, substituting the expansion \eqref{Rexpansion} into the second equation in \eqref{WI}, we find
\begin{equation}
 0=\left.\left(\partial_{\bar z}+\partial_{\bar \alpha} \right) \mathcal{F}^{(q)}\left(z, \bar{z}, \alpha, \bar{\alpha}\right)\right|_{\bar \alpha \rightarrow \bar z} = \sum_{m,n} \partial_{\bar z}  \mathcal F^{(q)}_{m\, n}(z,\bar z)\, \left(\frac{\alpha}{z}\right)^{\frac{m+n}{2}} =\partial_{\bar z}\mathcal{F}^{(q)}\left(z, \bar{z}, \alpha, \bar{z}\right).
\end{equation}
This implies that $\mathcal{F}^{(q)}(z,\bar z,\alpha,\bar z)$ does not depend on $\bar z$ and we can define
\begin{equation}\label{f1}
\mathcal{F}^{(q)}(z,\bar z,\alpha,\bar z) \colonequals f^{(q)}(z,\alpha)\,,
\end{equation}
for some holomorphic function $f^{(q)}(z,\alpha)$. Using the symmetries \eqref{symF}, we are led to the following Ansatz for the general solution of the Ward identities \eqref{WI},
\begin{eqnarray}\label{wardsolution}
&&\mathcal{F}^{(q)}(z,\bar z,\alpha,\bar\alpha) = \mathcal{C}^{(q)} + \frac{(z-\alpha)(\bar\alpha z-1)(\alpha\bar z-1)(\bar z-\bar\alpha)}{\alpha \bar \alpha \, z \bar z} F^{(q)}(z,\bar z,\alpha,\bar\alpha)\\
&&+\frac{ (\bar\alpha z-1)(\alpha\bar z-1)
\left[f^{(q)}(z,\alpha)+f^{(q)}(\bar z,\bar\alpha)\right] }{ (\alpha\bar\alpha-1)(z\bar z-1) }-\frac{ (z-\alpha)(\bar z-\bar\alpha) \left[f^{(q)}\!\left(z,\frac{1}{\bar\alpha}\right) +f^{(q)}\!\left(\bar{z},\frac{1}{\alpha}\right)\right] }{ (\alpha\bar\alpha-1)(z\bar z-1)}\,. \nonumber
\end{eqnarray}
As emphasized above, this solution applies independently to each Fourier sector of the correlator \eqref{Fourier}. Here $\mathcal{C}^{(q)}$ is a constant, while the function $F^{(q)}(z,\bar z,\alpha,\bar\alpha)$ encodes the unconstrained part of the correlator and satisfies the symmetry properties 
\begin{eqnarray}\label{symbigF}
F^{(q)}(z,\bar z,\alpha,\bar\alpha) &=& F^{(q)}(\bar z,z,\bar\alpha,\alpha)\,,\nonumber \\
F^{(q)}(z,\bar z,\alpha,\bar\alpha)
=F^{(q)}(\tfrac{1}{\bar z},\tfrac{1}{z},\alpha,\bar \alpha)&=&F^{(q)}\left(z,\bar z,\tfrac{1}{\bar\alpha},\tfrac{1}{\alpha}\right)=F^{(q)}\left(\tfrac{1}{\bar z},\tfrac{1}{z},\tfrac{1}{\bar\alpha},\tfrac{1}{\alpha}\right)\,.
\end{eqnarray}
The holomorphic function $f^{(q)}(z,\alpha)$ obeys
\begin{equation}
f^{(q)}(z,\alpha) = f^{(q)}\!\left(\frac{1}{z},\frac{1}{\alpha}\right), \qquad f^{(q)}(z,z) = -\mathcal{C}^{(q)}\,.
\end{equation}
These conditions ensure that the full expression \eqref{wardsolution} is consistent with both the Ward identities and the symmetries of the correlator \eqref{symF}.

\subsection{Chiral algebra and topological sector}
\label{sec:local}

Looking at equation \eqref{wardsolution} one immediately finds two interesting properties
\begin{align}\label{holotopo}
    \mathcal{F}^{(q)} (z,\bar z,\alpha,\bar{z})&= f^{(q)}(z,\alpha)\, , &  \mathcal{F}^{(q)} (z,\bar z,z,\bar{z})&= -\mathcal{C}^{(q)}\, .
\end{align}
The first identification reduces the correlator to a holomorphic function and it is known as the chiral algebra twist. In $\mathcal{N}=4$ SYM there is enough supersymmetry to perform a second twist which leads to a function that is independent of $z$ and $\bar z$. Remarkably, the class of operators selected by the second twist is the same as the one that can be probed using the localization techniques of \cite{Pestun:2009nn}. We review this construction in appendix \ref{twist}. Using these twists, we can extract non-trivial predictions for the two-point function \eqref{2pt} which we can use either as additional constraints for the bootstrap or as sanity checks of our results. Here we simply summarize these predictions. For simplicity, in this section we focus on the correlators of $\mathcal{O}_2$.

\subsubsection{Chiral algebra predictions} \label{chiralalgebra}
In the chiral algebra construction the surface defect is orthogonal to the chiral algebra plane. If it preserves a two-dimensional $\mathcal{N}=(2,2)$ superconformal algebra, it is in  $\mathbbmtt{Q}$-cohomology (see appendix \ref{twist} for a review of the chiral algebra construction) and defect operators form a module over the original chiral algebra \cite{Cordova:2017mhb,Pan:2017zie, Bianchi:2019sxz}.

The defect vacuum is associated with the defect identity operator $\hat{\id}$. Following the conventions of \cite{Bianchi:2019sxz}, we denote its cohomology class by
\begin{equation}
	\sigma(0) \colonequals [ \hat{\id}]_{\mathbbmtt{Q}}\, .
\end{equation}
Since the defect extends infinitely, it effectively intersects the chiral algebra plane both at the origin and at infinity and the four-dimensional $n$-point function of twisted-translated bulk Schur operators $\mathcal{O}_i$ in the presence of $\Sigma$ reduces, in the chiral algebra, to the following two-dimensional $(n+2)$-point function:
\begin{equation}
 \langle \sigma(\infty) \mathcal{O}_1(\z_1) \dots \mathcal{O}_n(\z_n) \sigma(0) \rangle \equalscolon  \langle \mathcal{O}_1(\z_1) \dots \mathcal{O}_n(\z_n) \rangle_\sigma\, ,
\end{equation}
where we have assumed the normalization $\langle\sigma(\infty)\sigma(0)\rangle=1$.

We first consider the chiral algebra image of the one-point function \eqref{1ptbroken}. The new element compared to \cite{Bianchi:2019sxz} is that we include the rotation-breaking sector. Performing the twist of appendix \ref{twist} with  $u_a=(1,\bar\z)$ and using
\begin{equation}\label{twistdict}
 	\frac{Y\!\circ\! Y}{P\!\circ\! P}=\frac{\tilde u_{\dot 1}\tilde u_{\dot 2}}{\z}\, ,
 	\qquad
 	e^{2i\eta}=e^{2i\theta}e^{2i\theta_R}=\z\,\frac{\tilde u_{\dot 1}}{\tilde u_{\dot 2}}\, ,
 \end{equation}
 which follow from \eqref{ptheta}, \eqref{etatheta} and \eqref{Ybroken}, the one-point function \eqref{1ptO2} becomes
\begin{equation}\label{1pttwisted}
	\langle \mathcal{O}_2(\z,\tilde u)\rangle_\Sigma
 	= a^{(+2)}_{\mathcal{O}_2}(\tilde u_{\dot 1})^2
	+ a^{(0)}_{\mathcal{O}_2}\frac{\tilde u_{\dot 1}\tilde u_{\dot 2}}{\z}
 	+ a^{(-2)}_{\mathcal{O}_2}\frac{(\tilde u_{\dot 2})^2}{\z^2}\, .
 \end{equation}
 Using \eqref{chiralcurrent} and passing to the Cartan-Weyl basis $J^{\pm}=J^1\pm iJ^2$, $J^3$ we obtain
\begin{equation}\label{onepointchiral}
\begin{split}
	\langle \sigma(\infty)J^{+}(\z)\sigma(0)\rangle&=\frac{4\,a^{(-2)}_{\mathcal{O}_2}}{\lambda_{\mathcal{B}_{[0,2,0]}}\z^{2}}\, , \\
	\langle \sigma(\infty)J^{3}(\z)\sigma(0)\rangle&=\frac{2a^{(0)}_{\mathcal{O}_2}}{\lambda_{\mathcal{B}_{[0,2,0]}}\z}\, ,
	\\
	\langle \sigma(\infty)J^{-}(\z)\sigma(0)\rangle&=-\frac{4\,a^{(+2)}_{\mathcal{O}_2}}{\lambda_{\mathcal{B}_{[0,2,0]}}}\, ,
\end{split}
\end{equation}
where from now on $\lambda_{\mathcal{B}_{[0,2,0]}} \colonequals \lambda_{\mathcal{O}_2\mathcal{O}_2\mathcal{O}_2}$. 
From the point of view of the affine $\mathfrak{su}(2)_{2k}$ algebra, the second of these correlators can be written as
\begin{equation}
\label{onepoint2d}
	\langle \sigma(\infty) J^3(\z) \sigma(0) \rangle =  \frac{1}{\z} \langle \sigma^{\dagger} | J^3_0 | \sigma \rangle \, ,
\end{equation}
where $J^3_0$ denotes the zero mode of the current.

The unequal powers of $\z$ in \eqref{onepointchiral} are the chiral algebra imprint of the breaking of $\mathcal{R}_+$. We return to this point below. First we focus on the rotation-preserving case.

\paragraph{Rotation-preserving case.} When $\mathcal{R}_+$ is preserved one has $a^{(\pm2)}_{\mathcal{O}_2}=0$ and only the neutral one-point function survives. All three currents then have at most a simple pole acting on the defect vacuum, so
\begin{equation}\label{affineprimary}
	J^A_n|\sigma\rangle = 0\, ,\qquad n\geq 1\, ,
\end{equation}
i.e.\ $|\sigma\rangle$ is an ordinary affine primary. The two-point function is then completely fixed by the current algebra
\begin{equation}\label{cartanalg}
	[J^3_m,J^3_n]=k\,m\,\delta_{m+n,0}\, ,\quad
	[J^3_m,J^{\pm}_n]=\pm J^{\pm}_{m+n}\, ,\quad
	[J^+_m,J^-_n]=2J^3_{m+n}+2k\,m\,\delta_{m+n,0}\, 
\end{equation}
and the zero mode data. Indeed, for $|\z_1|>|\z_2|$,
\begin{equation}
	\langle \sigma^\dagger|J^A(\z_1)J^B(\z_2)|\sigma\rangle
	=\sum_{m\geq 0}\z_1^{-m-1}\z_2^{\,m-1}\,
	\langle \sigma^\dagger|J^A_m J^B_{-m}|\sigma\rangle\, ,
\end{equation}
where the restriction to $m\geq0$ and to the total mode number being zero follows from \eqref{affineprimary} and from the $ L_0$ grading. For $m\geq1$ the second term of $J^A_mJ^B_{-m}=[J^A_m,J^B_{-m}]+J^B_{-m}J^A_m$ annihilates $|\sigma\rangle$, so that only the commutator survives, and
\begin{equation}
	\langle \sigma^\dagger|J^A_m J^B_{-m}|\sigma\rangle
	= i\,\varepsilon^{ABC} \delta^{C3}\langle \sigma^{\dagger} | J^3_0 | \sigma \rangle+ k\,m\,\delta^{AB}\, ,\qquad m\geq 1\, .
\end{equation}
Resumming with $\sum\limits_{m\geq1}\z_1^{-m-1}\z_2^{m-1}=(\z_1\z_{12})^{-1}$ and $\sum\limits_{m\geq1}m\,\z_1^{-m-1}\z_2^{m-1}=\z_{12}^{-2}$, and using the identity $\z_1^{-1}\z_{12}^{-1}+\tfrac12(\z_1\z_2)^{-1}=\tfrac12(\z_1+\z_2)(\z_1\z_2\z_{12})^{-1}$, we find
\begin{equation}\label{twoptchiralfinal}
	\langle \sigma(\infty)J^A(\z_1)J^B(\z_2)\sigma(0)\rangle
	=\frac{k\,\delta^{AB}}{\z_{12}^{2}}
	+ \frac{a_{\mathcal{O}_2}^{(0)}}{\lambda_{\mathcal{B}_{[0,2,0]}}}\, i\,\varepsilon^{ABC}\delta^{C3}\,
	\frac{\z_1+\z_2}{\z_1\z_2\,\z_{12}}
	+\frac{S^{AB}}{\z_1\z_2}\, ,
\end{equation}
where we used \eqref{onepointchiral} and
\begin{equation}\label{SABdef}
	S^{AB}\colonequals \tfrac12\,\langle \sigma^\dagger|\{J^A_0,J^B_0\}|\sigma\rangle
\end{equation}
is symmetric. Equation \eqref{twoptchiralfinal} is manifestly invariant under $(A,\z_1)\leftrightarrow(B,\z_2)$, as it must be. Its $\z_1\to\z_2$ expansion is controlled by the OPE \eqref{OPEchiral}, and reproduces
\begin{equation}\label{z1z2OPE}
    \langle \sigma(\infty) J^A(\z_1) J^B(\z_2) \sigma(0) \rangle\Big|_{\z_1 \to \z_2}
    = - \frac{2\, \delta^{AB}}{\lambda_{\mathcal{B}_{[0,2,0]}}^2 \, \z_{12}^2}
    + \frac{2\, a_{\mathcal{O}_2}^{(0)}}{\lambda_{\mathcal{B}_{[0,2,0]}}} \frac{i \, \varepsilon^{ABC}\delta^{C3}}{\z_2\,\z_{12}}+\dots \, .
\end{equation}
The other relevant OPE is the defect channel expansion, whose chiral algebra counterpart was found in \cite{Bianchi:2019sxz} and it reads 
\begin{align}\label{OPEdefectmodule}
    J^A(\z)\sigma(0)=\kappa_J \frac{ \sum_i b_{J\hat{\phi}_i}\hat{\phi}^A_i - a^{(0)}_{\mathcal{O}_2}\, \delta^{A 3}\sigma(0)}{\zeta}+\dots \, ,
\end{align}
where $\k_J$ is introduced in \eqref{chiralcurrent} and $\hat{\phi}^A_i$ are Schur defect operators that, from a chiral algebra point of view, are degenerate with the defect identity. As described in \cite{Bianchi:2019sxz} these operators have $\hat{\Delta}=-s=1$, where $s$ is the transverse spin. Translating to the $\mathcal{N}=4$ quantum numbers, as we do in equations \eqref{mapchiral} and \eqref{mapchiral_defect}, we find that such operators must have 
\begin{align}\label{hatphi}
  \hat{\phi}^+_i && &\to & \mathcal{R}&=0\, , & \bar {\mathcal{R}}&=0\, , & r_-&=-1\, ,\nn \\
  \hat{\phi}^3_i && &\to &  \mathcal{R}&=-1/2\, , & \bar {\mathcal{R}}& =1/2\, , & r_-&=0\, ,\\
  \hat{\phi}^-_i && &\to & \mathcal{R}&=-1\, , & \bar {\mathcal{R}}&=1\, , & r_-&=1\, ,\nn
\end{align}
where $\mathcal{R}\colonequals\mathcal{R}^1{}_1$ and $\bar{\mathcal{R}}\colonequals \bar{\mathcal{R}}^1{}_1$ are defined in \eqref{preservedgens}. In \eqref{unitaritydefshort}, we will see that the shortening condition for the superprimary of a defect short multiplet is $\hat\Delta=n+|r_-|$ with $n=2|\mathcal{R}|=2|\mathcal{\bar{R}}|$ for these specific cases.  Imposing this, the first line gives $n=0$ and $|r_-|=1$, hence $\hat\phi^+_i\to\widehat{\mathcal{B}}_{0,-1}$ (see appendix \ref{app:irreppsu} for a review of the relevant superconformal representations). The conjugate multiplet $\widehat{\mathcal{B}}_{0,+1}$ is in cohomology at $\infty$, but not at the origin. The second line gives $n=1$ and $r_-=0$, i.e.
\begin{equation}\label{B120}
	\hat\phi^{3}_i\;\to\;\widehat{\mathcal{B}}_{1/2,0}\,.
\end{equation}
The conditions in the third line cannot be fulfilled by any operator in the defect superconformal multiplets.
All these short multiplets are $\frac12$-BPS and they have interesting physical interpretations: $\widehat{\mathcal{B}}_{0,\mp1}$ is the displacement supermultiplet, while $\widehat{\mathcal{B}}_{1/2,0}$ contains exactly marginal defect operators and it is present only when there is a defect conformal manifold. Requiring compatibility of \eqref{OPEdefectmodule} with \eqref{twoptchiralfinal} provides a definition of $S^{AB}$ in terms of defect correlators
\begin{equation}\label{chiraldefectOPE}
   S^{AB}
    = \kappa_J^2 \left(\sum_i \tfrac{1}{2}b_{J\hat{\phi}_i}^2(\langle\hat{\phi}^A_i(\infty) \hat\phi^B_i(0)\rangle+\langle\hat{\phi}^B_i(\infty) \hat\phi^A_i(0)\rangle)+(a^{(0)}_{\mathcal{O}_2})^2 \delta^{A 3} \delta^{B 3}\right)\, .
\end{equation}
In section \ref{holomorphicorr} we will compare this expectation with our result for the holomorphic part of the correlator.
\paragraph{Breaking of $\mathcal{R}_+$ and spectral flow.} When $a^{(\pm2)}_{\mathcal{O}_2}\neq 0$ the analysis above must be modified, because \eqref{onepointchiral} shows that $J^{+}(\z)$ now has a double pole and $J^{-}(\z)$ a regular term when acting on the defect vacuum. Requiring consistency with the bulk-to-defect OPE and with \eqref{onepointchiral} gives
\begin{eqnarray}\label{OPEchiralbroken}
	J^{+}(\z)\sigma(0)&=&\frac{4\,a^{(-2)}_{\mathcal{O}_2}}{\lambda_{\mathcal{B}_{[0,2,0]}}}\,\frac{\sigma(0)}{\zeta^2}+\frac{\chi^{+}}{\zeta}+O(\z^{0})\, ,
	\nonumber\\
	J^{3}(\z)\sigma(0)&=&\frac{1}{\z}\left(\frac{2\, a^{(0)}_{\mathcal{O}_2}}{\lambda_{\mathcal{B}_{[0,2,0]}}}\,\sigma(0)+\chi^{3}\right)+O(1)\, ,
	\\
	J^{-}(\z)\sigma(0)&=&-\frac{4\,a^{(+2)}_{\mathcal{O}_2}}{\lambda_{\mathcal{B}_{[0,2,0]}}}\,\sigma(0)+ \psi^- + O(\z)\, ,\nonumber
\end{eqnarray}
where
\begin{equation}\label{chidef}
	\chi^{A}\colonequals \kappa_J\sum_i b_{J\hat{\phi}_i}\,\hat{\phi}^{A}_i\, ,
	\qquad A=+,3\, ,
\end{equation}
while $\psi^-$ is a term in the regular part of the OPE that is allowed also when the defect preserves transverse rotations (usually we only care about singular terms, but we will see that in this case also the regular part of $J^-$ is important). Mapping the quantum numbers to the four-dimensional ones, one can easily see that $\psi^-$ can receive contributions from both the $\mathcal{\hat{B}}_{1/2,0}$ and the $\mathcal{\hat{B}}_{1,0}$ multiplets. Since the $\hat{\phi}^{A}_i$'s and $\psi^-$ are distinct from the defect identity, the corresponding states satisfy $\langle\sigma^\dagger|\chi^{A}\rangle=\langle\sigma^\dagger|\psi^-\rangle=0$, which is what allows \eqref{onepointchiral} to read off the coefficient of $\sigma(0)$. Equivalently,
\begin{equation}\label{brokenannihilation}
	J^{+}_{n}|\sigma\rangle=0\ \ (n\geq2)\, ,\qquad
	J^{3}_{n}|\sigma\rangle=0\ \ (n\geq1)\, ,\qquad
	J^{-}_{n}|\sigma\rangle=0\ \ (n\geq0)\, .
\end{equation}
Here \eqref{OPEchiralbroken} fixes the leading power in each channel. The absence of more singular terms is the statement that the defect module is bounded from below with respect to the grading 
\begin{equation}\label{L0tilde}
	\widetilde L_0 = L_0 + J^3_0\, ,
\end{equation}
under which $J^{+}_{1}$, $J^3_0$ and $J^{-}_{-1}$ all have vanishing charge. This is precisely the Virasoro zero mode of the algebra spectrally flowed by one unit (up to a constant shift).
The importance of the spectral flow for the representation of monodromy defects in chiral algebra has been pointed out in \cite{Cordova:2017mhb}. Here we show that also the breaking of orthogonal rotations has the effect of introducing spectrally flowed representations. The spectral flow is an automorphism of the affine algebra  \eqref{cartanalg} consisting of the redefinitions
\begin{equation}\label{spectralflow}
	\widetilde J^{+}_{n}=J^{+}_{n+1}\, ,
	\qquad
	\widetilde J^{3}_{n}=J^{3}_{n}+k\,\delta_{n,0}\, ,
	\qquad
	\widetilde J^{-}_{n}=J^{-}_{n-1}\, .
\end{equation}
The module $\sigma$ is now an affine primary under the new currents
\begin{equation}\label{flowedprimary}
	\widetilde J^{A}_{n}|\sigma\rangle = 0\, ,\qquad A=+,3,-\, ,\qquad n\geq 1\, .
\end{equation}
Equation \eqref{spectralflow} is the standard spectral flow automorphism of the affine algebra, see e.g. \cite{Maldacena:2000hw}. In terms of the currents the flow reads
\begin{equation}\label{flowcurrents}
	\widetilde J^{\pm}(\z)=\z^{\pm1}J^{\pm}(\z)\, ,
	\qquad
	\widetilde J^{3}(\z)=J^{3}(\z)+\frac{k}{\z}\, 
\end{equation}
and for the Virasoro generators
\begin{equation}
    \widetilde L_n=L_n+J^3_n+\tfrac{k}{2}\,\delta_{n,0}\, ,
\end{equation}
so that the flowed Virasoro zero mode is \eqref{L0tilde} up to the additive constant $k/2$. One checks that $[\widetilde L_m,\widetilde J^{A}_n]=-n\,\widetilde J^{A}_{m+n}$ for all three components, i.e.\ the flowed currents are still weight-one primaries and the central charge is unchanged. 
We then conclude that a defect that preserves only the diagonal combination $\mathcal{R}_-$ defines a module of the algebra spectrally flowed by one unit, and $|\sigma\rangle$ is a primary of the flowed algebra. The amount of flow is fixed to one unit by the charges of $\mathcal{O}_2$: the allowed values $q=0,\pm2$ in \eqref{1ptO2} translate into the three powers $\z^{0},\z^{-1},\z^{-2}$ in \eqref{onepointchiral}, and hence into a shift of the modes of $J^{\pm}$ by one unit. For the higher chiral primaries $\mathcal{O}_p$ the same argument applies verbatim, since the twist correlates $\tilde u$ with $\z$ in a $p$-independent way; only the number of Fourier modes in \eqref{1ptfourier} grows.

In the flowed frame all three currents have a standard simple-pole one-point function,
\begin{equation}\label{flowedonept}
	\langle \sigma(\infty)\widetilde J^{A}(\z)\sigma(0)\rangle=\frac{\widetilde\jmath^{\,A}}{\z}\, ,
	\qquad
	\widetilde\jmath^{\,\pm}=\pm \frac{4\,a^{(\mp2)}_{\mathcal{O}_2}}{\lambda_{\mathcal{B}_{[0,2,0]}}}\, ,\qquad
	\widetilde\jmath^{\,3}=\frac{2\,a^{(0)}_{\mathcal{O}_2}}{\lambda_{\mathcal{B}_{[0,2,0]}}}+k\, ,
\end{equation}
and the derivation of \eqref{twoptchiralfinal} goes through, giving
\begin{equation}\label{flowedtwopt}
	\langle \sigma(\infty)\widetilde J^{A}(\z_1)\widetilde J^{B}(\z_2)\sigma(0)\rangle
	=\frac{k\,\kappa^{AB}}{\z_{12}^2}
	+\frac{C^{AB}}{2}\,\frac{\z_1+\z_2}{\z_1\z_2\z_{12}}
	+\frac{\widetilde S^{AB}}{\z_1\z_2}\, ,
\end{equation}
with non-vanishing entries 
\begin{equation}
\kappa^{33}=1\, , \qquad \kappa^{+-}=\kappa^{-+}=2\, , \qquad
    C^{+-}=-C^{-+}=2\,\widetilde\jmath^{\,3}\, ,\qquad
	C^{3\pm}=-C^{\pm3}=\pm\,\widetilde\jmath^{\,\pm}\, ,
\end{equation}
and with $\widetilde S^{AB}=\tfrac12\langle\sigma^\dagger|\{\widetilde J^{A}_0,\widetilde J^{B}_0\}|\sigma\rangle$. Undoing the flow gives the physical correlators
\begin{align}\label{unflowed}
	\langle J^{+}(\z_1)J^{+}(\z_2)\rangle_\sigma
	&=\frac{\widetilde S^{++}}{\z_1^2\z_2^2}\, , \nn\\
	\langle J^{-}(\z_1)J^{-}(\z_2)\rangle_\sigma
	&=\widetilde S^{--}\, , \nn\\
	\langle J^{3}(\z_1)J^{3}(\z_2)\rangle_\sigma
	&=\frac{k}{\z_{12}^2}
	+\frac{\widetilde S^{33}-2k\,\widetilde\jmath^{\,3}+k^2}{\z_1\z_2}\, , \\
	\langle J^{+}(\z_1)J^{-}(\z_2)\rangle_\sigma
	&=\frac{\z_2}{\z_1}\left[\frac{2k}{\z_{12}^2}
	+\widetilde\jmath^{\,3}\,\frac{\z_1+\z_2}{\z_1\z_2\z_{12}}
	+\frac{\widetilde S^{+-}}{\z_1\z_2}\right]\, ,\nn\\
	\langle J^{3}(\z_1)J^{+}(\z_2)\rangle_\sigma
	&=\frac{1}{\z_2}\left[\frac{\widetilde\jmath^{\,+}}{2}
	\frac{\z_1+\z_2}{\z_1\z_2\z_{12}}+\frac{\widetilde S^{3+}}{\z_1\z_2}\right]
	-\frac{k\,\widetilde\jmath^{\,+}}{\z_1\z_2^2}\, ,\nn\\
	\langle J^{-}(\z_1)J^{+}(\z_2)\rangle_\sigma
	&=\frac{\z_1}{\z_2}\left[\frac{2k}{\z_{12}^2}
	-\widetilde\jmath^{\,3}\,\frac{\z_1+\z_2}{\z_1\z_2\z_{12}}
	+\frac{\widetilde S^{+-}}{\z_1\z_2}\right]\, ,\nn\\
	\langle J^{3}(\z_1)J^{-}(\z_2)\rangle_\sigma
	&=-\frac{\widetilde\jmath^{\,-}}{2}\,\frac{\z_1+\z_2}{\z_1\z_{12}}
	+\frac{\widetilde S^{3-}-k\,\widetilde\jmath^{\,-}}{\z_1}\, .\nn
\end{align}
All of these are manifestly consistent with the exchange symmetry $(A,\z_1)\leftrightarrow(B,\z_2)$, which is a non-trivial check since the flow treats $J^{+}$ and $J^{-}$ asymmetrically. Notice that $\langle J^{\pm}J^{\pm}\rangle$ receives no contribution from the level or from the structure constants: it is pure zero-mode data. These components map to the maximally charged Fourier sector $q=\mp4$ of \eqref{Fourier}, and using \eqref{OPEchiralbroken} and the conjugated equation at $\infty$, it is easy to see that 
\begin{equation}\label{q4factorization}
	\widetilde S^{\pm\pm}=\big(\widetilde\jmath^{\,\pm}\big)^2\, ,
\end{equation}
i.e.\ the connected part vanishes. In section \ref{sec:breaking} we will see that this expectation is confirmed by four-dimensional arguments.

\subsubsection{Localization predictions}
The chiral algebra associated to $\mathcal{N}=4$ SYM is still supersymmetric. Performing a second twist on the $\z$-translation makes the resulting correlators topological (this procedure is summarized in appendix \ref{twist}). Remarkably, it was noted in \cite{Pestun:2009nn,Giombi:2009ds} that the same double twist leads to observables that are captured by supersymmetric localization. More specifically, correlators of operators in cohomology are mapped to observables in 2d Yang-Mills on a sphere.

A similar procedure works also in the presence of a Gukov-Witten surface defect \cite{Wang:2020seq,Choi:2024ktc}. In particular, the presence of a surface defect imposes non-trivial background values for the two-dimensional fields
\begin{equation}
-i*\boldsymbol{F} \colonequals \beta + \tilde{\varphi}, 
\qquad 
\langle -i*\boldsymbol{F} \rangle = \beta\,,
\end{equation}
where $\boldsymbol{F}$ is the curvature of the gauge field and $\tilde{\varphi}$ its fluctuation around the background value $\beta$, defined in \eqref{alphabetagammadef}. The two-dimensional Yang-Mills action can be expressed in terms of the fluctuation as 
\begin{equation}\label{2dYM}
S_{2d}
=
-\frac{1}{g_{2d}^2}
\int_{S^2}
\frac{\Tr[\tilde{\varphi}^2(\z,\bar \z)]}{(1+\z \bar \z)^2}\, d\z\, d\bar \z,
\qquad
g_{2d}^2=-\frac{\lambda}{2\pi N},
\end{equation}
where here $(\z,\bar \z)$ are complex coordinates on the two-sphere. Here we straightforwardly extend the computation of \cite{Choi:2024ktc} to the case of a bulk two-point function thus predicting the value of the constant $\mathcal{C}^{(q)}$ in \eqref{holotopo}. 

Each insertion of the operator $\mathcal O_2^{\mathrm{loc}}$, as defined in \eqref{Olocdef}, is mapped in the two-dimensional theory to
\begin{equation}
\mathcal O_2^{\mathrm{loc}}\quad\longrightarrow\quad
\Tr\left[\left(-i*\boldsymbol{F}\right)^2\right]=\Tr\left[\left(\beta+\tilde{\varphi}\right)^2\right].
\end{equation}
Accordingly, the normalized two-point function can be written as
\begin{equation}\label{2ptloc}
\frac{\bigl\langle \mathcal O_2^{\mathrm{loc}}\,\mathcal O_2^{\mathrm{loc}}\,\mathcal O_\Sigma\bigr\rangle}
{\langle \mathcal O_\Sigma\rangle\,\bigl\langle \mathcal O_2^{\mathrm{loc}}\,\mathcal O_2^{\mathrm{loc}}\bigr\rangle_0}
=
\frac{\left\langle
\Tr\left[\bigl(\beta+\tilde{\varphi}\bigr)^2\right]\,
\Tr\left[\bigl(\beta+\tilde{\varphi}\bigr)^2\right]
\right\rangle_{2d}}{\bigl\langle \mathcal O_2^{\mathrm{loc}}\,\mathcal O_2^{\mathrm{loc}}\bigr\rangle_0},
\end{equation}
where the denominator is the two-point function without the defect
\begin{equation}\label{eq:defect-free-2pt}
\langle \mathcal O_2^{\mathrm{loc}}\,\mathcal O_2^{\mathrm{loc}}\rangle_0
=
\frac{g_{2d}^4}{32\pi^2}N^2.
\end{equation}

Since the theory \eqref{2dYM} is Gaussian, the correlator \eqref{2ptloc} can be evaluated via Wick contractions. The relevant propagator for fluctuations within the same Levi block $l$ \eqref{levi}, evaluated at non-coincident points\footnote{At coincident points, additional contact terms render the propagator singular. These terms are relevant for the computation of one-point functions, see \cite{Choi:2024ktc} for further details.}, is \cite{Choi:2024ktc}
\begin{equation}
\bigl\langle \tilde{\varphi}_{ij}(x)\tilde{\varphi}_{km}(y)\bigr\rangle_l
=
\frac{g_{2d}^2}{8\pi}\delta_{im}\delta_{jk}\,, \quad x \neq y\,,
\end{equation}
where $(i,j)$ are fundamental indices of $U(N)$ belonging to the same Levi subgroup \eqref{levi}.

The two-point function contains a factorized part, arising from self-contractions within each insertion, and connected contributions from contractions between distinct insertions, involving both the linear and quadratic terms. The factorized piece is simply the square of the one-point function and can be read off from \eqref{1ptO2}. The contraction of the linear terms yields
\begin{equation}\label{linearwick} 
4\left\langle \Tr\bigl(\beta\tilde{\varphi}(x)\bigr)\,\Tr\bigl(\beta\tilde{\varphi}(y)\bigr)\right\rangle = 4\!\sum_{l=1}^M \sum_{i,j=1}^{N_l} \beta^2_l \, \langle \tilde{\varphi}_{ii}(x)\tilde{\varphi}_{jj}(y)\rangle_l = \frac{g_{2d}^2}{2\pi}\sum_{l=1}^M N_l\beta_l^2, 
\end{equation}
which follows from the fact that the propagator is diagonal in the matrix indices.
The contraction of the connected quadratic terms gives
\begin{equation}\label{quadraticwick} 
\left\langle \Tr\tilde{\varphi}(x)^2\,\Tr\tilde{\varphi}(y)^2\right\rangle = 2\sum_{l=1}^M \sum_{i,j=1}^{N_l} \left( \langle \tilde{\varphi}_{ij}(x)\tilde{\varphi}_{ji}(y)\rangle_l \right)^2 = \frac{g_{2d}^4}{32\pi^2}\left[N^2-\sum_{l=1}^M N_l(N-N_l)\right]\,, 
\end{equation}
where we used that the number of ordered pairs $(i,j)$ in the same Levi block is
\begin{equation}
\sum_{l=1}^M N_l^2 = N^2 - \sum_{l=1}^M N_l(N-N_l).
\end{equation}
Combining all contributions and exploiting
\begin{equation}
\beta_l^2
=
\frac{1}{2}\bigl(\beta_l^2+\gamma_l^2\bigr)
+
\frac{1}{4}(\beta_l+i\gamma_l)^2
+
\frac{1}{4}(\beta_l-i\gamma_l)^2,
\end{equation}
we obtain
\begin{eqnarray}
\frac{\bigl\langle \mathcal O_2^{\mathrm{loc}}\,\mathcal O_2^{\mathrm{loc}}\,\mathcal O_\Sigma\bigr\rangle}
{\langle \mathcal O_\Sigma\rangle\,\bigl\langle \mathcal O_2^{\mathrm{loc}}\,\mathcal O_2^{\mathrm{loc}}\bigr\rangle_0}
&=&
\frac{8 \pi^4}{\lambda^2}\left[\sum_{l=1}^M N_l\!\left(2\,(\beta^2_l\!+\!\gamma^2_l)\!+\!\frac{\lambda}{4\pi^2} \frac{N-N_l}{N}\!+\!(\beta_l\!+\! i \gamma_l)^2\!+\!(\beta_l-i \gamma_l)^2\right)\right]^2\nonumber\\
&&+1-\frac{16\pi^2}{\lambda N}\sum_{l=1}^M N_l(\beta_l^2+\gamma_l^2)
-\frac{1}{N^2}\sum_{l=1}^M N_l(N-N_l)
\nonumber\\
&&
-\frac{8\pi^2}{\lambda N}\sum_{l=1}^M N_l(\beta_l+i\gamma_l)^2
-\frac{8\pi^2}{\lambda N}\sum_{l=1}^M N_l(\beta_l-i\gamma_l)^2\,,
\end{eqnarray}
where the first line is the contribution of four times the squared one-point function.
Comparing the above expression with \eqref{2pt} and \eqref{wardsolution}, we deduce
\begin{eqnarray}\label{Cloc}
\mathcal{C}^{(0)}&=&-\left[a^{(0)}_{\mathcal{O}_2}\right]^2-2a^{(+2)}_{\mathcal{O}_2} a^{(-2)}_{\mathcal{O}_2} -\frac14+ \frac12\, \lambda_{\mathcal{B}_{[0,2,0]}}\,a^{(0)}_{\mathcal{O}_2}-\frac14\, \sum_{l=1}^M \frac{N_l(N-N_l)}{N^2}\,,\nonumber\\
\mathcal{C}^{(\pm 2)}&=&-2a^{(0)}_{\mathcal{O}_2} a^{(\pm 2)}_{\mathcal{O}_2} + \frac12 \lambda_{\mathcal{B}_{[0,2,0]}}\,a^{(\pm2)}_{\mathcal{O}_2}\,,\\
\mathcal{C}^{(\pm4)}&=& -\left[a^{(\pm 2)}_{\mathcal{O}_2}\right]^2\,,\nonumber
\end{eqnarray}
where $a^{(q)}_{\mathcal{O}_2}$ and $\lambda_{\mathcal{B}_{[0,2,0]}} \colonequals \lambda_{\mathcal{O}_2 \mathcal{O}_2 \mathcal{O}_2}$ are defined in \eqref{a2broken} and \eqref{lambdaO2}, respectively.

\subsection{Superconformal block decompositions}
The two-point function of $\mathcal{O}_p$ admits two distinct OPE
decompositions, corresponding to the bulk channel and the defect
channel. In this section, we outline their general structure, leaving many of the technical details and explicit expressions to appendix \ref{app:blocks}.

\subsubsection{Bulk channel}\label{sec:bulkOPE}
In the bulk channel, we decompose the product of the two $\mathcal{O}_p$ in terms of a sum of local bulk operators. The spectrum and OPE coefficients of bulk operators are intrinsic data of the bulk theory and are unaffected by the presence of the defect. The exchanged operators are organized into irreducible representations of the bulk superconformal algebra $\mathfrak{psu}(2,2|4)$. In particular, for the OPE of two $\mathcal{O}_p$ operators we have the following selection rules  \cite{Dolan:2002zh}
\begin{align}\nonumber
\mathcal{B}_{[0,p,0]}\times\mathcal{B}_{[0,p,0]}
&=
\sum_{\substack{0\le m\le n\le p\\ \ell=0}}
\mathcal{B}_{[\,n-m,\;2m,\;n-m\,]}
+
\sum_{\substack{0\le m\le n\le p-1\\ \ell\ge0}}
\mathcal{C}_{[\,n-m,\;2m,\;n-m\,],\ell}
\nonumber\\
&\quad+
\sum_{\substack{0\le m\le n\le p-2\\ \Delta, \,\ell\ge0}}
\mathcal{A}^{\Delta}_{[\,n-m,\;2m,\;n-m\,],\ell}\, ,
\label{bulkOPE}
\end{align}
where the $\mathcal{B}$'s are short multiplets (1/2 or 1/4 BPS), the $\mathcal{C}$'s are semi-short and the $\mathcal{A}$'s are long.
The superprimaries of the short multiplets $\mathcal{B}_{[a,b,a]}$ are scalar operators with protected scaling dimensions $\Delta=b+2a$. When $a=0$, these are $1/2$ BPS multiplets, while for $a>0$ they are $1/4$ BPS.  The CPO $\mathcal{O}_p$ is the superprimary of the $\mathcal{B}_{[0,p,0]}$ multiplet. The superprimaries of the semi-short multiplets $\mathcal{C}_{[a,b,a],\ell}$ have spin $\ell$ and protected scaling dimension
\begin{equation}
\Delta=\ell+2+2a+b\,, \qquad \ell\ge0\, .
\end{equation}
Finally, the superprimaries of the long multiplets $\mathcal{A}^\Delta_{[a,b,a],\ell}$ have unprotected scaling dimensions obeying the unitarity bound
\begin{equation}
\Delta\ge \ell+2+2a+b\,, \qquad \ell\ge0\, .
\end{equation}
In the presence of a $1/2$ BPS surface defect, a multiplet can contribute to the bulk channel only if it contains an  $SO(4)_{\mathcal R}$ singlet whose $\mathcal R_\perp$ charge can be compensated by transverse spin, so that the one-point function preserves $\mathcal R_-$. Contrary to what happens for interfaces and line defects \cite{Liendo:2016ymz}, all the multiplets appearing in \eqref{bulkOPE} admit non-vanishing one-point functions. As a concrete example, we specialize to the case $p=2$, when \eqref{bulkOPE} reduces to\footnote{An extra $\mathcal{C}_{[0,0,0], \ell}$ multiplet appears in the free theory, where it contains higher-spin conserved currents. The multiplets $\mathcal{B}_{[1,0,1]}$ and $\mathcal{B}_{[1,2,1]}$ are absent because they are incompatible with the symmetries of the correlator \eqref{symF}.} \cite{Beem:2016wfs} 
\begin{equation}\label{O2bulkope}
\mathcal{O}_2\times\mathcal{O}_2
\subset
\id
\,+\,\mathcal{B}_{[0,2,0]}
\,+\,\mathcal{B}_{[0,4,0]}
\,+\,\mathcal{B}_{[2,0,2]}
\,+\!\sum_{\text{even} \, \ell}\mathcal{C}_{[0,2,0],\ell}
\,\,+\!\sum_{\text{odd}\, \ell}\mathcal{C}_{[1,0,1],\ell}
\,+\!\!\!\!\!\sum_{\Delta,\, \text{even} \, \ell}\mathcal{A}^{\Delta}_{[0,0,0],\ell}\, .
\end{equation}
At the level of the correlator \eqref{2pt}, the bulk OPE leads to a decomposition in terms of bulk superconformal blocks,
\begin{equation}\label{bulkblockgen}
\mathcal{F}(z,\bar z,\alpha,\bar\alpha,\vartheta) =  \sum_{q}  \sum_{\mathcal{O}} (-2)^{n_{\mathcal{O}} - p}\lambda_{\mathcal{O}_p\mathcal{O}_p\mathcal{O}}\, a^{(q)}_{\mathcal{O}}\, e^{i q \frac{\vartheta}{2}}\, \left(\frac{\xi}{\xi_R}\right)^{-p} \mathcal{G}^{(q)}_{\mathcal{M}_\mathcal{O}}(z,\bar z,\alpha,\bar\alpha)\,,
\end{equation}
where $n_\mathcal{O} \colonequals a +b/2$ for a superprimary in the $SU(4)_R$ representation $[a,b,a]$. The coefficients $a^{(q)}_{\mathcal{O}}$ were introduced in \eqref{1ptfourier} and the functions $\mathcal{G}^{(q)}_{\mathcal{M}_\mathcal{O}}(z,\bar z,\alpha,\bar\alpha)$ are the bulk superconformal blocks of the charge $q$ sector. Each block resums the contribution of all superdescendants within a given supermultiplet $\mathcal{M}_\mathcal{O}$, with $\mathcal{O}$ being the superprimary of the multiplet. 

Superconformal blocks can be expressed as linear combinations of bosonic blocks associated to the conformal primaries that are contained in the supermultiplet and have a non-vanishing one-point function. Each bosonic block is a product of a spacetime and an R-symmetry block and the general ansatz takes the form
\begin{equation}\label{superbulkA}
\mathcal{G}_{\mathcal{M}}^{(q)}(z,\bar z,\alpha,\bar\alpha) = \sum_{\Delta,\ell,R} c^{(q)}_{\Delta,\ell,R}\, f^{(q)}_{\Delta,\ell}(z,\bar z)\, h^{(q)}_{R}(\alpha,\bar\alpha)\,,
\end{equation}
where the sum runs over the quantum numbers of the conformal primaries contained in the supermultiplet $\mathcal{M}$, namely the scaling dimension $\Delta$, the spin $\ell$, and the $SU(4)_R$ representation $R=[n-m,2m,n-m]$. Here $f^{(q)}_{\Delta,\ell}$ and $h^{(q)}_{R}$ denote the spacetime and R-symmetry conformal blocks, respectively. 
They take the form
\begin{equation}\label{noneqblock}
    f^{(q)}_{\Delta, \ell}(z,\bar z)=\frac{1}{2}\left[\tilde f^{(q)}_{\Delta, \ell}(z,\bar z)+\tilde f^{(-q)}_{\Delta, \ell}(z,\bar z) \right]\,, \quad h^{(q)}_{[n-m, \, 2m, \, n-m]}(\alpha,\bar{\alpha}) = f^{(q)}_{-m-n, n-m}\, (\alpha,\bar{\alpha})\,,
\end{equation}
with
\begin{eqnarray}\label{noneq}
    \tilde f^{(q)}_{\Delta, \ell}(z,\bar z) &=& \frac{(1-z)(1-\bar{z})}{1-z \bar{z}}(z \bar z)^{q/4}\left(k^{q,0}_{\Delta-\ell-2}(1-z) k^{q,0}_{\Delta+\ell}(1-\bar{z})+(-1)^{\ell}(z \leftrightarrow \bar{z})\right)\!,\nonumber \\
    k^{r,s}_\beta (z) &=& z^{\beta / 2}{ }_2 F_1\left(\frac{\beta+r}{2}, \frac{\beta-s}{2}, \beta, z\right).
\end{eqnarray}
The symmetrization under $q\leftrightarrow -q$ ensures that \eqref{bulkblockgen} is manifestly real. For $q=0$ these blocks have been known since \cite{Billo:2016cpy}. For $q\neq 0$ the blocks are new. Remarkably, they coincide with the conformal blocks of non-identical scalar operators \cite{Billo:2016cpy,Isachenkov:2018pef}, upon identifying $\Delta_1-\Delta_2 = \pm q$.  
We refer to appendix \ref{app:blocks} for further details on the derivation of these blocks.  Once the bosonic blocks are known, the coefficients $c^{(q)}_{\Delta,\ell,R}$ in \eqref{superbulkA} are fixed by imposing the Ward identities \eqref{WI}. 

We have explicitly constructed all the superconformal blocks associated with the operators appearing in the bulk OPE of two $\mathcal{O}_2$ operators \eqref{O2bulkope}. They are listed in tables \ref{tab:blocks} and \ref{tab:blocksq}. Since by construction these superconformal blocks satisfy the Ward identities \eqref{WI}, they can be expressed in the general form of the Ward identity solution \eqref{wardsolution}, namely
\begin{eqnarray}\label{blockWI}
    &&\mathcal{G}_{\mathcal{M}}^{(q)}(z,\bar z,\alpha,\bar\alpha) = \mathcal{C}^{(q)}_{\mathcal{M}} + \frac{(z-\alpha)(\bar\alpha z-1)(\alpha\bar z-1)(\bar z-\bar\alpha)}{\alpha \bar \alpha \, z \bar z} G^{(q)}_{\mathcal{M}}(z,\bar z,\alpha,\bar\alpha)\\
&&+\frac{ (\bar\alpha z-1)(\alpha\bar z-1)
\left[g^{(q)}_{\mathcal{M}}(z,\alpha)\!+\!g^{(q)}_{\mathcal{M}}(\bar z,\bar\alpha)\right] }{ (\alpha\bar\alpha-1)(z\bar z-1) }-\frac{ (z-\alpha)(\bar z-\bar\alpha) \left[g^{(q)}_{\mathcal{M}}\!\left(z,\frac{1}{\bar\alpha}\right) \!+\!g^{(q)}_{\mathcal{M}}\!\left(\bar{z},\frac{1}{\alpha}\right)\right] }{ (\alpha\bar\alpha-1)(z\bar z-1)}\,,
\nonumber
\end{eqnarray}
where $g^{(q)}_{\mathcal{M}}(z,\alpha) \colonequals \mathcal{G}_{\mathcal{M}}^{(q)}(z,\bar z,\alpha,\bar z)$ and $\mathcal{C}^{(q)}_{\mathcal{M}} \colonequals -g^{(q)}_{\mathcal{M}}(z,z)$. An important observation is that long multiplets $\mathcal{A}^{\Delta}_{[0,0,0],\ell}$ have a vanishing holomorphic contribution  $g^{(q)}_{\mathcal{A}^{\Delta}_{[0,0,0],\ell}}(z,\alpha)$, therefore they will only contribute to $F^{(q)}(z,\bar z,\alpha,\bar \alpha)$ in \eqref{wardsolution}. This is related to the chiral algebra discussion in section \ref{chiralalgebra}. Indeed, the holomorphic functions $g^{(q)}_{\mathcal{M}}(z,\alpha)$ correspond to exchanged operators in cohomology, whose coefficients can be predicted by the chiral algebra analysis. As we mentioned, the chiral algebra has only access to the protected sector of the theory, therefore long multiplets cannot contribute to the holomorphic sector.

Since the superconformal block associated with the $\mathcal{B}_{[0,2,0]}$ multiplet will play a distinguished role in the following discussion, we write it here explicitly,
\begin{eqnarray}\label{020superblock}
   \mathcal{G}^{(0)}_{\mathcal{B}_{[0,2,0]}} &=& f^{(0)}_{2,0}(z, \bar z) \, h^{(0)}_{[0,2,0]}(\alpha, \bar \alpha)-\frac{1}{12} f^{(0)}_{3,1}(z, \bar z)\, h^{(0)}_{[1,0,1]}(\alpha, \bar \alpha) -\frac{1}{180}f^{(0)}_{4,2}(z, \bar z) \, h^{(0)}_{[0,0,0]}(\alpha, \bar \alpha) \nonumber \\
   &=& \frac{\alpha +\bar \a}{(1-\alpha ) (1-\bar \a)} \frac{(z-1) (\bar z-1) \log (z \bar z)}{2 (z \bar z-1)} \nonumber\\
   &+&\frac{\alpha -\bar \a}{(1-\alpha ) (1-\bar \a)} \frac{(z-1) (\bar z+1) \log \bar z-(z+1) (\bar z-1) \log z}{2 (z \bar z-1)} \\
   &+&\frac{(z-1) ((\bar z-1) (z \bar z-1)+(\bar z-z \bar z) \log \bar z)-z (\bar z-1)^2 \log z}{(z-1) (\bar z-1) (z \bar z-1)}\,, \nonumber \\
   \mathcal{G}^{(\pm2)}_{\mathcal{B}_{[0,2,0]}} &=& f^{(\pm 2)}_{2,0}(z, \bar z) \, h^{(\pm2)}_{[0,2,0]}(\alpha, \bar \alpha)=\frac{(1-z) (1-\bar z) \sqrt{\alpha  \bar \a}}{\sqrt{z \bar z}(1-\alpha) (1-\bar \a)}\,, \nonumber \\
   \mathcal{G}^{(\pm4)}_{\mathcal{B}_{[0,2,0]}} &=&0\,.\nonumber
\end{eqnarray}
While the first line is a customary linear combination of bosonic blocks, the results for the charged sectors are considerably simpler and more surprising. For the $q=\pm2$ components only the superprimary contributes to the superblock. This is very unusual because the superconformal Ward identities (SCWI) \eqref{WI} are related to Ward identities for the one-point functions of the exchanged operators. In particular, an alternative way to fix the coefficients of the linear combination of bosonic primaries in \eqref{superbulkA} would be to find the relation between the one-point function of the superprimary operator in the supermultiplet and that of the bosonic superdescendants. We do this explicitly in appendix \ref{sec:WIs} and we verify that the one-point function of the charge-two operator is already supersymmetric. The other surprising fact of \eqref{020superblock} is that the charged bosonic block is extremely simple. This is again related to the simplicity of the charged one-point function \eqref{1ptO2}. In particular, expressing the one-point function in terms of complex coordinates in the orthogonal plane, one finds that the charged components are (anti)holomorphic. This leads to crucial simplifications in the OPE expansion, which can be easily resummed and leads to \eqref{020superblock}.

The $q=\pm4$ blocks in \eqref{020superblock} vanish identically because the $[0,2,0]$ representation does not contain states with charge $q=\pm4$ under the decomposition $SO(6)_R \rightarrow SO(4)_{\mathcal R} \times SO(2)_{\mathcal{R}_\perp}$. Similar simplifications occur for the other supermultiplets, as shown in appendix \ref{app:blocks}. 

Most importantly, we observe in table \ref{tab:blocksq} that long multiplets do not contribute to the bulk OPE in the sectors with $q\neq 0$. The underlying reason is that a non-vanishing charged one-point function requires the breaking of transverse rotations to be compensated by a suitable R-symmetry structure, in order to preserve a diagonal combination of the two symmetries. However, the superprimary of a long multiplet in \eqref{O2bulkope} is an R-symmetry singlet and therefore no such compensation is possible. Moreover, supersymmetry fixes the one-point functions of its superdescendants in terms of that of the superprimary (this is shown for the stress-tensor multiplet in appendix \ref{sec:WIs}). Consequently, the contribution of the entire long multiplet vanishes.

Finally, we observe again in table \ref{tab:blocksq} that the $q=\pm4$ sector receives contributions from a single non-vanishing block, namely the $\mathcal{B}_{[0,4,0]}$ block. Using its explicit expression given in table \ref{tab:blocksq}, we find that, after including the bulk identity prefactor $\left(\tfrac{\xi}{\xi_R}\right)^{-2}$, the entire contribution reduces to a constant. This constant contribution is simply the disconnected term given by the squared charge-two one-point function
\begin{equation}
    \mathcal{F}^{(\pm 4)}(z,\bar z,\alpha, \bar \alpha) = \left[a^{(\pm2)}_{\mathcal{O}_2}\right]^2\,.
\end{equation}

\subsubsection{Defect channel}
In the defect channel, each external bulk operator is expanded as an infinite tower of defect operators. 
Defect exchanged operators transform in representations of the defect superconformal algebra
$\mathfrak{psu}(1,1|2)\oplus \mathfrak{psu}(1,1|2)\oplus \mathfrak{u}(1)_-$, which we review in appendix \ref{app:irreppsu}.
For bulk scalar operators, the exchanged representations are labeled by the scaling dimension $\hat\Delta$, the $SO(4)_{\mathcal R}$ quantum numbers $\bigl[\tfrac n2,\tfrac n2\bigr]$, and the $U(1)_{-}$ central charge $r_-$. The supermultiplets appearing in the defect OPE are 
\begin{equation}\label{defectOPEreps}
    \mathcal{B}_{[0,p,0]} =
    \sum_{\substack{0\le n\le p\\ r_-}} \mathcal{\hat{B}}_{n/2,r_-}
    + \sum_{\substack{0\leq n \leq p-2 \\ \hat{\Delta}, \,r_-}}\mathcal{\hat{A}}^{\hat{\Delta}}_{n/2,r_-}\,,
\end{equation}
where the notation is summarized in \eqref{notationdefectmult} and for $n = p$ only $r_{-} = 0$ contributes (for $p = 2$, only $\mathcal{\hat{B}}_{1,0}$). Notice that, if $\mathcal{R}_+$ were preserved, we would classify representations also by their $r_+$ quantum numbers (or alternatively by their transverse spin $s$ and the orthogonal R-symmetry charge $r_\perp$). However, when $\mathcal{R}_+$ is broken we classify representations only by their $r_-=r_\perp-s$ eigenvalue. This implies that the sum on $r_-$ runs over all integers. 
The superprimaries of the short multiplets $\mathcal{\hat{B}}_{n/2,r_-}$ have protected scaling dimensions
\begin{equation}\label{unitaritydefshort}
    \hat\Delta=n+|r_-|\,,
\end{equation}
and transform in the $\left[\tfrac{n}{2},\tfrac{n}{2}\right]$ representation of $SO(4)_{\mathcal R}$, with $U(1)_{-}$ charge $r_-$. For $n=0$ or $r_{-} = 0$ these multiplets are $\tfrac12$-BPS, while for $n>0$ and $r_-\neq 0$ they are $\tfrac14$-BPS. A special case is
$\hat{\mathcal{B}}_{0,0} \colonequals \hat{\id}$,
which is simply the identity operator of the defect theory.

The decomposition also includes long multiplets $\mathcal{\hat{A}}^{\hat{\Delta}}_{n/2,r_-}$. By ordinary $SU(4)\to SU(2)\times SU(2) \times U(1)$ decomposition, operators in the $[\frac{n}{2},\frac{n}{2}]$ with $n\leq p$ would be allowed, but supersymmetry forces $n\leq p-2$. This can be seen as follows: the reduced correlator $F^{(0)}(z,\bar z,\alpha,\bar \alpha)$ in \eqref{wardsolution}  can be thought of as a correlator of two scalar operators in the $[0,p-2,0]$ representation. Indeed, for $p=2$ that function is independent of $\alpha$ and $\bar\alpha$. Since long supermultiplets do not contribute to the holomorphic part of the correlator, their properties are the same as a correlator of two operators in the $[0,p-2,0]$ representation. The same mechanism applies to the bulk channel \cite{Dolan:2001tt,Bissi:2015qoa} as one can see in \eqref{bulkOPE}. Operators in these multiplets are not protected and their scaling dimensions satisfy the unitarity bound
\begin{equation}
    \hat{\Delta} >n+|r_-|\,.
\end{equation}

 The details of the derivation of the superblock expansion are presented in appendix \ref{app:defect-superblocks}. Here we just outline the main steps. First, we notice that the bulk-to-defect couplings, when orthogonal rotations are broken, acquire a dependence on the parameter $\eta$ \eqref{etatheta} 
\begin{equation}
\begin{aligned}
  &\left\langle
  \mathcal O_p(P,Y)\,
  \widehat{\mathcal O}_{\widehat\Delta,n,r_-}
  (\widehat P,\widehat W)
  \right\rangle_\Sigma =
  \frac{
    (Y\circ Y)^{\frac12(p-n)}
    (Y\bullet\widehat W)^n
  }{
    (P\circ P)^{\frac12(p-\widehat\Delta)}
    (-2\, P\bullet\widehat P)^{\widehat\Delta}
  }\,
 e^{ir_-\theta}\,
   B_{\widehat\Delta,n,-r_-}(\eta)\,.
   \label{eq:bulk-defect-broken-rotations}
\end{aligned}
\end{equation}
 Here $\widehat P^M=(0,\widehat P^a)$ is the embedding-space position of the defect operator $\widehat{\mathcal O}_{\widehat\Delta,n,r_-}$, which has no transverse components. The $SO(4)_{\mathcal R}$ indices of the defect operator are contracted with a null
polarization vector $\widehat W^a$, $a=3,\dots,6$, with
$\widehat W\bullet\widehat W=0$, so that it transforms in the
$[\tfrac n2,\tfrac n2]$ representation, i.e. the rank-$n$ symmetric
traceless one. Finally, $r_-$ is the charge of the inserted defect operator. The coefficient $B_{\widehat\Delta,n,r_-}(\eta)$ is the bulk-to-defect coupling of the operator of charge $r_-$ in the defect OPE of $\mathcal{O}_p$. Its dependence on $\eta$ shows up in the bosonic block expansion \eqref{defblockOPE} and it can be traded for the two cross-ratios $\vartheta$ \eqref{theta} and $\omega$ \eqref{udef} (see \eqref{Beta1Beta2} for more details). The bosonic blocks
\begin{equation}
 \widehat g_{\hat\Delta,n,r_-}(r,r_R,w)
 =f_{\hat\Delta}(r)f_{-n}(r_R)w^{-r_-}\,,
 \label{eq:defect-bosonic-block-main}
\end{equation}
 with $f_a$ defined in \eqref{eq:defect-radial-function}, carry the dependence on the other three cross-ratios. The factor $w^{-r_-}$ follows from our convention in the bulk-to-defect couplings and is essential for the correct rotation-preserving limit as shown in appendix \ref{app:bosonicdefectblocks}. 

The defect superconformal block decomposition organizes the two-point function as
\begin{equation}
\mathcal F(z,\bar z,\alpha,\bar\alpha,\vartheta)
=
\sum_q \sum_{\hat{\mathcal O}}
\texttt{B}_{\hat{\mathcal O}}^{(q)}(\omega)\,
e^{i q \frac{\vartheta}{2}}\,
\widehat{\mathcal G}_{\widehat{\mathcal M}}
(z,\bar z,\alpha,\bar\alpha)\,.
\end{equation}
Here, each term in the sum is associated with a defect superconformal 
multiplet $\widehat{\mathcal M}$ whose superconformal primary is denoted by 
$\hat{\mathcal O}$, while 
$\widehat{\mathcal G}_{\widehat{\mathcal M}}$ denotes the corresponding 
defect superconformal block. The dynamical information entering this 
decomposition is encoded in the functions 
$\texttt{B}_{\hat{\mathcal O}}^{(q)}(\omega)$, with $\omega=\sqrt{\tfrac{z \bar \alpha}{\bar z \alpha}}$, that are obtained from the product of the bulk-to-defect OPE coefficients \eqref{eq:bulk-defect-broken-rotations} associated with the superprimary $\hat{\mathcal O}$ and its conjugate, see \eqref{Beta1Beta2} and \eqref{B2}.

A superconformal block is a finite linear combination
\begin{equation}
 \widehat{\mathcal G}_{\widehat{\mathcal M}}
 =\sum_I c_I(\omega)\,
 \widehat g_{\hat\Delta_I,n_I,r_-}(r,r_R,w)\,,
 \label{eq:defect-superblock-main}
\end{equation}
with the important caveat that the coefficients $c_I(\omega)$ are finite Laurent polynomials in $\omega$. The SCWI fix the relative coefficients in this expansion, leaving an arbitrary overall function of $\omega$ undetermined. We remove this ambiguity by normalizing the coefficient of the superprimary bosonic block to one. The resulting blocks with $r_-\geq0$ are collected in table~\ref{tab:defblocks}. Once more, we refer to appendix \ref{app:defect-superblocks} for a thorough derivation. For completeness, we note that it is sufficient to determine the superconformal blocks and the coefficients $\texttt{B}_{\hat{\mathcal O}}^{(q)}(\omega)$ for $r_->0$. Their counterparts in the $r_-<0$ sector are then obtained by applying the transformation
\begin{equation}\label{minusplus}
  (z,\bar z,\alpha,\bar\alpha,r_-)
  \longleftrightarrow
  (\bar z,z,\bar\alpha,\alpha,-r_-).
\end{equation}
For the two-point function of $\mathcal{O}_2$, the individual Fourier sectors can be expressed explicitly as
\begin{eqnarray}\label{defectOPEexpl}
    \mathcal F^{(0)}(z,\bar z, \alpha, \bar \alpha)&=& \left[ a^{(0)}_{\mathcal{O}_2}\right]^2+a^{(+2)}_{\mathcal{O}_2} a^{(-2)}_{\mathcal{O}_2} \left(\frac{\alpha \bar z}{\bar{\alpha} z}+\frac{\bar{\alpha} z}{\alpha \bar z}\right)\nonumber\\
&+&\texttt{B}_{\widehat B_{1/2,0}}^{(0)}\!(\omega)\,
 \widehat{\mathcal G}_{\widehat B_{1/2,0}}\!(z,\bar z, \alpha, \bar \alpha)
+\texttt{B}_{\widehat B_{1,0}}^{(0)}\!(\omega)\,
 \widehat{\mathcal G}_{\widehat B_{1,0}}\!(z,\bar z, \alpha, \bar \alpha) \nonumber
\\
&+&\sum_{r_->0}\Big[
\texttt{B}_{\widehat B_{0,r_-}}^{(0)}\!\!\!(\omega)\,
 \widehat{\mathcal G}_{\widehat B_{0,r_-}}\!\!(z,\bar z, \alpha, \bar \alpha)
+\texttt{B}_{\widehat B_{0,-r_-}}^{(0)}\!\!\!(\omega)\,
 \widehat{\mathcal G}_{\widehat B_{0,-r_-}}\!\!(z,\bar z, \alpha, \bar \alpha) \nonumber
\\
&+&\texttt{B}_{\widehat B_{1/2,r_-}}^{(0)}\!\!\!(\omega)\,
 \widehat{\mathcal G}_{\widehat B_{1/2,r_-}}\!\!(z,\bar z, \alpha, \bar \alpha)
+\texttt{B}_{\widehat B_{1/2,-r_-}}^{(0)}\!\!\!(\omega)\,
 \widehat{\mathcal G}_{\widehat B_{1/2,-r_-}}\!\!(z,\bar z, \alpha, \bar \alpha)
\Big] \nonumber
\\
&+&\sum_{r_-}\sum_{\hat\Delta}
\texttt{B}_{\widehat A^{\hat\Delta}_{0,r_-}}^{(0)}\!\!\!(\omega)\,
\widehat{\mathcal G}_{\widehat A^{\hat\Delta}_{0,r_-}}\!\!(z,\bar z, \alpha, \bar \alpha)\,, \\
\mathcal F^{(\pm2)}(z,\bar z,\alpha, \bar \alpha)
&=&a^{(0)}_{\mathcal{O}_2} a^{(\pm2)}_{\mathcal{O}_2} \left(\sqrt{\frac{\bar{\alpha} z}{\alpha \bar z}}+\sqrt{\frac{\alpha \bar z}{\bar{\alpha} z}}\right)+\texttt{B}_{\widehat B_{1/2,0}}^{(\pm2)}\!(\omega)\,
 \widehat{\mathcal G}_{\widehat B_{1/2,0}}\!(z,\bar z,\alpha, \bar \alpha) \nonumber
\\
&+&\sum_{r_->0}\Big[
\texttt{B}_{\widehat B_{0,r_-}}^{(\pm2)}\!(\omega)\,
 \widehat{\mathcal G}_{\widehat B_{0,r_-}}\!\!(z,\bar z,\alpha, \bar \alpha)
+\texttt{B}_{\widehat B_{0,-r_-}}^{(\pm2)}\!\!(\omega)\,
 \widehat{\mathcal G}_{\widehat B_{0,-r_-}}\!\!(z,\bar z,\alpha, \bar \alpha)
\Big]\,,\nonumber\\
\mathcal F^{(\pm4)}(z,\bar z,\alpha,\bar \alpha)&=&[a^{(\pm2)}_{\mathcal{O}_2}]^2\,.\nonumber
\end{eqnarray}
Here the blocks and coefficients with charge $-r_-$ are obtained from those with charge $r_-$ using \eqref{minusplus}. Moreover, we used that the bulk-defect coefficient of the identity operator is the one-point function of the external operator,
\begin{equation}
    \sum_q \texttt{B}^{(q)}_{\hat{\id}}(\omega) e^{i q \frac{\vartheta}{2}} =A_{\mathcal{O}_2}(\eta_1)A_{\mathcal{O}_2}(\eta_2)\,.
\end{equation}

\section{Analytic bootstrap of \texorpdfstring{$\mathbf{\langle \mathcal{O}_2 \mathcal{O}_2\rangle_\Sigma}$}{<O2 O2> Sigma}}\label{sec:bootstrapO2}

In this section we analytically bootstrap the two-point function of $\mathcal{O}_2$ in the presence of a half-BPS surface defect, at the first non-trivial order in the large-$N$ and strong ’t Hooft coupling expansion. More precisely, we retain the leading contribution of order $N^{-1}$, while corrections of order $N^{-2}$ are neglected. In particular, we focus on the probe brane regime discussed in section \ref{sec:holo}. Remarkably, part of the result turns out to be exact in both $N$ and the coupling. At the end of this section, we extract the CFT data from the correlator, which must be interpreted as averages over perturbatively degenerate operators. The analysis presented here can be straightforwardly extended to the two-point function of $\mathcal{O}_p$ for arbitrary $p$. 

\subsection{Conformal dispersion relation}\label{sec:dispersion}
Our main tool for determining the correlator is a conformal dispersion relation\footnote{See also \cite{Caron-Huot:2017vep,Carmi:2019cub,Bissi:2019kkx, Ferragatta:2026ilp} for related techniques in the absence of a defect.} \cite{Bianchi:2022ppi,Barrat:2022psm}. This formula reconstructs the full correlator from its singularities in a suitable complex kinematic variable, in the spirit of Cauchy's theorem. The key observation is that these singularities are fixed by a subset of the OPE data. 

The singularities entering the dispersion relation arise from Lorentzian kinematics. More precisely, the two-point function of $\mathcal{O}_p$ develops singularities in the lightcone limit, $z \to 1$ or $\bar z \to 1$, where one operator crosses the lightcone of the other \cite{Lemos:2017vnx}. In terms of the radial variables defined in \eqref{radialcoord}, this limit corresponds to $w\to r$ or $w\to 1/r$, where the correlator develops branch points in the complex $w$-plane. Each Fourier mode of the correlator $\mathcal{F}^{(q)}(r,w,r_R,w_R)$ \footnote{With an abuse of notation we use $\mathcal{F}^{(q)}(r,w,r_R,w_R)$ and $\mathcal{F}^{(q)}(z,\bar z,\alpha,\bar \alpha)$ to indicate the same object upon the identifications \eqref{radialcoord}.} exhibits these singularities and can be reconstructed from them via a dispersion relation, which takes the form \cite{Bianchi:2022ppi, Barrat:2022psm} 
\begin{equation}\label{dispersion}
\begin{split}
\mathcal{F}^{(q)}(r,w,r_R,w_R) &\cong \int_{0}^{r} \frac{dw'}{2\pi i}\, \frac{\text{Disc}_{w<r}\mathcal{F}^{(q)}(r,w',r_R,w_R)}{w'-w}\\
&+ \int_{1/r}^{\infty} \frac{dw'}{2\pi i}\, \frac{\text{Disc}_{w>1/r}\mathcal{F}^{(q)}(r,w',r_R,w_R)}{w'-w}\, .
\end{split}
\end{equation}
The symbol $\cong$ indicates that the equality holds up to possible ambiguities, which will be discussed below. The presence of two distinct integrals reflects the fact that, for codimension-$2$ defects, the correlator is not required to be invariant under $w\leftrightarrow 1/w$ (see the discussion around equation \eqref{symF}). As a result, the regions $w<r$ and $w>1/r$ give independent contributions \cite{Gimenez-Grau:2021wiv}. Note that the radial variable $r$ and the R-symmetry cross-ratios $r_R$ and $w_R$ enter in \eqref{dispersion} only as spectator variables.

The dispersion relation reconstructs the correlator from its discontinuities across the branch cuts in the $w$ variable:
\begin{equation}
\text{Disc}\mathcal{F}^{(q)}(r,w,r_R,w_R) \colonequals \mathcal{F}^{(q)}(r, w+i \epsilon,r_R,w_R)-\mathcal{F}^{(q)}(r,w-i\epsilon,r_R,w_R)\, .
\end{equation}
These discontinuities receive contributions from operators appearing in the bulk OPE channel. This is because the bulk OPE expansion converges in the vicinity of the branch points $w=r$ and $w=1/r$, allowing one to express the discontinuity in terms of the OPE data.
Explicitly, 
\begin{equation}\label{Discwithblocks}
    \text{Disc}\mathcal{F}^{(q)}(r,w,r_R, w_R) =  \sum_{ \mathcal{O}}(-2)^{n_\mathcal{O} -p} \lambda_{\mathcal{O}_p\mathcal{O}_p\mathcal{O}}\, a^{(q)}_{\mathcal{O}} \,\text{Disc}\left[\left(\frac{\xi}{\xi_R}\right)^{-p} \mathcal{G}^{(q)}_{\mathcal{M}_\mathcal{O}}(z,\bar z,\alpha,\bar \alpha) \right]\,.
\end{equation}
The structure of the branch points at $w=r$ and $w=1/r$ (i.e. $z=1$ and $\bar z=1$) is manifest from the explicit form of the (super)conformal blocks, which exhibit non-integer power-law behavior, see \eqref{noneqblock} and \eqref{noneq}. Schematically, for $w \to r$,
\begin{equation}\label{singblock}
    f^{(q)}_{\Delta, \ell}(z,\bar z)|_{z=rw,\bar z=r/w} \sim (w-r)^{\frac{\Delta-\ell}{2}} \times (\text{regular at $w=r$})\,.
\end{equation}
The same power law behavior arises in the limit where $w \to 1/r$.
In some circumstances, as we will see below, the discontinuity kills many contributions reducing the result to a limited set of CFT data. For these cases, the dispersion relation is a powerful tool to reconstruct the full correlator. 

Even when the exponent in \eqref{singblock} is an integer, the discontinuity can contribute to the correlator when $\Delta-\ell<2p$, once the prefactor $\xi^{-p}$ is included. Indeed, the discontinuity should be interpreted in a distributional sense
\begin{equation}\label{discdelta}
\text{Disc}_{z<0}\left[z^{-n}\right] = \frac{2\pi i\,(-1)^n}{(n-1)!}\, \partial^{\,n-1}\delta(z)\, , 
\end{equation}
where $\delta(z)$ denotes the Dirac delta function. 

The derivation of \eqref{dispersion} relies on a contour deformation argument, which assumes that the correlator has no pole at $w=0$ and exhibits sufficiently fast decay at large $w$. More precisely, one requires
\begin{equation}
\begin{split}
    \mathcal{F}^{(q)}(r,w,\alpha, \bar \alpha) &\sim w^{s^*}, \qquad w\to\infty, \qquad s^*<0\,, \\
    \mathcal{F}^{(q)}(r,w,\alpha, \bar \alpha) &\sim w^{-s^*}, \qquad w\to0, \qquad s^*<0\, .
    \end{split}
\end{equation}
If instead $s^*\ge 0$, the dispersion relation may fail to reconstruct the full correlator. This is the origin of the approximate equality symbol in \eqref{dispersion}. The missing contributions correspond to the exchange of defect operators with low transverse spin $|s|\le s^*$ in the defect OPE (we use the absolute value because for codimension-two defects the transverse spin can have
either sign) \cite{Lemos:2017vnx}. The ambiguity associated with the low-spin contributions can often be resolved by combining the dispersion relation with crossing symmetry and the superconformal Ward identities \cite{Barrat:2021yvp}. This strategy will enable us to reconstruct the full two-point function in the following sections.

\subsection{Computation of the discontinuity}
For the charge-$q$ components of the two-point function of $\mathcal{O}_2$, the discontinuity at $w=r$ or $w=1/r$ is given by \eqref{Discwithblocks}:
\begin{equation}
\text{Disc}\mathcal{F}^{(q)}(r,w,r_R, w_R) =  \sum_{\mathcal{O}} (-2)^{n_\mathcal{O} -2} \lambda_{\mathcal{O}_2\mathcal{O}_2\mathcal{O}}\, a^{(q)}_{\mathcal{O}}\, \text{Disc}\left[\left(\frac{\xi}{\xi_R}\right)^{-2} \mathcal{G}^{(q)}_{\mathcal{M}_\mathcal{O}}(z,\bar z,\alpha,\bar \a) \right]\,, 
\end{equation}
where $q=0,\pm2,\pm4$ and the sum runs over the supermultiplets listed in \eqref{O2bulkope}. The explicit form of the corresponding superconformal blocks is given by \eqref{blockWI} and tables \ref{tab:blocks} and \ref{tab:blocksq}. Building on the discussion around \eqref{singblock} and taking into account the prefactor $\xi^{-2}=\tfrac{r^2w^2}{(1-rw)^2(w-r)^2}$, we find that a block has a non-vanishing discontinuity only in two cases. The first occurs when the exchanged primary operator has integer dimension $\Delta<\ell+4$, in which case the block develops a pole. The resulting discontinuity must then be understood in the distributional sense, as in \eqref{discdelta}. From the explicit structure of the supermultiplets \cite{Dolan:2004mu} and of the corresponding superconformal blocks (see appendix \ref{app:blocks}), we see that in the present setup discontinuities of this type arise only from the bulk identity and the $\mathcal{B}_{[0,2,0]}$ supermultiplet.

The second case arises when the exchanged primary has non-integer scaling dimension, leading to branch cuts in the block. This occurs for operators belonging to long supermultiplets. As discussed in section \ref{sec:bulkOPE}, such operators do not appear in the OPE in the charged sectors with $q\neq 0$. Moreover, at large $N$ the correlator factorizes and it reduces to disconnected contributions, i.e. to the two-point function of a dimension-two scalar field in generalized free field (GFF) theory without any defect plus the square of the one-point function. Therefore, in the strict large $N$ limit the only exchanged operators are double trace operators with integer dimension $\Delta=4+2n+\ell+\mathcal{O}(1/N^2)$, where $\ell$ is even. The first subleading correction appears at order $1/N^2$ \cite{DHoker:1999kzh,Arutyunov:2000py,Dolan:2001tt,Goncalves:2014ffa}
\begin{equation}\label{dtdim}
\Delta = 4+2n+\ell +\frac{1}{N^2}\left(c_1+\frac{c_2}{\lambda^{3/2}}+\ldots\right) +\mathcal{O}(1/N^{4})\,,
\end{equation}
which implies that at order $1/N$ these operators enter in the OPE with their classical (integer) dimensions $\Delta=4+2n+\ell$ and do not contribute to the discontinuity\footnote{In order to avoid contributions to the discontinuity from long operators, it is crucial that they satisfy $\Delta-\ell \geq 2p$, with $\Delta-\ell$ positive even integer. This condition is satisfied at strong coupling up to $\mathcal{O}(1/N^2)$, see \eqref{dtdim}. At weak coupling, however, this bound can be violated, and long operators contribute to the discontinuity at all orders in the weak-coupling expansion.}, see \eqref{singblock}. At the order we are working, this means that only protected operators contribute to the discontinuity. 

For the two-point function of $\mathcal{O}_2$, this singles out the contribution of the bulk identity and of the $\mathcal{B}_{[0,2,0]}$ supermultiplet, whose one-point function is the only new piece of data we need to reconstruct the correlator (up to low spin ambiguities). As a result, the discontinuity around $w=r$ or $w=1/r$ takes the form
\begin{equation}\label{discO2}
\text{Disc}\mathcal{F}^{(q)}=\frac{\delta_{q,0}}{4}\text{Disc}\left[\left(\frac{\xi}{\xi_R}\right)^{-2}\right]\!- \frac12\, \lambda_{\mathcal{O}_2\mathcal{O}_2\mathcal{O}_2}\, a^{(q)}_{\mathcal{O}_2}\, \text{Disc}\left[\left(\frac{\xi}{\xi_R}\right)^{-2} \!\!\mathcal{G}^{(q)}_{\mathcal{B}_{[0,2,0]}}\right]\!+\mathcal{O}(1/N^2)\,,
\end{equation}
where the OPE coefficient $\lambda_{\mathcal{O}_2\mathcal{O}_2\mathcal{O}_2}$ is given in \eqref{lambdaO2}, while $a^{(q)}_{\mathcal{O}_2}$, for the case of a Gukov-Witten defect, is given in  \eqref{a2broken}.

Substituting the explicit form of the superconformal block $\mathcal{G}^{(q)}_{\mathcal{B}_{[0,2,0]}}$ \eqref{020superblock} and the definition of $\xi$ \eqref{crossratio} into \eqref{discO2}, the discontinuity can be expressed in radial variables as 
\begin{align}\label{discO2comp}
&\operatorname{Disc}_{w<r}
\mathcal{F}^{(0)}\!\left(r,w,r_R,w_R\right)\nonumber
\\
&\quad=
\frac{2\pi i\,r^3\xi_R}{(1-r^2)^2}
\Bigg[
\frac{\xi_R}{4}\left(r\partial_w-\frac{2}{1-r^2}\right)
+\frac12 a\lambda_{\mathcal{B}_{[0,2,0]}}^{(0)}
\Bigg\{
\frac{(1+r^2)\xi_R}{2(1-r^2)}
+w_R-\frac{1}{w_R}\nonumber
\\
&\qquad\qquad
-\frac{2\log r}{w_R(1-r^2)^2}
\left(r^2w_R-r_R\right)
\left(r^2w_R-\frac{1}{r_R}\right)
\Bigg\}
\Bigg]\delta(w-r)
+\mathcal{O}\!\left({1}/{N^2}\right),\nonumber
\\[8pt]
&\operatorname{Disc}_{w>1/r}
\mathcal{F}^{(0)}\!\left(r,w,r_R,w_R\right)\nonumber
\\
&\quad=
\frac{2\pi i\,\xi_R}{(1-r^2)^2}
\Bigg[
\frac{\xi_R}{4}\left(\partial_w+\frac{2r^3}{1-r^2}\right)
-\frac12 a\lambda_{\mathcal{B}_{[0,2,0]}}^{(0)}r
\Bigg\{
\frac{(1+r^2)\xi_R}{2(1-r^2)}
-w_R+\frac{1}{w_R}\nonumber
\\
&\qquad\qquad
-\frac{2\log r}{w_R(1-r^2)^2}
\left(r^2-r_Rw_R\right)
\left(r^2-\frac{w_R}{r_R}\right)
\Bigg\}
\Bigg]\delta\!\left(w-\frac{1}{r}\right)
+\mathcal{O}\!\left({1}/{N^2}\right),\nonumber
\\[8pt]
&\operatorname{Disc}_{w<r}
\mathcal{F}^{(\pm2)}\!\left(r,w,r_R,w_R\right)
=
\pi i\,a\lambda_{\mathcal{B}_{[0,2,0]}}^{(\pm2)}
\frac{r^2\xi_R}{1-r^2}\,\delta(w-r),\nonumber
\\[8pt]
&\operatorname{Disc}_{w>1/r}
\mathcal{F}^{(\pm2)}\!\left(r,w,r_R,w_R\right)
=
-\pi i\,a\lambda_{\mathcal{B}_{[0,2,0]}}^{(\pm2)}
\frac{\xi_R}{1-r^2}\,
\delta\!\left(w-\frac{1}{r}\right),\nonumber
\\[8pt]
&\operatorname{Disc}_{w<r}
\mathcal{F}^{(\pm4)}\!\left(r,w,r_R,w_R\right)
=
\operatorname{Disc}_{w>1/r}
\mathcal{F}^{(\pm4)}\!\left(r,w,r_R,w_R\right)
=0\, ,
\end{align}
where we used $\xi_R$ defined in \eqref{crossratio} and we have introduced the shorthand notation 
\begin{equation}\label{alambda}
a\lambda_{\mathcal{M}_\mathcal{O}}^{(q)}\colonequals\lambda_{\mathcal{O}_2\mathcal{O}_2\mathcal{O}}\, a^{(q)}_{\mathcal{O}}\, ,
\end{equation}
with $\mathcal{O}$ being the superprimary of the $\mathcal{M}_\mathcal{O}$ supermultiplet.
Notice that the discontinuities in the charged sectors are exact, because the correlator receives no contributions from long multiplets in
these channels, as follows from the explicit form of the superconformal blocks in appendix \ref{app:blocks}. On the other hand, the expression for the discontinuity in the $q=0$ sector is valid only at large $N$ and strong coupling. 

\subsection{Rotation-invariant sector}\label{sec:q0}

In the rotation-invariant sector $q=0$, inserting the discontinuity \eqref{discO2comp} into the dispersion relation
\eqref{dispersion}, we obtain
\begin{eqnarray}\label{resdisp}
    \mathcal{F}^{(0)}(z,\bar z,\a,\bar \a)  &\cong&- \frac12 a\lambda_{\mathcal{B}_{[0,2,0]}}^{(0)}\frac{(1-\alpha )^2 (1-\bar \a)^2 }{\alpha  \bar \a} \left[\frac{z \bar z \left(1-z^2 \bar z^2+2 z \bar z \log (z \bar z)\right)}{2 (1-z) (1-\bar z)(1-z \bar z)^3} \right.\nonumber\\
    &+&\!\left. \frac{\alpha \!-\!\bar \a}{(1\!-\!\alpha ) (1\!-\!\bar \alpha)} \frac{z \bar z  ((z-2) \bar z+1) (2(1- z \bar z)+(1+\!z \bar z) \log (z \bar z))}{2 (1-z) (1-\bar z) (1-z \bar z)^3} \right.\\
   &-&\left. \frac{\alpha \!+\!\bar \alpha}{(1\!-\!\alpha ) (1\!-\!\bar \alpha)} \frac{z \bar z \log (z \bar z)}{2 (1-z) (1-\bar z) (1-z \bar z)}\right]\!+\!\frac{(1\!-\!\alpha )^2 (1\!-\!\bar \a)^2 z \bar z}{4\, (1\!-\!z)^2 (1\!-\!\bar z)^2 \alpha  \bar \alpha}\!+\!\mathcal{O}\!\left(1/N^2\right)\!\, , \nonumber
\end{eqnarray}
where we expressed the result in lightcone coordinates \eqref{lightcone}. The last term corresponds to the bulk identity contribution, namely the two-point function in the absence of the defect.

The result obtained from the dispersion relation \eqref{resdisp} does not satisfy either the symmetry properties \eqref{symF} or the Ward identities \eqref{WI}, signaling that additional contributions associated with low transverse spin are missing, as expected from the discussion in subsection \ref{sec:dispersion}. The full result for $\mathcal{F}^{(0)}(z,\bar z,\a,\bar \a)$ must have the form
\begin{eqnarray}\label{Fansatz}
      \mathcal{F}^{(0)}(z,\bar z,\alpha,\bar \alpha)&=& -\frac{a\lambda_{\mathcal{B}_{[0,2,0]}}^{(0)} }{2}\left[\frac{(1-\alpha )^2 (1-\bar \alpha)^2}{ \alpha  \bar \alpha}\left(\!\mathcal F^{(0)}_0(z,\bar z)+\frac{z \bar z \left(1-z^2 \bar z^2+2 z \bar z \log (z \bar z)\right)}{2 (1-z) (1-\bar z) (1-z \bar z)^3}\!\right) \right.\nonumber\\
    &\!\!\!\!\!\!\!\!\!\!\!\!\!\!\!\!\!\!\!\!\!\!\!\!\!\!\!\!\!\!\!\!\!\!\!\!\!\!\!\!\!\!\!\!\!\!\!\!\!\!\!\!\!\!\!\!\!\!\!\!-&\!\!\!\!\!\!\!\!\!\!\!\!\!\!\!\!\!\!\!\!\!\!\!\!\!\!\!\!\!\!\! \left. \frac{(1-\alpha ) (1-\bar \alpha)(\alpha -\bar \alpha)}{\alpha \bar \alpha} \left(\! \mathcal F^{(0)}_1(z,\bar z)+\frac{z \bar z ((2-z) \bar z-1) (2(1- z \bar z)+(1+z \bar z) \log (z \bar z))}{2(1-z) (1-\bar z) (1-z \bar z)^3}\!\right) \right.\nonumber \\
   &\!\!\!\!\!\!\!\!\!\!\!\!\!\!\!\!\!\!\!\!\!\!\!\!\!\!\!\!\!\!\!\!\!\!\!\!\!\!\!\!\!\!\!\!\!\!\!\!\!\!\!\!\!\!\!\!\!\!\!\!+&\!\!\!\!\!\!\!\!\!\!\!\!\!\!\!\!\!\!\!\!\!\!\!\!\!\!\!\!\!\!\!\left. \frac{(1-\alpha ) (1-\bar \a)(\alpha +\bar \alpha)}{\alpha \bar \alpha} \left(\mathcal F^{(0)}_2(z,\bar z)-\frac{z \bar z \log (z \bar z)}{2 (1-z) (1-\bar z)(1-z \bar z)}\right)+\frac{(\alpha -\bar \a)^2}{\alpha \bar \alpha} \mathcal F^{(0)}_3(z,\bar z)\right. \nonumber \\
   &\!\!\!\!\!\!\!\!\!\!\!\!\!\!\!\!\!\!\!\!\!\!\!\!\!\!\!\!\!\!\!\!\!\!\!\!\!\!\!\!\!\!\!\!\!\!\!\!\!\!\!\!\!\!\!\!\!\!\!\!+&\!\!\!\!\!\!\!\!\!\!\!\!\!\!\!\!\!\!\!\!\!\!\!\!\!\!\!\!\!\!\! \left. \frac{(\alpha -\bar \alpha) (\alpha +\bar \alpha) }{\alpha \bar \alpha} \mathcal F^{(0)}_4(z,\bar z)+\frac{(\alpha +\bar \alpha)^2}{\alpha \bar \alpha} \mathcal F^{(0)}_5(z,\bar z)\right]+\frac{(1-\alpha )^2 (1-\bar \alpha)^2 z \bar z}{4(1-z)^2 (1-\bar z)^2 \alpha  \bar \alpha}\nonumber\\
   &\!\!\!\!\!\!\!\!\!\!\!\!\!\!\!\!\!\!\!\!\!\!\!\!\!\!\!\!\!\!\!\!\!\!\!\!\!\!\!\!\!\!\!\!\!\!\!\!\!\!\!\!\!\!\!\!\!\!\!\!+&\!\!\!\!\!\!\!\!\!\!\!\!\!\!\!\!\!\!\!\!\!\!\!\!\!\!\!\!\!\!\!\left[a^{(0)}_{\mathcal{O}_2}\right]^2+a^{(+2)}_{\mathcal{O}_2} a^{(-2)}_{\mathcal{O}_2} \left(\frac{\alpha \bar z}{\bar{\alpha} z}+\frac{\bar{\alpha} z}{\alpha \bar z}\right)+\mathcal{O}\left(1/N^2\right)\,.
\end{eqnarray}
Here we have organized the answer according to the six allowed
R-symmetry structures. To count them one can simply enumerate singlet representations in the decomposition of the product $\mathbf{20'}\times\mathbf{20'}$ under the $SO(6)\to SO(4)\times U(1)$ symmetry breaking. The functions $\mathcal F^{(0)}_i (z,\bar z)$ encode the missing low-spin contributions. In the following sections, we will determine these functions exploiting the general solution of the Ward identities \eqref{wardsolution} and other properties of the correlator. The last term in \eqref{Fansatz} corresponds to the charge $q=0$ component of the product of the two one-point functions,
\begin{eqnarray}\label{squared1ptcharge}
  A_{\mathcal{O}_2}(\eta_1)A_{\mathcal{O}_2}(\eta_2) &=& \left[ a^{(0)}_{\mathcal{O}_2}\right]^2+a^{(+2)}_{\mathcal{O}_2} a^{(-2)}_{\mathcal{O}_2} \left(\frac{\alpha \bar z}{\bar{\alpha} z}+\frac{\bar{\alpha} z}{\alpha \bar z}\right) \nonumber \\
  &+&a^{(0)}_{\mathcal{O}_2} a^{(+2)}_{\mathcal{O}_2} e^{i \vartheta} \left(\sqrt{\frac{\bar{\alpha} z}{\alpha \bar z}}+\sqrt{\frac{\alpha \bar z}{\bar{\alpha} z}}\right)+a^{(0)}_{\mathcal{O}_2} a^{(-2)}_{\mathcal{O}_2} e^{-i \vartheta} \left(\sqrt{\frac{\bar{\alpha} z}{\alpha \bar z}}+\sqrt{\frac{\alpha \bar z}{\bar{\alpha} z}}\right)\nonumber \\
  &+&\left[a^{(+2)}_{\mathcal{O}_2} \right]^2 e^{2 i \vartheta }+\left[a^{(-2)}_{\mathcal{O}_2}\right]^2 e^{-2 i \vartheta }\,,
\end{eqnarray}
where the $a^{(q)}_{\mathcal{O}_2}$ coefficients are given in \eqref{a2broken}, in the specific case of a Gukov-Witten defect. Together with the bulk identity contribution, it accounts for the disconnected part of the two-point function.

\subsubsection{Holomorphic correlator}\label{holomorphicorr}
The two-point function \eqref{Fansatz} must be consistent with the Ward identity solution \eqref{wardsolution}. In particular, setting $\bar z=\bar \alpha$, it  has to reduce to a holomorphic function $f^{(0)}(z,\alpha)$. Further restricting to $z=\bar z=\alpha=\bar \alpha$, the correlator simplifies to the constant $-\mathcal{C}^{(0)}$. In the case of a Gukov-Witten defect, $\mathcal{C}^{(0)}$ is given in \eqref{Cloc}, for arbitrary values of $N$ and of the coupling. More generally, the superconformal blocks in tables \ref{tab:blocks} and \ref{tab:defblocks}
show that $\mathcal{C}^{(0)}$ can be expressed in terms of protected OPE data either in the bulk or defect channel,
\begin{align}
    \mathcal{C}^{(0)}&=-1/4+ \frac 12 a \lambda_{\mathcal{B}_{[0,2,0]}}^{(0)}-a \lambda_{\mathcal{B}_{[0,4,0]}}^{(0)}\\
    &=-\left[a^{(0)}_{\mathcal{O}_2}\right]^2-
2a^{(+2)}_{\mathcal{O}_2}a^{(-2)}_{\mathcal{O}_2}-\texttt{B}_{\widehat{\mathcal B}_{1/2,0}}^{(0)}
    \left(\omega\right)\big|_{\omega=1}-\texttt{B}_{\widehat{\mathcal B}_{1,0}}^{(0)}
    \left(\omega\right)\big|_{\omega=1}\, .\nonumber
\end{align}
The holomorphic correlator $f^{(0)}(z,\alpha)$ can be fixed non-perturbatively, i.e. to all orders in $1/N$ and in the coupling $\lambda$. It can be reconstructed from a small subset of protected CFT data via Cauchy's theorem\footnote{See \cite{Rigatos:2025sar} for an alternative approach based on an ansatz constrained by crossing symmetry.}, or equivalently through a holomorphic version of the dispersion relation \eqref{dispersion}. This is equivalent to what we did in section \ref{chiralalgebra} using the chiral algebra language.

Our strategy is to compute the holomorphic function exactly and use it to fix part of the ambiguities in the full result \eqref{Fansatz}. To this end, we assume that the only spacetime singularities of $f^{(0)}(z,\alpha)$ occur at $z=0$ and $z=1$\footnote{This assumption is justified by the chiral algebra
construction of section \ref{chiralalgebra}.}.
The limit $z\to 0$ is governed by the defect OPE \eqref{defectOPEexpl}, while $z\to 1$ is controlled by the bulk OPE \eqref{bulkblockgen}. 

Explicitly, the bulk channel expansion of the holomorphic correlator reads
\begin{eqnarray}\label{bulkOPEf}
    f^{(0)}(z,\alpha)&=&\frac{z (1\!-\!\alpha)^2}{4\alpha (1\!-\!z)^2}- a \lambda_{\mathcal{B}_{[0,2,0]}}^{(0)} \frac{z (1\!-\!\alpha)^2}{2\alpha (1\!-\!z)^2} g^{(0)}_{\mathcal{B}_{[0,2,0]}} (z,\alpha)+a \lambda_{\mathcal{B}_{[0,4,0]}}^{(0)} \frac{z (1\!-\!\alpha)^2}{\alpha (1\!-\!z)^2}g^{(0)}_{\mathcal{B}_{[0,4,0]}} (z,\alpha) \nonumber\\
    &+&a \lambda_{\mathcal{B}_{[2,0,2]}}^{(0)} \frac{z (1\!-\!\alpha)^2}{\alpha (1\!-\!z)^2}g^{(0)}_{\mathcal{B}_{[2,0,2]}} (z,\alpha) - \frac12 \sum_{\ell \ \text{even}} a \lambda_{\mathcal{C}_{[0,2,0], \ell}}^{(0)} \frac{z (1\!-\!\alpha)^2}{\alpha (1\!-\!z)^2} g^{(0)}_{\mathcal{C}_{[0,2,0]},\ell} (z,\alpha)\nonumber \\
    &-\frac12&\sum_{\ell \ \text{odd}} a \lambda_{\mathcal{C}_{[1,0,1],\ell}}^{(0)}\frac{z (1\!-\!\alpha)^2}{\alpha (1\!-\!z)^2} g^{(0)}_{\mathcal{C}_{[1,0,1]}, \ell} (z,\alpha)\,,
\end{eqnarray}
where the holomorphic blocks $g^{(0)}_{\mathcal{M}_\mathcal{O}} (z,\alpha)$ are given in table \ref{tab:blocks}. Notice that long multiplets do not appear in the expansion. The only singular contributions in the limit $z\to 1$ arise from the bulk identity and the $g^{(0)}_{\mathcal{B}_{[0,2,0]}}$ block, leading to
\begin{equation}\label{fsing1}
f^{(0)}(z,\alpha)
\xrightarrow{z \to 1}
\frac{z(1-\alpha)^2}{4 \alpha (1-z)^2}
-\frac12
a\lambda_{\mathcal{B}_{[0,2,0]}}^{(0)} \frac{1-\alpha ^2}{2 \alpha (1-z)}
+\mathcal{O}\!\bigl((1-z)^0\bigr)\, .
\end{equation}
The defect channel expansion reads 
\begin{eqnarray}
    f^{(0)}(z,\alpha) &=&
\left[a^{(0)}_{\mathcal{O}_2}\right]^2
+
a^{(+2)}_{\mathcal{O}_2}a^{(-2)}_{\mathcal{O}_2}
\left(
\frac{z}{\alpha}
+
\frac{\alpha}{z}
\right)\nonumber\\
&+&\tilde b^2_{1,1,0,0,0}
\left(\sqrt{\frac z\alpha}+\sqrt{\frac\alpha z}\,\right)
\hat g_{\widehat B_{1/2,0}}\!(z,\alpha)
+\tilde b^2_{2,2,0,0,0}\,\hat g_{\widehat B_{1,0}}(z,\alpha) 
\\
&+&\sum_{r_->0}\Bigg[
\left(\tilde b^2_{r_-,0,-r_-,0,1}
+\frac\alpha z\,\tilde b^2_{r_-,0,-r_-,0,0}\right)
\hat g_{\widehat B_{0,-r_-}}\!\!(z,\alpha)
\Bigg]\nonumber\\
&+&\sum_{r_->0}\Bigg[\sqrt{\frac\alpha z}\,
\tilde b^2_{r_-+1,1,-r_-,0,0}\,
\hat g_{\widehat B_{1/2,-r_-}}\!\!(z,\alpha)\Bigg]\, ,\nonumber
\end{eqnarray}
with $\tilde b^2_{\hat\Delta,n,r_-,q,m}$ defined in \eqref{Beta1Beta2}. Here $\hat g_{\widehat {\mathcal{M}}}(z,\alpha) \colonequals \widehat {\mathcal{G}}_{\widehat {\mathcal{M}}}(z,\bar z,\alpha,\bar z)$ and the defect superconformal blocks are given in table \ref{tab:defblocks}. From the explicit form of the blocks, we observe that
\begin{equation}\label{fsing0}
     f^{(0)}(z,\alpha) \xrightarrow{z \to 0} a^{(+2)}_{\mathcal{O}_2} a^{(-2)}_{\mathcal{O}_2} \, \frac{\alpha}{z}+\hat{f}(\alpha) +\mathcal{O}(z) \, .
\end{equation}
Here $\hat{f}(\alpha)$ stands for an unknown function that encodes the $\mathcal{O}(z^0)$ terms in the defect block expansion. 
Using the symmetry $f^{(0)}(z,\alpha)=f^{(0)}(1/z,1/\alpha)$, we can also control the behavior at large $z$: 
\begin{equation}\label{fbound}
     f^{(0)}(z,\alpha) \xrightarrow{z \to + \infty} a^{(+2)}_{\mathcal{O}_2} a^{(-2)}_{\mathcal{O}_2} \, \frac{z}{\alpha}+\hat{f}(1/\alpha)+\mathcal{O}(1/z) \,.
\end{equation}
Assuming no additional singularities, Cauchy’s theorem then implies
\begin{equation}
f^{(0)}(z,\alpha) = \oint \frac{dz'}{2\pi i}\frac{f^{(0)}(z',\alpha)}{z'-z} = \hat{f}(1/\alpha)\, + \, a^{(+2)}_{\mathcal{O}_2} a^{(-2)}_{\mathcal{O}_2} \left(\frac{\alpha}{z}+\frac{ z}{\alpha}\right) + \int_{1}^{+\infty}\frac{dz'}{2\pi i}
\frac{\text{Disc}_{z'>1}\left[f^{(0)}\right]}{z'-z}\,,
\end{equation}
where the discontinuity is understood in a distributional sense, as
in \eqref{discdelta}. Alternatively, since the singularity is a pole, we can replace the last term with a residue. We stress that the discontinuity is fixed exactly in $N$ and in the coupling by \eqref{fsing1}.  Unlike in the case of the full correlator, it cannot receive contributions from non-protected operators in long multiplets. Moreover, the $z \leftrightarrow 1/z$ symmetry constrains the asymptotic behavior, which in turn completely fixes the possible ambiguities. Using the bulk OPE input \eqref{fsing1} to evaluate the discontinuity, we finally obtain
\begin{eqnarray}\label{smallf0}
f^{(0)}(z,\alpha)&=& -\frac14-2a^{(+2)}_{\mathcal{O}_2} a^{(-2)}_{\mathcal{O}_2}-\mathcal{C}^{(0)}+a^{(+2)}_{\mathcal{O}_2} a^{(-2)}_{\mathcal{O}_2} \left(\frac{\alpha}{z}+\frac{ z}{\alpha}\right)\nonumber\\
&+&\frac{z(1-\alpha)^2}{4\alpha(1-z)^2} -a\lambda_{\mathcal{B}_{[0,2,0]}}^{(0)}\frac{(\alpha+1)(z-\alpha)}
{4\alpha(1-z)}\,,
\end{eqnarray}
where we fixed the value of $\hat{f}(1/\alpha)$ imposing $f^{(0)}(z,z)=-\mathcal{C}^{(0)}$. The explicit expressions for the coefficients $a_{\mathcal O_2}^{(q)}$, $a\lambda_{\mathcal B_{[0,2,0]}}^{(0)}$ and $\mathcal{C}^{(0)}$ are given in \eqref{a2broken}, \eqref{lambdaO2} and \eqref{Cloc} for the case of a Gukov-Witten defect. 
As a non-trivial check, at weak coupling this result agrees with the diagrammatic computation of \cite{Holguin:2025bfe}, see \eqref{smallfweak}. At strong coupling, it can be reproduced from Witten diagrams, as discussed in section \ref{sec:holo-witten}.

\paragraph{Comparison with the chiral algebra result.} Equation \eqref{smallf0} can be compared directly with the chiral algebra result \eqref{twoptchiralfinal} (setting for a moment $a^{(\pm2)}_{\mathcal{O}_2}\to0$). The relevant contractions of polarization vectors are 
\begin{align}
 	Y_1\!\circ\! Y_2 & = \frac{1}{4} \left( (\bar \z_1 - \bar \z_2)\tilde u^{12} + (\bar \z_1 + \bar \z_2)\tilde u^{12}_{\perp} \right) \, , \\
 	Y_1\!\bullet\! Y_2 & = \frac{1}{4} \left( (\bar \z_1 - \bar \z_2)\tilde u^{12} - (\bar \z_1 + \bar \z_2)\tilde u^{12}_{\perp} \right) \, ,
 \end{align}
 where, in addition to $\tilde u^{ij}$ defined in \eqref{uijtilde}, we introduced
\begin{equation}
	\tilde u^{ij}_{\perp} \colonequals \tilde u^i_{\dot 1} \tilde u^j_{\dot 2} + \tilde u^i_{\dot 2} \tilde u^j_{\dot 1} \, .
 \end{equation}
 With these identifications, the lightcone cross-ratios defined in appendix \ref{app:cross} reduce to
\begin{equation}
	\label{crosschiral}
	z = \frac{\z_1}{\z_2} \,, \quad
 	\alpha = \frac{(\tilde u^{12} - \tilde u^{12}_{\perp})^2}{\tilde u^{11}_{\perp}\tilde u^{22}_{\perp}} \,, \quad	\bar z = \bar \alpha = \frac{\bar \z_1}{\bar \z_2} \, .
\end{equation}
 Using the identity $(\tilde u^{12}_{\perp})^2-(\tilde u^{12})^2=\tilde u^{11}_{\perp}\tilde u^{22}_{\perp}$, the R-symmetry cross-ratio takes the remarkably simple form
\begin{equation}\label{crosschiralsimple}
	\alpha=\frac{\tilde u^1_{\dot 2}  \tilde u^2_{\dot 1}}{\tilde u^1_{\dot 1}\tilde u^2_{\dot 2}}\, .
 \end{equation}
 Using \eqref{2pt}, \eqref{holotopo} and the identification \eqref{chiralcurrent} we obtain the following relation
\begin{equation}
\label{twopointchiral2}
 	\tilde u^1_{\dot a} \tilde u^1_{\dot b}
 	\tilde u^2_{\dot c} \tilde u^2_{\dot d}
 	\langle \sigma(\infty) J^{\dot a \dot b}(z) J^{\dot c \dot d}(1) \sigma(0) \rangle
 	= \kappa_J^2\frac{\tilde u^{11}_{\perp} \tilde u^{22}_{\perp}}{4 z} f^{(0)}(z, \alpha)
 	= \kappa_J^2\frac{\tilde u^1_{\dot 1}\tilde u^1_{\dot 2}\tilde u^2_{\dot 1}\tilde u^2_{\dot 2}}{ z} f^{(0)}(z, \alpha) ,
 \end{equation}
 where the left hand side is the two-dimensional correlator \eqref{twoptchiralfinal} evaluated at $\zeta_1=z$ and $\zeta_2=1$ and with indices contracted with the $\tilde u$'s. 
 Since $\alpha$ satisfies \eqref{crosschiralsimple}, the Laurent coefficients of $f^{(0)}$ in $\alpha$ are in one-to-one correspondence with the components of the current correlator: the $\alpha^0$, $\alpha^{+1}$ and $\alpha^{-1}$ terms multiply $\tilde u^1_{\dot 1}\tilde u^1_{\dot 2}\tilde u^2_{\dot 1}\tilde u^2_{\dot 2}$, $(\tilde u^1_{\dot 2})^2(\tilde u^2_{\dot 1})^2$ and $(\tilde u^1_{\dot 1})^2(\tilde u^2_{\dot 2})^2$ respectively. Hence using the decomposition \begin{equation}\label{Jcartanvsdotted}
	\widetilde u_{\dot a}\widetilde u_{\dot b}J^{\dot a\dot b}
	=\tfrac{1}{2}(\widetilde u_{\dot 1})^2\,J^{-}
 	-\widetilde u_{\dot 1}\widetilde u_{\dot 2}\,J^{3}
	-\tfrac{1}{2}(\widetilde u_{\dot 2})^2\,J^{+}\, ,
 \end{equation}
 we see immediately that 
\begin{align}
	\frac{z}{\k^2_J}\langle \sigma(\infty)J^3(z)J^3(1)\sigma(0)\rangle=f^{(0)}(z,\alpha)\Big|_{\alpha^{0}}
	&=-\frac14-\mathcal{C}^{(0)}+ \frac14 a\lambda^{(0)}_{\mathcal{B}_{[0,2,0]}}
	-\frac{z}{2(1-z)^2}\, ,\nn\\
	-\frac14\frac{z}{\k^2_J}\langle \sigma(\infty)J^+(z)J^-(1)\sigma(0)\rangle=f^{(0)}(z,\alpha)\Big|_{\alpha^{+1}}
	&=\frac{z}{4(1-z)^2}+\frac{a\lambda^{(0)}_{\mathcal{B}_{[0,2,0]}}}{4(1-z)}\, ,\\
	-\frac14 \frac{z}{\k^2_J}\langle \sigma(\infty)J^-(z)J^+(1)\sigma(0)\rangle=f^{(0)}(z,\alpha)\Big|_{\alpha^{-1}}
	&=\frac{z}{4(1-z)^2}-\frac{z\,a\lambda^{(0)}_{\mathcal{B}_{[0,2,0]}}}{4(1-z)}\, , \nn
\end{align}
and the comparison with \eqref{twoptchiralfinal} is immediate. It yields three independent statements. The first two are consistency checks,
\begin{equation}\label{checks}
	k=-\frac{\kappa_J^2}{2}\, ,
	\qquad
	\frac{\, a^{(0)}_{\mathcal{O}_2}}{\lambda_{\mathcal{B}_{[0,2,0]}}} = - \frac{k}{2}\;a\lambda^{(0)}_{\mathcal{B}_{[0,2,0]}}\, ,
\end{equation}
which reproduce respectively the level as fixed in \eqref{OPEchiral} and the identifications \eqref{k1sulambda} and \eqref{alambda}. The third one determines $S^{AB}$ in \eqref{SABdef} as 
\begin{equation}\label{Sresult}
	S^{33}-\left(2 \frac{a^{(0)}_{\mathcal{O}_2}}{\lambda_{\mathcal{B}_{[0,2,0]}}}\right)^{2}
	=\kappa_J^2\Big(-\frac14-\mathcal{C}^{(0)}
	+\frac14 a\lambda^{(0)}_{\mathcal{B}_{[0,2,0]}}
	-\big[a^{(0)}_{\mathcal{O}_2}\big]^2\Big)
	=\kappa_J^2\; \hat{\texttt{B}}^{(0)}_{\widehat B_{1/2,0}}\, ,
	\quad
	S^{+-}=- \frac{2\, a^{(0)}_{\mathcal{O}_2}}{\lambda_{\mathcal{B}_{[0,2,0]}}}\, ,
\end{equation}
where in the last step we introduced the coefficient $\hat{\texttt{B}}^{(0)}_{\widehat B_{1/2,0}}$ defined by the corresponding entry of table \ref{tab:defOPEdata1}, with $\texttt{B}^{(0)}_{\widehat B_{1/2,0}}(\omega)=\hat{\texttt{B}}^{(0)}_{\widehat B_{1/2,0}} (\omega+\omega^{-1})$. This is essentially saying that $S^{33}$ is given by the product of one-point functions of $J^3$ plus the contribution of a single defect multiplet. In particular, this multiplet is precisely the one found in \eqref{B120}, confirming the expectation that these operators in chiral algebra are degenerate with the defect identity.

\subsubsection{Fixing the low-spin ambiguities}
Substituting the holomorphic function $f^{(0)}(z,\alpha)$ given in \eqref{smallf0} into the Ward identity solution \eqref{wardsolution}, and comparing it with \eqref{Fansatz}, we obtain 
\begin{eqnarray}\label{fi}
   \mathcal F^{(0)}_1(z,\bar z) &=&-\frac{(z-\bar z) (z \bar z+1) \mathcal F^{(0)}_0(z,\bar z)}{2 z \bar z}+\frac{z \bar z}{(z \bar z-1)^2}-\frac{z \bar z(1+z \bar z) \log (z \bar z)}{2 (z \bar z-1)^3}\,, \nonumber \\
   \mathcal F^{(0)}_2(z,\bar z) &=&-\frac{(z \bar z (z+\bar z-4)+z+\bar z) \mathcal F^{(0)}_0(z,\bar z)}{2 z \bar z}+ \frac{z \bar z}{(z \bar z-1)^2}-\frac{z \bar z(1+z \bar z) \log (z \bar z)}{2 (z \bar z-1)^3}\,, \nonumber\\
  \mathcal F^{(0)}_3(z,\bar z) &=&-\frac{(z-1) (z+1) (\bar z-1) (\bar z+1) \mathcal F^{(0)}_0(z,\bar z)}{4 z \bar z} -\frac{(z+1) (\bar z+1) (z \bar z+1)}{8 (z \bar z-1)^2}\nonumber\\
   &+&\frac{z (z+1) \bar z (\bar z+1) \log (z \bar z)}{4 (z \bar z-1)^3}-\frac{\frac14 + \left[a^{(0)}_{\mathcal{O}_2}\right]^2+2a^{(+2)}_{\mathcal{O}_2} a^{(-2)}_{\mathcal{O}_2} - \frac12 a\lambda_{\mathcal{B}_{[0,2,0]}}^{(0)} + \mathcal{C}^{(0)}}{2 a\lambda_{\mathcal{B}_{[0,2,0]}}^{(0)}}\nonumber\\
   \mathcal F^{(0)}_4(z,\bar z) &=&\frac{(z-1) (\bar z-1) (z-\bar z) \mathcal F^{(0)}_0(z,\bar z)}{2 z \bar z}+\frac{(z-\bar z) (z \bar z+1)}{4 (z \bar z-1)^2}\\
   &-&\frac{z \bar z (z-\bar z) \log (z \bar z)}{2 (z \bar z-1)^3}\,, \nonumber\\
   \mathcal F^{(0)}_5(z,\bar z) &=&\frac{(z-1)^2 (\bar z-1)^2 \mathcal F^{(0)}_0(z,\bar z)}{4 z \bar z}+\frac{(z-1) (\bar z-1) (z \bar z+1)}{8 (z \bar z-1)^2}\nonumber\\
   &-&\frac{(z-1) z (\bar z-1) \bar z \log (z \bar z)}{4 (z \bar z-1)^3}+\frac{\frac14 +\left[a^{(0)}_{\mathcal{O}_2}\right]^2+2a^{(+2)}_{\mathcal{O}_2} a^{(-2)}_{\mathcal{O}_2} - \frac12 a\lambda_{\mathcal{B}_{[0,2,0]}}^{(0)} + \mathcal{C}^{(0)}}{2 a\lambda_{\mathcal{B}_{[0,2,0]}}^{(0)}}\, .\nonumber
\end{eqnarray}
In other words, by exploiting the Ward identity solution and isolating the protected contribution, we reduced the low-spin ambiguity to a single function of the spacetime cross-ratios, $\mathcal F^{(0)}_0(z,\bar z)$. This remaining ambiguity can be fixed by making an ansatz and constraining it using the general properties of the correlator. In particular, we assume that, at the order we are working, $\mathcal F^{(0)}_0(z,\bar z)$ is given by a combination of logarithmic and rational functions, i.e. the correlator has maximal transcendentality one. We further impose the following constraints:
\begin{itemize}
\item \textbf{Finite transverse spin.} The function admits a decomposition in defect conformal blocks with finite support in transverse spin $|s| \le s^*$.
\item \textbf{Symmetry.} It is invariant under $z \leftrightarrow \bar z$ and $z \leftrightarrow 1/\bar z$, as required by \eqref{symF}.
\item \textbf{Bulk OPE.} It is regular at $z=1$ and $\bar z=1$ (equivalently, at $w=r$ and $w=1/r$ in radial coordinates \eqref{radialcoord}), and therefore does not contribute to the dispersion relation \eqref{dispersion}.
\item \textbf{Defect OPE.} Its behavior as $r \to 0$ is consistent with the defect OPE of the two-point function. In particular, the leading term of the expansion of the full correlator matches the contribution of the defect identity operator, given in \eqref{squared1ptcharge}. The correlator may exhibit a logarithmic branch point at $r=0$, arising from the contribution of non-protected defect operators that acquire anomalous dimensions. Finally, the low-spin ambiguity must have a small $r$ expansion compatible with the order $N^0$ spectrum (GFF).
\item \textbf{Analyticity.} The function is regular everywhere except at the expected physical singularities. In particular, it must be regular at $r=1$ or $w=1$ \cite{Lemos:2017vnx}.
\item \textbf{Large-dimension growth and flat-space limit.}
We require the defect anomalous dimensions to have the mildest growth at large classical dimension compatible with the remaining constraints. This selects the weakest allowed singularity at $r=-1$. For the Gukov-Witten defect at $\beta=\gamma=0$, the condition admits a more stringent interpretation: the Lorentzian singularity at $r=-1$ probes the flat-space limit, and its strength constrains the high-energy behavior of the corresponding scattering amplitude \cite{Penedones:2010ue,Pufu:2023vwo,Alday:2024srr,Chen:2025cod}.\footnote{This interpretation has been derived for defects admitting a holographic description as an $AdS_{p_{\text{def}}+1}$ brane embedded in $AdS_{d+1}$, such as the Gukov-Witten defect at $\beta=\gamma=0$. We do not assume its validity for more general defects.}
\end{itemize}

An ansatz consistent with these requirements is
\begin{equation}
\mathcal F^{(0)}_0(z=rw,\bar z=r/w)=\sum_{s=0}^{s^*}R(r,s)\left(w^{s}+w^{-s}\right)\, ,
\end{equation}
where
\begin{equation}
    R(r,s) = \sum_{n=0}^{N_{ans}} \left[ \frac{a_{n,s}+b_{n,s}(r+r^{-1})}{(r-r^{-1})^{2n}}+\frac{[c_{n,s}+d_{n,s}(r+r^{-1})]\log(r)}{(r-r^{-1})^{2n+1}}\right]\, .
\end{equation}
Here $s^*$ and $N_{ans}$ are arbitrary non-negative integers and the numerical coefficients $a_{n,s}$, $b_{n,s}$, $c_{n,s}$ and $d_{n,s}$ are fixed by imposing the above constraints. We have verified that, for a wide range of values of $s^*$ and $N_{ans}$, the constraints above select a unique spin $s^*=0$ ambiguity,
\begin{eqnarray}\label{f0ans}
\mathcal F^{(0)}_0(z,\bar z)
&=& - \frac{z \bar{z} \left(z^2 \bar{z}^2\!+\!10 z \bar{z}+1\right) \left(-\tfrac{1}{2}a\lambda_{\mathcal{B}_{[0,2,0]}}^{(0)}\!\!+\!2 [a_{\mathcal{O}_2}^{(0)}]^2\!+\!4 a_{\mathcal{O}_2}^{(+2)}a_{\mathcal{O}_2}^{(-2)}\!\!+\!2 \mathcal{C}^{(0)}\!+\!\tfrac{1}{2}\right)}{ a\lambda_{\mathcal{B}_{[0,2,0]}}^{(0)} \left(1-z \bar{z}\right)^4}\\
&\!-&\!\!\frac{6\,  z^2 \bar{z}^2 \left(z \bar{z}\!+\!1\right) \log (z \bar{z}) \left(- \tfrac{1}{2} a\lambda_{\mathcal{B}_{[0,2,0]}}^{(0)}\!\!+\!2 [a_{\mathcal{O}_2}^{(0)}]^2\!+\!4 a_{\mathcal{O}_2}^{(+2)}a_{\mathcal{O}_2}^{(-2)}\!\!+\!2  \mathcal{C}^{(0)}\!+\!\tfrac{1}{2}\right)}{a\lambda_{\mathcal{B}_{[0,2,0]}}^{(0)} \left(1-z \bar{z}\right)^5}\!+\!\mathcal{O}\left(\tfrac{1}{N^2}\right). \nonumber
\end{eqnarray}
Substituting \eqref{f0ans} into \eqref{fi} and \eqref{Fansatz}, we obtain the final expression for the rotation-preserving part of the correlator at large $N$ and strong coupling, 
\begin{align}\label{q0result}
\mathcal F^{(0)}(z,\bar z,\alpha,\bar\alpha)
= &
\frac{z\bar z(1-\alpha)^2(1-\bar\alpha)^2}
{4\alpha\bar\alpha(1-z)^2(1-\bar z)^2}
+\bigl[a_{\mathcal O_2}^{(0)}\bigr]^2
+a_{\mathcal O_2}^{(+2)}a_{\mathcal O_2}^{(-2)}
\left(\frac{\bar z\alpha}{z\bar\alpha}
     +\frac{z\bar\alpha}{\bar z\alpha}\right)\nonumber\\
    &-\frac{a\lambda_{\mathcal B_{[0,2,0]}}^{(0)}}{4}
\Biggl[
-1+\frac{(1-\alpha)(1-\bar\alpha)}
{(1-z)(1-\bar z)}
\left(\frac{z}{\alpha}+\frac{\bar z}{\bar\alpha}\right)\nonumber\\
&+\frac{(z-\alpha)(\bar z-\bar\alpha)
(1-\bar z\alpha)(1-z\bar\alpha)}
{\alpha\bar\alpha(1-z)(1-\bar z)}
\frac{1-z^2\bar z^2+2z\bar z\log(z\bar z)}
{(1-z\bar z)^3}
\Biggr]
\\
&+\left(
\frac14 +\mathcal C^{(0)}
+\bigl[a_{\mathcal O_2}^{(0)}\bigr]^2
+2a_{\mathcal O_2}^{(+2)}a_{\mathcal O_2}^{(-2)}
- \frac{a\lambda_{\mathcal B_{[0,2,0]}}^{(0)}}{4}
\right)
\nonumber \\
&\times\Biggl[
\frac{(z-\alpha)(\bar z-\bar\alpha)
(1-\bar z\alpha)(1-z\bar\alpha)}
{\alpha\bar\alpha}\frac{
1+10z\bar z+z^2\bar z^2}
{(1-z\bar z)^4}\nonumber \\
&+\frac{(z-\alpha)(\bar z-\bar\alpha)
(1-\bar z\alpha)(1-z\bar\alpha)}
{\alpha\bar\alpha}\frac{6z\bar z(1+z\bar z)\log(z\bar z)}
{(1-z\bar z)^5}
-1\Biggr]+\mathcal{O}(1/N^2)\, .\nonumber
\end{align}
The result depends on the defect-specific coefficients $a_{\mathcal O_2}^{(0)}$, $a_{\mathcal O_2}^{(\pm2)}$, $a\lambda_{\mathcal B_{[0,2,0]}}^{(0)}$ and $\mathcal{C}^{(0)}$, whose values for Gukov-Witten defects are given in \eqref{a2broken}, \eqref{lambdaO2} and \eqref{Cloc}, respectively. At first glance, the prefactor multiplying the last term in
\eqref{q0result} depends on the one-point functions and thus appears to be
of order $N^0$. However, $\mathcal{C}^{(0)}$ is the topological limit of
the full two-point function and therefore contains both disconnected and
connected contributions. Schematically, for any defect,
\begin{equation}
    \mathcal{C}^{(0)}
    = -1/4-\bigl[a_{\mathcal O_2}^{(0)}\bigr]^2
    -2a_{\mathcal O_2}^{(+2)}a_{\mathcal O_2}^{(-2)}
    - \text{connected terms of order $\mathcal{O}(1/N)$}\, .
\end{equation}
Its disconnected part therefore precisely cancels the contributions from
the one-point functions, so that the combined coefficient starts at order
$1/N$. This cancellation is manifest in the Gukov-Witten case, see
\eqref{Cloc}. As shown in table \ref{tab:defOPEdata1}, the
prefactor corresponds to the single defect OPE coefficient that appeared in \eqref{Sresult}:
\begin{equation}
    1/4+\mathcal C^{(0)}
    +\bigl[a_{\mathcal O_2}^{(0)}\bigr]^2
    +2a_{\mathcal O_2}^{(+2)}a_{\mathcal O_2}^{(-2)}
    -\frac{a\lambda_{\mathcal B_{[0,2,0]}}^{(0)}}{4}
    =
    -\hat{\texttt{B}}^{(0)}_{\widehat B_{1/2,0}}\, .
\end{equation}

In terms of the holomorphic and non-holomorphic components appearing in the Ward identity solution \eqref{wardsolution}, the correlator \eqref{q0result} decomposes as
\begin{eqnarray}
    f^{(0)}(z,\alpha)&=& -1/4-2a^{(+2)}_{\mathcal{O}_2} a^{(-2)}_{\mathcal{O}_2}-\mathcal{C}^{(0)}+a^{(+2)}_{\mathcal{O}_2} a^{(-2)}_{\mathcal{O}_2} \left(\frac{\alpha}{z}+\frac{ z}{\alpha}\right)\nonumber\\
&+&\frac{z(1-\alpha)^2}{4\alpha(1-z)^2} - a\lambda_{\mathcal{B}_{[0,2,0]}}^{(0)}\frac{(\alpha+1)(z-\alpha)}
{4\, \alpha\, (1-z)}\,,
\nonumber\\
F^{(0)}(z,\bar z)
&=&\frac{z\bar z}{4(1-z)^2(1-\bar z)^2}-\frac{a\lambda_{\mathcal B_{[0,2,0]}}^{(0)}
\,z\bar z(1+z\bar z)}
{4(1-z)(1-\bar z)(1-z\bar z)^2}
\\
&-&\frac{a\lambda_{\mathcal B_{[0,2,0]}}^{(0)}
\,z^2\bar z^2\log(z\bar z)}
{2 (1-z)(1-\bar z)(1-z\bar z)^3}
- \frac{a\lambda_{\mathcal B_{[0,2,0]}}^{(0)}}{2}\mathcal F^{(0)}_0(z,\bar z)+\mathcal{O}(1/N^2)\, .\nonumber
\end{eqnarray}
Finally, we specialize to the Gukov-Witten defect at $\beta=\gamma=0$ in the single probe brane regime, to compare with the holographic computation in section \ref{sec:holo-witten}. In this case, the low-spin ambiguity $\mathcal F^{(0)}_0(z,\bar z)$ vanishes and the correlator reads
\begin{eqnarray}\label{finalbg0}
\mathcal{F}^{(0)}(z,\bar z,\alpha,\bar \alpha)&=&\frac12+\frac{(\alpha -1)^2 (\bar{\alpha}-1)^2 z \bar{z}}{4\alpha  \bar{\alpha} (z-1)^2 (\bar{z}-1)^2}\nonumber\\
&-& \frac{1}{4N}\left[\frac{2 \left(\bar{\alpha} z^3 (\bar{z}-1) \bar{z} (2 \alpha  \bar{z}-\bar{\alpha})+\alpha  (\bar{z}-1) (\alpha  \bar{z}-2 \bar{\alpha})\right)}{\alpha  \bar{\alpha} (z-1) (\bar{z}-1) (z \bar{z}-1)^2}\right. \\
&+&\left.\frac{2 z \left(-\bar{\alpha} (2 \alpha \!+\!\bar{\alpha})+\alpha ^2 \bar{z}^3+2 \alpha  \bar{z}^2 (-2 \alpha  \bar{\alpha}\!+\!\alpha +2 \bar{\alpha}-2)+\bar{z} (\alpha  \bar{\alpha}+1)^2\right)}{\alpha  \bar{\alpha} (z-1) (\bar{z}-1) (z \bar{z}-1)^2} \right.\nonumber\\
&+&\left.\frac{2 z^2 \left(\bar{\alpha}^2-\alpha  \bar{z}^3 (\alpha +2 \bar{\alpha})+2 \bar{\alpha} \bar{z} (-2 \alpha  \bar{\alpha}+2 \alpha +\bar{\alpha}-2)+(\alpha  \bar{\alpha} \bar{z}+\bar{z})^2\right)}{\alpha  \bar{\alpha} (z-1) (\bar{z}-1) (z \bar{z}-1)^2}\right.\nonumber\\
&-&\left.\frac{4 z \bar{z} (z-\alpha ) (\bar{\alpha} z-1) (\alpha  \bar{z}-1) (\bar{z}-\bar{\alpha}) \log (z \bar{z})}{\alpha  \bar{\alpha} (z-1) (\bar{z}-1) (z \bar{z}-1)^3}\right]+\mathcal{O}\left(1/N^2\right)\,.\nonumber
\end{eqnarray}

\subsection{Rotation-breaking sector}\label{sec:breaking}

We can follow the same procedure as in the previous section to determine the charge $q \neq 0$ components of the two-point function. This case turns out to be considerably simpler. Using the discontinuities in \eqref{discO2comp} together with the dispersion relation \eqref{dispersion}, we obtain
\begin{eqnarray}
\mathcal{F}^{(\pm2)}(z,\bar z,\alpha,\bar \alpha)&\cong& - \frac{a\lambda_{\mathcal{B}_{[0,2,0]}}^{(\pm 2)}}{2}\frac{\sqrt{z \bar{z}}(1-\alpha) (1-\bar{\alpha}) }{(1-z) (1-\bar{z}) \sqrt{\alpha  \bar{\alpha}}}\,,\nonumber \\
    \mathcal{F}^{(\pm4)}(z,\bar z,\alpha,\bar \alpha)&\cong&0\,.
\end{eqnarray}
Remarkably, these results are exact up to possible low-spin ambiguities. This follows from the fact that the discontinuity itself is exact, since long multiplets do not contribute to the bulk OPE in sectors with charge $q \neq 0$. In particular, the result of the dispersion relation in the $q=\pm 2$ sector corresponds to a single bulk superconformal block, namely the $\mathcal{B}_{[0,2,0]}$ block \eqref{020superblock}. Since this contribution preserves all symmetries of the correlator, including the Ward identities \eqref{WI}, we are led to conjecture that no additional low-spin ambiguities arise, aside from the disconnected contribution associated with the squared one-point function \eqref{squared1ptcharge}. The latter encodes the contributions of infinitely many bulk operators belonging to the short and semi-short multiplets in \eqref{O2bulkope}. 

This conjecture can be tested by computing the exact form of the holomorphic function $f^{(\pm2)}(z,\alpha)$. Proceeding as in the rotation-invariant case, the singular behavior of $f^{(\pm2)}(z,\alpha)$ at $z=1$ and $z=0$, together with the large-$z$ asymptotics, can be extracted from the bulk and defect OPEs. Explicitly, using the blocks in table \ref{tab:blocksq}, we find
\begin{eqnarray}
    \sqrt{\frac{\alpha}{z}} f^{(\pm2)}(z,\alpha)&\xrightarrow{z\to 1}& - \frac{a\lambda_{\mathcal{B}_{[0,2,0]}}^{(\pm 2)}}{2}\frac{1-\alpha }{1-z} +\mathcal{O}((1-z)^0)\,,\nonumber\\
    \sqrt{\frac{\alpha}{z}}f^{(\pm2)}(z,\alpha)&\xrightarrow{z\to 0}& a^{(0)}_{\mathcal{O}_2} a^{(\pm 2)}_{\mathcal{O}_2} \frac{\alpha}{z}+\hat{f}(\alpha) +\mathcal{O}(z)\,, \\
    \sqrt{\frac{\alpha}{z}}f^{(\pm2)}(z,\alpha)&\xrightarrow{z\to + \infty}& a^{(0)}_{\mathcal{O}_2} a^{(\pm 2)}_{\mathcal{O}_2} +\mathcal{O}(1/z)\,, \nonumber
\end{eqnarray}
where $\hat{f}(\alpha)$ once again stands for $\mathcal{O}(z^0)$ terms in the defect OPE expansion.
Applying Cauchy’s theorem, we obtain
\begin{eqnarray}\label{smallf2}  
f^{(\pm2)}(z,\alpha)&=&a^{(0)}_{\mathcal{O}_2} a^{(\pm 2)}_{\mathcal{O}_2} \left(\sqrt{\frac{z}{\alpha}}+\sqrt{\frac{\alpha}{z}}\ \right)-\frac{a\lambda_{\mathcal{B}_{[0,2,0]}}^{(\pm 2)}}{2}\frac{\sqrt{z}(1-\alpha) }{(1-z) \sqrt{\alpha}}\,. 
\end{eqnarray}
This result is consistent with the expectation that the only low-spin ambiguity arises from the defect identity contribution. Since the $q=\pm2$ sector does not receive contributions from non-protected operators, the OPE data extracted from $f^{(\pm2)}(z,\alpha)$ are sufficient to reconstruct the full correlator. One indeed finds that the complete result is given by the contribution obtained from the dispersion relation together with the squared one-point function.

A similar conclusion holds for the $q=\pm 4$ sector, which receives no contribution from the dispersion relation. In this case, the OPE analysis of section \ref{sec:bulkOPE} and the superconformal blocks listed in table \ref{tab:blocksq} imply that the only low-spin ambiguity is the contribution of the defect identity. 

All in all, this gives the two-point function of $\mathcal{O}_2$ in the $q=\pm 2,\pm 4$ sectors,
\begin{eqnarray}\label{chargedres}
\mathcal{F}^{(\pm2)}(z,\bar z,\alpha,\bar \alpha)&=& a^{(0)}_{\mathcal{O}_2} a^{(\pm 2)}_{\mathcal{O}_2} \left(\sqrt{\frac{\bar{\alpha} z}{\alpha \bar z}}+\sqrt{\frac{\alpha \bar z}{\bar{\alpha} z}}\right)-\frac{a\lambda_{\mathcal{B}_{[0,2,0]}}^{(\pm 2)}}{2}\frac{\sqrt{z \bar{z}}(1-\alpha) (1-\bar{\alpha}) }{(1-z) (1-\bar{z}) \sqrt{\alpha  \bar{\alpha}}}\,, \nonumber \\
    \mathcal{F}^{(\pm4)}(z,\bar z,\alpha,\bar \alpha)&=&\left[a^{(\pm 2)}_{\mathcal{O}_2}\right]^2\,.
\end{eqnarray}
We conjecture that this expression is exact in both $N$ and the coupling, given the exact expressions for $a^{(q)}_{\mathcal{O}_2}$ and $a\lambda_{\mathcal{B}_{[0,2,0]}}^{(q)}$ in \eqref{a2broken}, \eqref{alambda} and \eqref{lambdaO2}.
A highly non-trivial check of our conjecture comes from the comparison with the weak coupling result of \cite{Holguin:2025bfe} for the case of a Gukov-Witten defect. We perform this check in appendix \ref{app:weak} and we find perfect agreement.

\paragraph{Comparison with the chiral algebra result.} The charged sectors can be matched against the chiral algebra prediction \eqref{unflowed} exactly as we did for $f^{(0)}$ in section \ref{holomorphicorr}. On the chiral algebra locus the two angular cross-ratios reduce to
\begin{equation}\label{omegachiral}
\omega=\sqrt{\frac{z}{\alpha}}\, ,\qquad e^{2i\vartheta}=\z_1\z_2\,\frac{\tilde u^1_{\dot 1}\tilde u^2_{\dot 1}}{\tilde u^1_{\dot 2}\tilde u^2_{\dot 2}}\, ,
\end{equation}
so that the nine components of $\langle J^AJ^B\rangle_\sigma$ distribute over the Fourier sectors according to their flavor charge: the sector $q=0$ collects $\langle J^3J^3\rangle_\sigma$ and $\langle J^{\pm}J^{\mp}\rangle_\sigma$, as used in section \ref{holomorphicorr}, while
\begin{equation}\label{qassignment}
	q=\mp2:\quad \langle J^{3}J^{\pm}\rangle_\sigma\, ,\ \ \langle J^{\pm}J^{3}\rangle_\sigma\, ,
	\qquad\qquad
	q=\mp4:\quad \langle J^{\pm}J^{\pm}\rangle_\sigma\, .
\end{equation}
The latter is the simplest because the four-dimensional result simply consists of a disconnected term as predicted in \eqref{q4factorization}. For the charge-two result we notice that $\alpha=z/\omega^{2}$ and the result \eqref{smallf2} is a Laurent polynomial in $\omega$ of degree one,
\begin{equation}\label{fpm2omega}
	f^{(\pm2)}(z,\alpha)=
	\left(a^{(0)}_{\mathcal{O}_2}a^{(\pm2)}_{\mathcal{O}_2}-\frac{a\lambda^{(\pm 2)}_{\mathcal{B}_{[0,2,0]}}}{2(1-z)}\right)\omega
	+\left(a^{(0)}_{\mathcal{O}_2}a^{(\pm2)}_{\mathcal{O}_2}+\frac{z\,a\lambda^{(\pm 2)}_{\mathcal{B}_{[0,2,0]}}}{2(1-z)}\right)\omega^{-1}\, ,
\end{equation}
and its two coefficients are matched by the two orderings in \eqref{qassignment}. As in the neutral sector, the matching yields one consistency check and one new piece of data. The check is that both orderings reproduce the same charged one-point function,
\begin{equation}\label{checkscharged}
	\widetilde\jmath^{\,\pm}= \pm \, \kappa_J^2\,a\lambda^{(\mp2)}_{\mathcal{B}_{[0,2,0]}}\, ,
\end{equation}
which reduces to \eqref{flowedonept} upon using \eqref{k1sulambda} and \eqref{alambda}. This is the charged counterpart of the second relation in \eqref{checks}, and it is non-trivial because the two orderings have different $\z$-dependence. The new information is the mixed zero-mode data,
\begin{equation}\label{S3pmresult}
	\widetilde S^{3\pm}=\widetilde\jmath^{\,3}\,\widetilde\jmath^{\,\pm}-\tfrac12\,\widetilde\jmath^{\,\pm}\, ,
	\qquad\text{i.e.}\qquad
\langle \sigma^\dagger|\widetilde J^{3}_0\widetilde J^{+}_0|\sigma\rangle=\widetilde\jmath^{\,3}\,\widetilde\jmath^{\,+}\, ,
	\quad
 	\langle \sigma^\dagger|\widetilde J^{-}_0\widetilde J^{3}_0|\sigma\rangle=\widetilde\jmath^{\,3}\,\widetilde\jmath^{\,-}\, ,
 \end{equation}
 where we used $[\widetilde J^{3}_0,\widetilde J^{\pm}_0]=\pm\widetilde J^{\pm}_0$. Finally, repeating the neutral-sector matching of section \ref{holomorphicorr} without setting $a^{(\pm2)}_{\mathcal{O}_2}\to0$ upgrades \eqref{Sresult} to the flowed frame,
\begin{equation}\label{Sresultflowed}
	\widetilde S^{33}-\big(\widetilde\jmath^{\,3}\big)^{2}=\kappa_J^2\,\hat{\texttt{B}}^{(0)}_{\widehat B_{1/2,0}}\, ,
	\qquad
	\widetilde S^{+-}=\widetilde\jmath^{\,3}-4\,\kappa_J^2\,a^{(+2)}_{\mathcal{O}_2}a^{(-2)}_{\mathcal{O}_2}\, ,
\end{equation}
where now $\hat{\texttt{B}}^{(0)}_{\widehat B_{1/2,0}}$ is the full entry of table \ref{tab:defOPEdata1}, including the $-2a^{(+2)}_{\mathcal{O}_2}a^{(-2)}_{\mathcal{O}_2}$ term. The rotation-breaking sector therefore does not modify the identification of the defect multiplet in \eqref{B120}; it only dresses its coefficient.

Through \eqref{unflowed}, $\widetilde S^{+-}$ determines the two-point function of the superprimary $\hat\phi^+$ of the displacement multiplet $\widehat{\mathcal B}_{0,-1}$, which appears in $J^+_0|\sigma\rangle$ according to \eqref{OPEdefectmodule}. Reading off the coefficient of $(\z_1\z_2)^{-1}$ in $\langle J^-(\z_1)J^+(\z_2)\rangle_\sigma$, and noticing that \eqref{Sresultflowed} implies $\widetilde S^{+-}-\widetilde\jmath^{\,+}\widetilde\jmath^{\,-}=\widetilde\jmath^{\,3}$, we find
\begin{equation}\label{Dmultiplet2pt}
	\langle\hat\phi^-(\infty)\,\hat\phi^+(0)\rangle=-\frac{4\,a^{(0)}_{\mathcal{O}_2}}{\lambda_{\mathcal{B}_{[0,2,0]}}}\, ,
\end{equation}
where $\hat\phi^+$ is normalized as $|\hat\phi^+\rangle\colonequals J^+_0|\sigma\rangle$, i.e.\ $\kappa_J\,b_{J\hat\phi}=1$ in \eqref{OPEdefectmodule}. That the two-point function of the displacement multiplet is fixed by the one-point function of $\mathcal{O}_2$ is a consequence of the supersymmetric relation between the two-point function of the displacement operator and the one-point function of the stress tensor \cite{Bianchi:2018zpb,Bianchi:2019sxz,Bachas:2025brp}. We have checked that \eqref{Dmultiplet2pt} reproduces exactly the prediction of \cite{Bianchi:2019sxz}.

\begin{table}[h!]
\begin{center}
\renewcommand{\arraystretch}{1.6}
\begin{tabular}{|c|c|}
\hline
$\mathcal{M}$ & $a\lambda_{\mathcal{M}}^{(0)}$ \\
\hline\hline

$\id$ 
& $1$ \\
\hline

$\mathcal{B}_{[0,2,0]}$ 
& $\begin{aligned}
    \\
    &a\lambda_{\mathcal{B}_{[0,2,0]}}^{(0)}\\
    &
    \end{aligned}$ \\
\hline

$\mathcal{B}_{[0,4,0]}$ 
& $\begin{aligned}
    \\
    
    &-\frac14-\mathcal{C}^{(0)}+ \frac12 a\lambda^{(0)}_{\mathcal B_{[0,2,0]}}\\
    &
    
    \end{aligned}$ \\
\hline

$\mathcal{B}_{[2,0,2]}$ 
& $\begin{aligned}
\\
&
\frac{1}{6}\left(\frac14+\mathcal{C}^{(0)}+6a^{(+2)}_{\mathcal O_2}
                    a^{(-2)}_{\mathcal O_2}\right)\\
&
\end{aligned}$ \\
\hline

$\mathcal{C}_{[0,2,0],\ell}$ 
& $\begin{aligned}
\\
& 
-\frac{\sqrt{\pi}\,\Gamma(\ell+4)}
 {2^{2\ell+4}\Gamma(\ell+\tfrac72)}
 \left[\frac14+\mathcal{C}^{(0)}- \frac12 a\lambda^{(0)}_{\mathcal B_{[0,2,0]}}
 -(\ell+2)(\ell+5)a^{(+2)}_{\mathcal O_2}a^{(-2)}_{\mathcal O_2}\right]\\
&
\end{aligned}$ \\
\hline

$\mathcal{C}_{[1,0,1],\ell}$ 
& $\begin{aligned}
\\
&
 -\frac{\sqrt{\pi}\,\Gamma(\ell+4)}
 {2^{2\ell+5}\Gamma(\ell+\tfrac72)}
 \left[\frac14+\mathcal{C}^{(0)}
 +(\ell+3)(\ell+4)a^{(+2)}_{\mathcal O_2}a^{(-2)}_{\mathcal O_2}\right]\\
&
\end{aligned}$ \\
\hline

$\mathcal{A}^{\Delta}_{[0,0,0],\ell}$ 
& $\begin{aligned}
\\
&
\tfrac{\pi\Gamma(n+3)\Gamma(n+\ell+4)}
 {2^{2\ell+4n+7}\Gamma(n+\tfrac52)\Gamma(n+\ell+\tfrac72)}
 \Bigg\{\!\!\bigl[a^{(0)}_{\mathcal O_2}\bigr]^2\!+\!(-1)^n\Bigg[(\ell\!+\!1)(2n\!+\!\ell\!+\!6)
 a^{(+2)}_{\mathcal O_2}a^{(-2)}_{\mathcal O_2}
 \!+\!\tfrac{a\lambda^{(0)}_{\mathcal B_{[0,2,0]}}}{4}\\
 &\qquad \!+\!(n^2\!+\!5n\!+\!5)\left(
 \frac14\!+\!\mathcal{C}^{(0)}\!+\!\bigl[a^{(0)}_{\mathcal O_2}\bigr]^2
 \!+\!2a^{(+2)}_{\mathcal O_2}a^{(-2)}_{\mathcal O_2}
 \!-\!\tfrac{a\lambda^{(0)}_{\mathcal B_{[0,2,0]}}}{4}
 \right)\!\Bigg]\!\Bigg\}\!+\!\mathcal{O}(\tfrac{1}{N^2})
 \\
&
\end{aligned}$ \\
\hline

\end{tabular}
\renewcommand{\arraystretch}{1}
\end{center}
\caption{Bulk OPE data in the $q=0$ sector. The long operators appear with their leading-order dimension $\Delta=4+2n+\ell$, with $n\in \mathbb{N}$ and $\ell$ even. The semi-short operators in $\mathcal{C}_{[0,2,0],\ell}$ and $\mathcal{C}_{[1,0,1],\ell}$ have even and odd $\ell$, respectively.}
\label{tab:bulkOPEdata}
\end{table}

\begin{table}[h!]
\begin{center}
\renewcommand{\arraystretch}{1.6}

\makebox[\textwidth]{
\begin{minipage}[t]{0.49\textwidth}
\vspace{-3pt}
\centering
\begin{tabular}{|c|c|}
\hline
$\mathcal{M}$ & $a\lambda_{\mathcal{M}}^{(\pm2)}$ \\
\hline\hline
$\id$ 
& $0$ \\
\hline

$\mathcal{B}_{[0,2,0]}$ 
& $a\lambda_{\mathcal{B}_{[0,2,0]}}^{(\pm 2)}$ \\
\hline
$\mathcal{B}_{[0,4,0]}$ & $2 a^{(0)}_{\mathcal{O}_2}a^{(\pm2)}_{\mathcal{O}_2}$ \\
\hline
$\mathcal{B}_{[2,0,2]}$ 
& $0$ \\
\hline
$\mathcal{C}_{[0,2,0],\ell}$ 
& $\textstyle \frac{\sqrt{\pi}\,\Gamma(\ell+5)}{2^{2\ell+4}\Gamma\!\left(\ell+\tfrac{7}{2}\right)}
\,a^{(0)}_{\mathcal{O}_2}a^{(\pm2)}_{\mathcal{O}_2}$ \\
\hline
$\mathcal{C}_{[1,0,1],\ell}$ 
& $0$ \\
\hline
$\mathcal{A}^{\Delta}_{[0,0,0],\ell}$ 
& $0$ \\
\hline
\end{tabular}
\vfill
\end{minipage}
\hspace{0.005\textwidth}
\begin{minipage}[t]{0.49\textwidth}
\vspace{0pt}
\centering
\begin{tabular}{|c|c|}
\hline
$\mathcal{M}$ & $a\lambda_{\mathcal{M}}^{(\pm4)}$ \\
\hline\hline
$\id$ 
& $0$ \\
\hline

$\mathcal{B}_{[0,2,0]}$ 
& $0$ \\
\hline
$\mathcal{B}_{[0,4,0]}$ & $\big[a^{(\pm2)}_{\mathcal{O}_2}\big]^2$ \\
\hline
$\mathcal{B}_{[2,0,2]}$ 
& $0$ \\
\hline
$\mathcal{C}_{[0,2,0],\ell}$ & $0$ \\
\hline
$\mathcal{C}_{[1,0,1],\ell}$ 
& $0$ \\
\hline
$\mathcal{A}^{\Delta}_{[0,0,0],\ell}$ 
& $0$ \\
\hline
\end{tabular}
\vfill
\end{minipage}
}

\renewcommand{\arraystretch}{1}
\end{center}
\caption{Bulk OPE data in the $q=\pm2, \pm4$ sectors.}
\label{tab:bulkOPEdata2}
\end{table}

\begin{table}[!htbp]
\centering
\small
\renewcommand{\arraystretch}{2.2}
\setlength{\tabcolsep}{10pt}
\begin{tabular}{|c|c|}
\hline
$\widehat{\mathcal M}$ & $\texttt{B}_{\widehat{\mathcal M}}^{(0)}\left(\omega\right)$ \\
\hline
$\hat{\id}$
& $\begin{aligned}
\\
&[a^{(0)}_{\mathcal{O}_2}]^2
+a^{(+2)}_{\mathcal{O}_2}a^{(-2)}_{\mathcal{O}_2}\left(\omega^2+\omega^{-2}\right) \\
&
\end{aligned}$ \\
\hline
$\widehat B_{1/2,0}$
& $
\begin{aligned}
\\
&\Big(-\frac14 -\left[a^{(0)}_{\mathcal{O}_2}\right]^2-2a^{(+2)}_{\mathcal{O}_2} a^{(-2)}_{\mathcal{O}_2} +\tfrac{a\lambda_{\mathcal{B}_{[0,2,0]}}^{(0)}}{4} - \mathcal{C}^{(0)}\Big)
\left(\omega+\omega^{-1}\right)\\
&
\end{aligned}$\\
\hline
$\widehat B_{1,0}$
& $
\begin{aligned}
\\
&\frac12 +\left[a^{(0)}_{\mathcal{O}_2}\right]^2+2a^{(+2)}_{\mathcal{O}_2} a^{(-2)}_{\mathcal{O}_2} - \frac12 a\lambda_{\mathcal{B}_{[0,2,0]}}^{(0)}+ \mathcal{C}^{(0)}\\
&
\end{aligned}$ \\
\hline
$\begin{gathered}\widehat B_{0,r_-},\\[-2pt]
\widehat B_{1/2,r_-}\end{gathered}$
& $\begin{aligned}
\\
&-\frac{r_-+1}{4}+\frac14a\lambda^{(0)}_{\mathcal{B}_{[0,2,0]}}+\left(\frac{r_--1}{4}+\frac14a\lambda^{(0)}_{\mathcal{B}_{[0,2,0]}}\right)\omega^2\\
&
\end{aligned}$ \\[5pt]
\hline
$\widehat B_{0,r_-}$
& $ 
\begin{aligned}
\\
&\left(\frac{r_--1}{4}+\frac14a\lambda^{(0)}_{\mathcal{B}_{[0,2,0]}}\right)\omega^2+\mathcal{O}(1/N^2)\\
&
\end{aligned}$\\
\hline
$\widehat B_{1/2,r_-}$
& $ 
\begin{aligned}
\\
&\left(-\frac{r_-+1}{4}+\frac14a\lambda^{(0)}_{\mathcal{B}_{[0,2,0]}}\right)\omega+\mathcal{O}(1/N^2)\\
&
\end{aligned}$\\
\hline
$\widehat A^{\hat\Delta}_{0,r_-}$
& $
\begin{aligned}
\\
&\frac{|r_-|\!+\!2k\!+\!3}{4}
 \!-\!\frac{2k\!+\!3}{4}a\lambda^{(0)}_{\mathcal B_{[0,2,0]}}+\delta_{r_-,0}(6k^2\!+\!18k\!+\!13)\\
 & \times \left(\frac14\!+\!\mathcal{C}^{(0)}\!+\!\bigl[a^{(0)}_{\mathcal O_2}\bigr]^2
 \!+\!2a^{(+2)}_{\mathcal O_2}a^{(-2)}_{\mathcal O_2}
 \!-\!\tfrac{a\lambda^{(0)}_{\mathcal B_{[0,2,0]}}}{4}\right) \!+\!\mathcal{O}(1/N^2) \\
 &
 \end{aligned}$ \\
\hline
\end{tabular}
\caption{Defect OPE data in the $q=0$ sector, for $r_-\geq0$. The $r_-<0$ coefficients are obtained by conjugation, $\omega \rightarrow \omega^{-1}$, with $\omega$ as in \eqref{udef}. In the fourth row, we report the combination $\texttt{B}_{\widehat B_{0,r_-}}^{(0)}(\omega) +\omega^{-1}\, \texttt{B}_{\widehat B_{1/2,r_-}}^{(0)}(\omega)$, that can be extracted exactly from the protected correlator. The individual coefficients can be extracted from the full correlator. The long family has classical dimension $\hat\Delta^{(0)}=|r_-|+2+2k$, with $k = 0,1,2,\dots$, and its anomalous dimension at order $1/N$ is given in \eqref{defgamma}.}
\label{tab:defOPEdata1}
\end{table}

\begin{table}[!htbp]
\centering
\small
\renewcommand{\arraystretch}{2.15}
\setlength{\tabcolsep}{7pt}
\begin{tabular}[t]{|c|c|}
\hline
$\widehat{\mathcal M}$ & $\texttt{B}_{\widehat{\mathcal M}}^{(\pm2)}(\omega)$ \\
\hline
$\hat{\id}$
& $a^{(0)}_{\mathcal{O}_2}a^{(\pm2)}_{\mathcal{O}_2}(\omega+\omega^{-1})$ \\
\hline
$\widehat B_{1/2,0}$ & $-\frac{a\lambda^{(\pm2)}_{\mathcal{B}_{[0,2,0]}}}{2}$ \\
\hline
$\widehat B_{1,0}$ & $0$ \\
\hline
$\widehat B_{0,r_-}$ & $\frac{a\lambda^{(\pm2)}_{\mathcal{B}_{[0,2,0]}}}{2} \omega$ \\
\hline
$\widehat B_{1/2,r_-}$ & $0$ \\
\hline
$\widehat A^{\hat\Delta}_{0,r_-}$
& $0$ \\
\hline
\end{tabular}%
\hspace{2em}%
\begin{tabular}[t]{|c|c|}
\hline
$\widehat{\mathcal M}$ & $\texttt{B}_{\widehat{\mathcal M}}^{(\pm4)}(\omega)$ \\
\hline
$\hat{\id}$ & $[a^{(\pm2)}_{\mathcal{O}_2}]^2$ \\
\hline
$\widehat B_{1/2,0}$ & $0$ \\
\hline
$\widehat B_{1,0}$ & $0$ \\
\hline
$\widehat B_{0,r_-}$ & $0$ \\
\hline
$\widehat B_{1/2,r_-}$ & $0$ \\
\hline
$\widehat A^{\hat\Delta}_{0,r_-}$
& $0$ \\
\hline
\end{tabular}
\caption{Defect OPE data in the $q=\pm2, \pm4$ sectors.}
\label{tab:defOPEdata2}
\end{table}

\subsection{OPE data}
Given the correlators \eqref{q0result} and \eqref{chargedres}, we can extract the OPE data in both the bulk and defect channels. In the bulk channel, this is achieved by matching the correlators with the OPE \eqref{bulkblockgen}, with the blocks in tables \ref{tab:blocks} and \ref{tab:blocksq}. This determines the OPE data reported in tables \ref{tab:bulkOPEdata} and \ref{tab:bulkOPEdata2}\footnote{We choose to report the product of one- and three-point functions of the superprimaries. The one-point function coefficients can then be extracted using the bulk OPE coefficients in table $3$ of \cite{Beem:2016wfs}, modulo degeneracies and the different normalization of $\mathcal{O}_2$.}, as a function of the defect-specific coefficients $a_{\mathcal O_2}^{(0)}$, $a_{\mathcal O_2}^{(\pm 2)}$, $a\lambda_{\mathcal B_{[0,2,0]}}^{(0)}$ and $\mathcal{C}^{(0)}$, whose values for Gukov-Witten defects are given in \eqref{a2broken}, \eqref{lambdaO2} and \eqref{Cloc}. The coefficients associated to the superprimaries of short and semi-short multiplets are exact in both $N$ and the coupling.

In the charge $q=0$ sector, the full correlator in \eqref{q0result} allows us to extract new OPE data also for non-protected operators at large $N$ and strong coupling. In particular, we obtain
\begin{align}\label{alambdalong}
 &\left\langle a\lambda^{(0)}_{
 \mathcal A^{\Delta=4+2n+\ell}_{[0,0,0],\ell}}\right\rangle
 =\frac{\pi\Gamma(n+3)\Gamma(n+\ell+4)}
 {2^{2\ell+4n+7}\Gamma(n+\tfrac52)\Gamma(n+\ell+\tfrac72)}
 \Bigg\{\bigl[a^{(0)}_{\mathcal O_2}\bigr]^2\nonumber \\
 &\quad+(-1)^n\Bigg[(\ell+1)(2n+\ell+6)
 a^{(+2)}_{\mathcal O_2}a^{(-2)}_{\mathcal O_2}
 +\frac14a\lambda^{(0)}_{\mathcal B_{[0,2,0]}}\\
 &\qquad +(n^2+5n+5)\left(
 \frac14+\mathcal{C}^{(0)}+\bigl[a^{(0)}_{\mathcal O_2}\bigr]^2
 +2a^{(+2)}_{\mathcal O_2}a^{(-2)}_{\mathcal O_2}
 -\frac14a\lambda^{(0)}_{\mathcal B_{[0,2,0]}}
 \right)\Bigg]\Bigg\}+\mathcal{O}(1/N^2)\,,\nonumber
\end{align}
where $\ell$ is even and the last line encodes the contribution of the $\mathcal{F}^{(0)}_0(z,\bar z)$ ambiguity.
These coefficients should be understood as sums over degenerate operators. Disentangling the individual contributions would require solving a mixing problem, analogous to the one encountered in $\mathcal{N}=4$ SYM without defects \cite{Aprile:2017xsp}. We leave this analysis for future work.

In the defect channel, we can extract bulk-defect couplings for both protected and non-protected operators, by matching \eqref{q0result} and \eqref{chargedres} with the defect superconformal block expansions \eqref{defectOPEexpl}. The resulting OPE data are collected in tables \ref{tab:defOPEdata1} and \ref{tab:defOPEdata2}.  For each superprimary operator, we list the corresponding coefficient function $\texttt{B}^{(q)}_{\hat{\mathcal O}}(\omega)$, defined in \eqref{B2}. The bulk-defect coefficients can be obtained by comparing with \eqref{Beta1Beta2} and \eqref{eq:B-eta-corrected}. Some of the OPE data of short multiplets are exact in $N$ and in the coupling as detailed in the corresponding tables. We present explicitly only the multiplets with $r_-\geq 0$, since the data for their conjugate counterparts are not independent and can be obtained by the replacement $\omega\to\omega^{-1}$, with $\omega$ defined in \eqref{udef}.

The logarithmic terms proportional to $\log(z\bar z)$ in the $q=0$ sector
signal the presence of long defect operators with anomalous dimensions. We
parametrize their dimensions as
\begin{equation}
    \hat\Delta_{\hat{\mathcal O}}
    =\hat\Delta^{(0)}
    +\frac{\gamma_{\hat{\mathcal O}}}{N}
    +\mathcal O\!\left({1}/{N^2}\right),
\end{equation}
where $\hat\Delta^{(0)}=2+2k+|r_-|$ is the order $N^0$ (GFF) classical dimension extracted from the disconnected piece of the two-point function.\footnote{The integer label $k$ here should not be confused with the affine level of the chiral algebra used in subsection \ref{chiralalgebra}.}
To extract the anomalous dimensions, we expand the defect superconformal
blocks in \eqref{defectOPEexpl} around $\hat\Delta^{(0)}$ and match the
resulting $\log(z\bar z)$ terms with those of the $q=0$ sector. Schematically,
\begin{eqnarray}
\left.
\texttt{B}_{\hat{\mathcal O}}\,
\widehat{\mathcal G}_{\widehat{A}}
\right|_{\hat\Delta=\hat\Delta^{(0)}+\!\frac{\gamma_{\hat{\mathcal O}}}{N}}
\!&=&
\left.\texttt{B}_{\hat{\mathcal O}}\right|_{\mathcal O(1)}
\widehat{\mathcal G}_{\widehat{A}}
\big|_{\hat\Delta=\hat\Delta^{(0)}}
\\
&+&
\frac{1}{N}
\left[
\left.\texttt{B}_{\hat{\mathcal O}}\right|_{\mathcal O(1/N)}
\widehat{\mathcal G}_{\widehat{A}}
\big|_{\hat\Delta=\hat\Delta^{(0)}}
+
\left.\!\texttt{B}_{\hat{\mathcal O}}\right|_{\mathcal O(1)}
\gamma_{\hat{\mathcal O}}\,
\partial_{\hat\Delta}
\widehat{\mathcal G}_{\widehat{A}}
\big|_{\hat\Delta=\hat\Delta^{(0)}}
\right]
+\cdots \, .\nonumber
\end{eqnarray}
We obtain
\begin{align}\label{defgamma}
    \frac{\langle\gamma_{\widehat{\mathcal O}}\rangle}{N}
 &=4(k+1)(k+2)\Bigg[ - 
 \frac{a\lambda^{(0)}_{\mathcal B_{[0,2,0]}}}{2 \left(|r_-|+2k+3\right)}
 \nonumber\\
 &\qquad+2\delta_{r_-,0}
 \left(\frac14+\mathcal{C}^{(0)}+\bigl[a^{(0)}_{\mathcal O_2}\bigr]^2
 +2a^{(+2)}_{\mathcal O_2}a^{(-2)}_{\mathcal O_2}
 -\frac{a\lambda^{(0)}_{\mathcal B_{[0,2,0]}}}{4}\right)\Bigg].
\end{align}
The non-logarithmic terms in the same block expansion determine the
corresponding bulk-to-defect OPE coefficients,
\begin{align}\label{longdefcoeff}
 \left\langle\!\texttt{B}^{(0)}_{
 \widehat{\mathcal A}^{\widehat\Delta}_{0,r_-}}\!\!\!\!\!(\omega)\!\!\right\rangle
 &=\frac{|r_-|\!+\!2k\!+\!3}{4}
 \!-\!\frac{2k\!+\!3}{4}a\lambda^{(0)}_{\mathcal B_{[0,2,0]}}
 \\
 &\quad+\delta_{r_-,0}(6k^2\!+\!18k\!+\!13)
 \left(\frac14\!+\!\mathcal{C}^{(0)}\!+\!\bigl[a^{(0)}_{\mathcal O_2}\bigr]^2
 \!+\!2a^{(+2)}_{\mathcal O_2}a^{(-2)}_{\mathcal O_2}
 \!-\!\frac{a\lambda^{(0)}_{\mathcal B_{[0,2,0]}}}{4}\right) \!+\!\mathcal{O}\left(\tfrac{1}{N^2}\right)\, . \nonumber
\end{align}
Here, the brackets indicate that the extracted CFT data are averaged over
operators that are degenerate at leading order, as in the case of the non-protected bulk operators. The last lines in both \eqref{defgamma} and \eqref{longdefcoeff} are the contributions of the low-spin ambiguity.

\section{Holographic Witten diagram bootstrap}\label{sec:holo-witten}
The results of the previous sections admit an independent holographic derivation. As discussed in section \ref{sec:holo}, the Gukov-Witten surface defect can be realized, at large $N$ and strong 't Hooft coupling, by a probe D3-brane in $AdS_5\times S^5$. The two-point function of chiral primary operators is mapped to the corresponding correlator of the dual Kaluza-Klein modes of type IIB supergravity in the probe D3-brane background. At leading order, this correlator is given by a sum of tree-level Witten diagrams with a single brane insertion. Reproducing the correlator \eqref{q0result} from this holographic description provides a non-trivial strong-coupling check of our result and elucidates its geometric origin.

In principle, such a computation could be carried out from first principles by deriving all the relevant interaction vertices from the supergravity and probe-brane actions and evaluating the corresponding tree-level Witten diagrams.\footnote{A first-principles analysis of this type has been carried out for the holographic Wilson line in $\mathcal{N}=4$ SYM in \cite{Gimenez-Grau:2023fcy}.} This would require detailed knowledge of the cubic couplings of type IIB supergravity on $AdS_5\times S^5$, the linear couplings of the supergravity fields to the D3-brane, as well as the quadratic and cubic interaction vertices of the DBI and Wess--Zumino brane actions. Even in the probe approximation, deriving these couplings is technically challenging. Moreover, the supergravity action is affected by contact-term ambiguities originating from field redefinitions \cite{Lee:1998bxa,Arutyunov:1998hf}.

Instead of performing the explicit supergravity computation, we adopt the position-space Witten diagram bootstrap introduced in \cite{Rastelli:2017udc} and generalized to holographic defects in \cite{Gimenez-Grau:2023fcy}. The idea is to write the correlator as the most general linear combination of Witten diagrams compatible with the spectrum of bulk\footnote{Throughout this section, the term ``bulk'' is used in two distinct senses. Depending on the context, it may denote either the bulk of the AdS spacetime or a bulk operator in the ambient CFT, namely an operator that is not localized on the defect. The intended meaning will be clear from the context.} and brane fields, leaving their coefficients undetermined. 
 
 The unknown coefficients encode the bulk and brane interaction vertices and are fixed by imposing the superconformal Ward identities \eqref{WI}, together with matching the leading term in the bulk OPE expansion. In this way, the holographic correlator can be reconstructed from the knowledge of the spectrum of bulk and defect supergravity fields, the corresponding Witten diagrams, and supersymmetry, without requiring the explicit interaction vertices.

For simplicity, in this section we restrict to the rotation-invariant case
\begin{equation}
    \beta=\gamma=0\,,
\end{equation}
or equivalently $u_0=0$ in the holographic description, see \eqref{u0def}. As discussed in section \ref{sec:holo}, this corresponds to the limit where the circle inside $AdS_5$ collapses, its radius being proportional to $\sinh u_0$. The D3-brane worldvolume geometry remains $AdS_3 \times S^1$, with the $AdS_3$ factor embedded in $AdS_5$ and, crucially, the $S^1$ entirely contained in $S^5$. As a consequence, spacetime transverse rotations and $SO(2)_{\mathcal R_\perp}$ are no longer geometrically identified, and the full $SO(2)_{\mathcal M_\perp} \times SO(2)_{\mathcal R_\perp}$ symmetry is preserved.

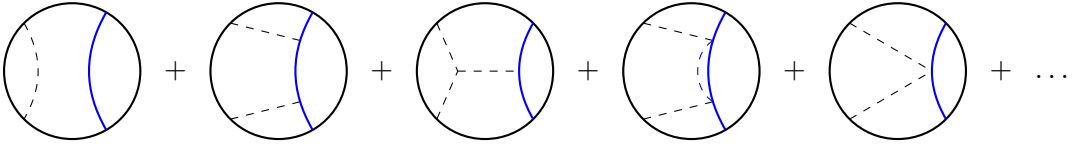
\begin{figure}
 \begin{align*}
 \begin{tikzpicture}[valign,scale=0.9]
    \tikzstyle{every node}=[font=\scriptsize]
    \pgfmathsetmacro{\x}{sqrt(1)/2}
    \pgfmathsetmacro{\y}{sqrt(3)/2}
    \pgfmathsetmacro{\xx}{sqrt(2)/2}
    \pgfmathsetmacro{\yy}{sqrt(2)/2}
    \pgfmathsetmacro{\z}{0.5}
    \pgfmathsetmacro{\rx}{0.32}
    \pgfmathsetmacro{\ry}{0.45}
    \draw [thick] (0,0) circle [radius=1];
    \draw [dashed]      (-\xx,+\yy) to[out=-60,in=60] (-\xx,-\yy);
    \draw [thick, blue] (\x,+\y) to[out=240,in=120] (\x,-\y);
 \end{tikzpicture}
 \;\;+\;\;
 \begin{tikzpicture}[valign,scale=0.9]
    \tikzstyle{every node}=[font=\scriptsize]
    \pgfmathsetmacro{\x}{sqrt(1)/2}
    \pgfmathsetmacro{\y}{sqrt(3)/2}
    \pgfmathsetmacro{\xx}{sqrt(2)/2}
    \pgfmathsetmacro{\yy}{sqrt(2)/2}
    \pgfmathsetmacro{\z}{0.5}
    \pgfmathsetmacro{\rx}{0.32}
    \pgfmathsetmacro{\ry}{0.45}
    \draw [thick] (0,0) circle [radius=1];
    \draw [dashed]      (-\xx,+\yy) -- (\rx,\ry);
    \draw [dashed]      (-\xx,-\yy) -- (\rx,-\ry);
    \draw [thick, blue] (\x,+\y) to[out=240,in=120] (\x,-\y);
 \end{tikzpicture}
 \;\;+\;\;
   \begin{tikzpicture}[valign,scale=0.9]
    \tikzstyle{every node}=[font=\scriptsize]
    \pgfmathsetmacro{\x}{-sqrt(2)/2}
    \pgfmathsetmacro{\y}{sqrt(2)/2}
    \pgfmathsetmacro{\z}{-0.5}
    \pgfmathsetmacro{\r}{0.4}
    \draw [thick] (0,0) circle [radius=1];
    \draw [thick, blue] (-\x,+\y) to[out=240,in=120] (-\x,-\y);
    \draw [dashed] (+\x,+\y) -- (-\r,0);
    \draw [dashed] (+\x,-\y) -- (-\r,0);
    \draw [dashed] (-\r,  0) -- (-\z,0);
 \end{tikzpicture}
 \;\;+\;\;
 \begin{tikzpicture}[valign,scale=0.9]
    \tikzstyle{every node}=[font=\scriptsize]
    \pgfmathsetmacro{\x}{sqrt(1)/2}
    \pgfmathsetmacro{\y}{sqrt(3)/2}
    \pgfmathsetmacro{\xx}{sqrt(2)/2}
    \pgfmathsetmacro{\yy}{sqrt(2)/2}
    \pgfmathsetmacro{\z}{0.5}
    \pgfmathsetmacro{\rx}{0.32}
    \pgfmathsetmacro{\ry}{0.45}
    \draw [thick] (0,0) circle [radius=1];
    \draw [dashed]      (-\xx,+\yy) -- (\rx,\ry);
    \draw [dashed]      (-\xx,-\yy) -- (\rx,-\ry);
    \draw [dashed,blue] (\rx,\ry) to[out=210,in=150] (\rx,-\ry);
    \draw [thick, blue] (\x,+\y) to[out=240,in=120] (\x,-\y);
 \end{tikzpicture}
 \;\;+\;\;
 \begin{tikzpicture}[valign,scale=0.9]
    \tikzstyle{every node}=[font=\scriptsize]
    \pgfmathsetmacro{\x}{sqrt(2)/2}
    \pgfmathsetmacro{\y}{sqrt(2)/2}
    \pgfmathsetmacro{\z}{0.5}
    \pgfmathsetmacro{\r}{-0.75}
    \draw [thick] (0,0) circle [radius=1];
    \draw [thick, blue] (+\x,+\y) to[out=240,in=120] (+\x,-\y);
    \draw [dashed]      (-\x,+\y) -- (\z,0);
    \draw [dashed]      (-\x,-\y) -- (\z,0);
  \end{tikzpicture}
  \;\;+\;\;
  \ldots
\end{align*}
\caption{The Witten diagrams for the two-point function of $\mathcal{O}_p$ in the presence of a holographic surface defect. The solid blue line denotes the brane, while the dashed black lines represent propagators. The figure is taken from \cite{Gimenez-Grau:2023fcy}.}
\label{fig:diagrams}
\end{figure}

At leading order in the supergravity expansion, the two-point function of the $[0,p,0]$ chiral primary $\mathcal{O}_p$, in the presence of the holographic surface defect, can be written as\footnote{The correlator $\mathcal{F}(z,\bar z,\alpha,\bar\alpha,\vartheta)$ collapses to $\mathcal{F}^{(0)}(z,\bar z,\alpha,\bar\alpha)$, since in this section we are considering the case when the spacetime and R-symmetry transverse rotations are independently preserved.}
\begin{eqnarray}\label{ansatz1}\nonumber
\mathcal{F}^{(0)}(z,\bar z,\alpha,\bar\alpha)
&=&
\sum_{\mathcal X\in\mathrm{bulk}} A_{\mathcal X }
\,E^{\Delta_\mathcal{X},\ell}_{p,p}(z,\bar z)\,
\xi_R^p \,h^{(0)}_{R_\mathcal{X}}(\alpha,\bar{\alpha})
+
\sum_{\widehat{\mathcal X}\in\mathrm{defect}}
\widehat{A}_{\widehat{\mathcal X}}\,
\widehat{E}^{\hat\Delta_{\widehat{\mathcal X}},s}_{p,p}(z,\bar z)\,
\hat h^{(0)}_{\hat R_{\widehat{\mathcal X}}}(\alpha,\bar{\alpha})\\
&&+
C_{p,p}(z,\bar z)
\sum_{n,m=-p}^{p} B_{n,m}\,
\bigg(\frac{1+\alpha\bar\alpha}{\sqrt{\alpha\bar\alpha}}\bigg)^n
\bigg(\frac{\alpha}{\bar\alpha}\bigg)^{m/2} + \,\text{disconnected\,.}
\end{eqnarray}
The coefficients $A_{\mathcal X}$, $\widehat A_{\widehat{\mathcal X}}$ and $B_{n,m}$ encode the unknown bulk and brane interaction vertices. The functions $h^{(0)}$ and $\hat h^{(0)}$, defined in appendix \ref{app:blocks}, are the R-symmetry blocks associated with the representations of the exchanged fields. The cross-ratio $\xi_R$ is defined in \eqref{crossratio}, while $E$, $\widehat E$ and $C$ denote the last three Witten diagrams of figure \ref{fig:diagrams}, respectively.
Their explicit expressions were derived in \cite{Gimenez-Grau:2023fcy} and are collected in appendix \ref{app:witten}. More precisely, the bulk-exchange diagrams $E^{\Delta_\mathcal{X},\ell}_{p,p}(z,\bar z)$ contain an intermediate supergravity field $\mathcal{X}$ with scaling dimension $\Delta_\mathcal{X}$ and spin $\ell$, propagating in $AdS_5$ and ending on the brane through a one-point coupling. The defect-exchange diagrams $\widehat{E}^{\hat\Delta_{\widehat{\mathcal X}},s}_{p,p}(z,\bar z)$ contain an intermediate probe-brane fluctuation field $\widehat{\mathcal{X}}$ with scaling dimension  $\hat\Delta_{\widehat{\mathcal{X}}}$ and transverse spin $s$, propagating along the defect worldvolume. The contact diagrams $C_{p,p}(z,\bar z)$ are instead local bulk-brane vertices with the two external bulk fields of equal conformal dimension $p$ inserted directly on the brane. The first two diagrams of figure \ref{fig:diagrams} are the disconnected contribution, which is simply given by $(-2)^{-p}(\xi_R/\xi)^{p}$, with $\xi$ and $\xi_R$ defined in \eqref{crossratio}, together with the squared one-point function $\left[a^{(0)}_{\mathcal{O}_p}\right]^2$. The external legs of all Witten diagrams correspond to bulk-to-boundary propagators of the scalar Kaluza-Klein mode $s_p$, dual to $\mathcal{O}_p$.

Following the analysis of \cite{Gimenez-Grau:2023fcy}, it is convenient to repackage the bulk exchange diagrams into \emph{Polyakov-Regge blocks}, denoted by $P^{\Delta_{\mathcal X},\ell}_{p,p}(z,\bar z)$. These are obtained by supplementing the bulk exchange Witten diagrams with a suitable combination of contact diagrams, chosen such that the resulting object is Regge superbounded\footnote{By Regge superboundedness we mean that $|P^{\Delta_{\mathcal X},\ell}_{pp}(r,w)|\leq C \, |w|^{-\epsilon}$ as $w\to\infty$, for some $C,\epsilon>0$.}. We refer to appendix \ref{app:witten} for explicit examples. The main advantage of this basis is that each Polyakov-Regge block is in one-to-one correspondence with the conformal dispersion relation applied to a single bulk conformal block of the same scaling dimension and spin, see again \cite{Gimenez-Grau:2023fcy}.

The spectrum of bulk and defect fields appearing in the Witten diagram expansion was determined in \cite{Arutyunov:1998hf, Arutyunov:1999en} and \cite{Constable:2002xt}, respectively. The bulk spectrum consists of the Kaluza-Klein modes of type IIB supergravity that appear in the quadratic supergravity action and can couple to the external chiral primary operators through non-vanishing cubic interactions. The defect spectrum is instead given by the fluctuations of the probe D3-brane around its classical embedding, obtained by expanding the DBI and Wess--Zumino worldvolume actions to quadratic order. At this stage, we remain agnostic as to which of these fields actually possess non-vanishing couplings to the $s_p$ KK modes.

Let us now specialize to the two-point function of $\mathcal{O}_2$. The relevant bulk and defect fields in the spectrum are listed in table \ref{tab:fields}. In particular, $s_2$ is the Kaluza-Klein mode dual to $\mathcal{O}_2$, while $A_\mu$ and $\varphi_{(\mu,\nu)}$ are its superdescendants. The defect excitations $\hat y^A$ and $\{\hat x^i, \hat z^i\}$ describe fluctuations of the probe D3-brane inside $S^5$ and $AdS_5$, respectively. The complete spectrum of bulk and probe-brane fluctuations is considerably larger, see \cite{Arutyunov:1999en,Constable:2002xt}. However, only the fields listed in table \ref{tab:fields} can contribute to the correlator of two $\mathcal{O}_2$ operators. This follows from the R-symmetry selection rules and from the fact that only defect fluctuations with vanishing parallel spin can be exchanged, since the external operators are scalars.
\begin{table}[t]
\begin{center}
\renewcommand{\arraystretch}{1.6}

\makebox[\textwidth]{
\begin{minipage}[t]{0.49\textwidth}
\centering
\begin{tabular}{|c|c|c|c|}
\hline
$\text{Bulk field}$ & $\Delta$ & $\ell$ & $\text{R-symm.}$ \\
\hline\hline
$s_2$ & $2$ & $0$ & $[0,2,0]$ \\
\hline
$A_{\mu}$ & $3$ & $1$ & $[1,0,1]$ \\
\hline
$\varphi_{(\mu,\nu)}$ & $4$ & $2$ & $[0,0,0]$ \\
\hline
\end{tabular}
\vfill
\end{minipage}
\begin{minipage}[t]{0.49\textwidth}
\centering
\begin{tabular}{|c|c|c|c|}
\hline
$\text{Defect field}$ & $\hat{\Delta}$ & $s$ & $\text{R-symm.}$ \\
\hline\hline
$\hat y^A$ & $2$ & $0$ & $\left[\tfrac{1}{2},\tfrac{1}{2}\right]_{\pm1}$ \\
\hline
$\hat x^i$ & $3$ & $\pm1$ & $\left[0,0\right]_{0}$ \\
\hline
$\hat z^i$ & $1$ & $\pm1$ & $\left[0,0\right]_{\pm2}$ \\
\hline
\end{tabular}
\vfill
\end{minipage}
}

\renewcommand{\arraystretch}{1}
\end{center}
\caption{Supergravity fields that couple to the brane and to $\mathcal{O}_2$.}
\label{tab:fields}
\end{table}
The most general Witten diagram ansatz consistent with this spectrum is
\begin{eqnarray}\label{ansatz2}\nonumber
\mathcal{F}^{(0)}(z,\bar z,\alpha,\bar\alpha)
&=&
A_{s}
\,P^{2,0}_{2,2}(z,\bar z)\,
\xi_R^2 \, h^{(0)}_{[0,2,0]}(\alpha,\bar{\alpha})
+A_{A}
\,P^{3,1}_{2,2}(z,\bar z)\,
\xi_R^2 \, h^{(0)}_{[1,0,1]}(\alpha,\bar{\alpha})
\nonumber\\
&+& A_{\varphi}
\,P^{4,2}_{2,2}(z,\bar z)\,
\xi_R^2 \, h^{(0)}_{[0,0,0]}(\alpha,\bar{\alpha})
+ \widehat{A}_{\hat x}\,
(\widehat{E}^{3,1}_{2,2}(z,\bar z)+\widehat{E}^{3,-1}_{2,2}(z,\bar z)) \nonumber
\\
&+&\widehat{A}_{\hat y}\,
\widehat{E}^{2,0}_{2,2}(z,\bar z)\,
(\hat h^{(0)}_{[1/2,1/2]_{1}}(\alpha,\bar{\alpha})+\hat h^{(0)}_{[1/2,1/2]_{-1}}(\alpha,\bar{\alpha}))
\\
&+&\widehat{A}_{\hat z}\,
(\widehat{E}^{1,1}_{2,2}(z,\bar z)\,
\hat h^{(0)}_{[0,0]_{2}}(\alpha,\bar{\alpha})+\widehat{E}^{1,-1}_{2,2}(z,\bar z)\,
\hat h^{(0)}_{[0,0]_{-2}}(\alpha,\bar{\alpha}))\nonumber
\\
&+&
C_{2,2}(z,\bar z)
\sum_{n,m=-2}^{2}
B_{n,m}
\left(\frac{1+\alpha\bar\alpha}{\sqrt{\alpha\bar\alpha}}\right)^n
\left(\frac{\alpha}{\bar\alpha}\right)^{m/2}
+\frac14 (\xi_{R}/\xi)^{2}+\left[a^{(0)}_{\mathcal{O}_2}\right]^2\!\!. \nonumber
\end{eqnarray}
The explicit form of each of these diagrams is collected in appendix \ref{app:witten}. In particular, we use the Polyakov blocks in the form given in \eqref{polyakovb}, without extra normalization factors.
The superconformal Ward identities \eqref{WI} impose linear relations among the unknown coefficients. Solving these constraints, we find
\begin{equation}\label{coeffans}
A_\varphi = -\frac{A_s}{180}\,, \quad
A_A = -\frac{A_s}{12}\,, \quad
\hat A_{\hat z} = -\frac{2A_s}{\pi}\,, \quad
B_{0,-2} = \frac{2 A_s}{\pi}\,, \quad
B_{1,-1} = -\frac{2 A_s}{\pi}\,.
\end{equation}
All the other coefficients are set to zero by the Ward identities.
The remaining overall normalization $A_s$ can be fixed by considering the bulk OPE limit $z,\bar z\rightarrow1$. In this limit, the Witten diagram ansatz must reproduce the leading singular behavior dictated by the bulk conformal block expansion \eqref{bulkblockgen}. Explicitly,
\begin{equation}\label{normalans}
\mathcal{F}^{(0)}(z,\bar z,\alpha, \bar \alpha)\xrightarrow{z,\zb\to1} \frac{\xi^2_R}{\xi^2}\left(\frac14 - a \lambda_{\mathcal{B}_{[0,2,0]}}^{(0)}\frac{(1-z) (1-\bar z) (\alpha  (\bar \alpha+2)+2 \bar \alpha+1)}{12 (\alpha -1) (\bar \alpha-1)}+\ldots\right)\,, 
\end{equation}
where $\xi$ and $\xi_R$ are defined in \eqref{crossratio} and $a \lambda_{\mathcal{B}_{[0,2,0]}}^{(0)}$ is defined in \eqref{alambda}.
Matching \eqref{ansatz2} with \eqref{normalans} uniquely determines $A_s$ in terms of the protected three- and one-point function coefficients $\lambda_{\mathcal{O}_2 \mathcal{O}_2 \mathcal{O}_2}$ and $a^{(0)}_{\mathcal{O}_2}$, given in \eqref{lambdaO2} and \eqref{a2broken}. We obtain
\begin{equation}\label{normans}
    A_s= - \frac12 a \lambda_{\mathcal{B}_{[0,2,0]}}^{(0)}= - \frac12 \lambda_{\mathcal{O}_2 \mathcal{O}_2 \mathcal{O}_2} a_{\mathcal{O}_2}^{(0)} = -\frac{1}{N} +\mathcal{O}(1/N^2)\,,
\end{equation}
where in the last step we used the explicit value of the Gukov-Witten coefficient, see \eqref{lambdaO2} and \eqref{a2broken}. Substituting the coefficients \eqref{coeffans} and \eqref{normans} into the Witten diagram ansatz \eqref{ansatz2}, we recover precisely the correlator $\mathcal{F}^{(0)}(z,\bar z,\alpha,\bar\alpha)$ given in \eqref{finalbg0}.

\section{Conclusions}\label{sec:concl}

In this work we have bootstrapped the two-point function of the stress-tensor multiplet superprimary $\mathcal{O}_2$ in $\mathcal{N}=4$ SYM in the presence of a half-BPS surface defect. The holomorphic sector and the charged sectors $q\neq0$ are fixed exactly in $N$ and in the 't~Hooft coupling, while the neutral sector is reconstructed from the conformal dispersion relation at leading non-trivial order at large $N$ and strong coupling. Along the way we have constructed the bulk and defect superconformal blocks in all charged sectors, identified the defect with a spectrally flowed module of the chiral algebra, and extracted infinitely many new pieces of defect CFT data. Most of these results rely only on the symmetry preserved by the defect, and not on its explicit realization.

Our analysis admits several natural extensions. The same methods should apply to two-point functions of higher chiral primaries $\mathcal{O}_p$, for which the only additional input is a set of protected one-point functions, some of which are already available in \cite{Choi:2024ktc}. It would be interesting to push the neutral sector to subleading orders in $1/N$,\footnote{A different but interesting set-up where this is carried out is presented for example in \cite{Chen:2026fnf}.} which requires resolving the mixing of degenerate long operators, in analogy with what is done for four-point functions in the absence of defects \cite{Aprile:2017xsp}. On the holographic side, our result provides a target for an explicit probe brane computation in the dual supergravity background \cite{Drukker:2008wr,Gimenez-Grau:2023fcy}, and the one-loop extension of the Witten diagram bootstrap appears feasible. It would also be valuable to apply our analysis to more general half-BPS surface defects in $\mathcal{N}=4$ SYM \cite{Harvey:2008zz,Arav:2024exg,Conti:2025wwf}, as well as to non-supersymmetric codimension-two defects in free and critical scalar theories \cite{Lauria:2020emq,Trepanier:2023tvb,Raviv-Moshe:2023yvq,Giombi:2023dqs,Diatlyk:2024ngd,Soderberg:2017oaa,Bianchi:2021snj}. Particularly close to our setup are the defects constructed in \cite{Kravchuk:2025evf} since they also preserve a combination of transverse rotations and a global symmetry.

A more ambitious direction is to move beyond two-point functions. Multipoint correlators in defect CFT have become accessible in recent years, mostly for the half-BPS Wilson line of $\mathcal{N}=4$ SYM, where correlators of several defect operators have been computed and bootstrapped both at weak and at strong coupling \cite{Barrat:2021tpn,Barrat:2022eim,Artico:2024wut}. More recently, a lightcone bootstrap has been developed for defect correlators involving bulk and defect operators at the same time \cite{Bianchi:2026orb}: imposing crossing in the lightcone limit constrains the defect spectrum at large transverse spin and predicts new twist-accumulating families of defect operators. The simplest observable of this kind, the three-point function of two bulk operators and one defect operator, is a natural continuation of the present analysis, since its defect channel is controlled by the same bulk-to-defect couplings that we have extracted here, while its bulk channel probes genuinely new data. For a surface defect the kinematics is richer than for a line: the breaking of transverse rotations introduces additional angular cross-ratios in the multipoint case as well, and the associated blocks still have to be constructed. We expect both the dispersive methods and the chiral algebra techniques used in this paper to have a counterpart in that setting.

\acknowledgments
We would like to thank A. Chalabi, G. Cuomo, R. Izquierdo García, M. Meineri, C. Meneghelli and Y. Wang for useful discussions. LB's research is partially supported
by the MUR PRIN contract 2022N9CTAE ``Constraining strongly coupled quantum field theories using symmetry''. DB's research is  funded by the Deutsche Forschungsgemeinschaft (DFG, German Research Foundation) under the Collaborative Research Center - SFB 1624 “Higher structures, moduli spaces and integrability” - 506632645. VF is supported by the DFG via the Heisenberg Program 506208580.  The authors would like to thank the Isaac Newton Institute for Mathematical Sciences, Cambridge, for support and hospitality during the programme ``Quantum Field Theory with Boundaries, Impurities, and Defects'' where work on this paper was undertaken. This work was supported by EPSRC grant no EP/R014604/1.

\appendix

\section{Embedding space and superconformal algebra}\label{app:conv}

\subsection*{Embedding space conventions}
In the embedding space formalism four-dimensional spacetime is realized as the projective null cone in $\mathbb{R}^{2,4}$. In this description, conformal transformations act linearly on the embedding coordinates, making conformal covariance manifest. A point $x^\mu \in \mathbb{R}^{1,3}$ is represented by the null vector
\begin{equation}
x^\mu \;\longmapsto\; P^M = (P^\mu,P^+,P^-)
    =(x^\mu,1,x^2)\,,
    \qquad
    P \cdot P =0\,,
\end{equation}
with the projective identification $P^M \sim \alpha P^M$.

In the presence of a flat surface defect,  it is convenient to decompose the embedding coordinates into directions transverse and parallel to the defect,
\begin{equation}
P^M=(P^i,P^a)\,,
\qquad
i=1,2\,,
\qquad
a=3,4,+,-\,.
\end{equation}
The subgroup $SO(2,2)$ preserved by the defect acts linearly on the parallel coordinates $P^a$. Following \cite{Billo:2016cpy}, we distinguish contractions in the various subspaces by using different symbols:
\begin{equation}
P\cdot Q \,,
\qquad
P\circ Q \,,
\qquad
P\bullet Q\,,
\end{equation}
for scalar products in the full embedding space, in the transverse directions, and in the directions parallel to the defect, respectively.

An analogous construction is useful for the $SO(6)_R$ R-symmetry space, where the polarization vectors $Y^M$ satisfy
\begin{equation}
Y\cdot Y =0\,.
\end{equation}
When the R-symmetry is broken to
\begin{equation}
SO(4)\times SO(2)\,,
\end{equation}
we similarly decompose
\begin{equation}
Y^M=(Y^i,Y^a)\,,
\qquad
i=1,2\,,
\qquad
a=3,4,5,6\,.
\end{equation}
We then define
\begin{equation}
Y\circ Y = Y^i Y_i\,,
\qquad
Y\bullet Y = Y^a Y_a\,.
\end{equation}
It is often convenient to introduce polar coordinates in the orthogonal planes with
\begin{equation}\label{ptheta}
P^i
=
(\rho\cos\theta,\rho\sin\theta)\,,
\qquad
Y^i
=
(\rho_R\cos\theta_R,\rho_R\sin\theta_R)\,.
\end{equation}
In these variables, we define the mixed angle
\begin{equation}\label{etatheta}
\eta = \theta + \theta_R\,.
\end{equation}
 
 \subsection*{Spinor conventions}
The spinor $\sigma$-matrices are defined as
\begin{align}
 \sigma^{\mu}_{\alpha\dot \beta}&=(\sigma^a,i\mathbf{1})\, , &  (\bar \sigma^{\mu})^{\dot \alpha \beta}&=(\sigma^a,-i\mathbf{1})\, ,
\end{align}
so that $\bar \sigma^\mu=(\sigma^\mu)^\dagger$. Here $\mu,\nu=1,\ldots,4$ denote spacetime indices, while $\alpha,\beta=1,2$ denote spinor indices.

Spinor indices are raised and lowered using epsilon tensors according to
\begin{align}
\xi^{\alpha}=\epsilon^{\alpha \beta} \xi_{\beta}\, ,  \quad \bar{\xi}_{\dot{\alpha}}=\bar{\xi}^{\dot{\beta}}  \epsilon_{\dot{\beta} \dot{\alpha} }\, ,\quad  \e_{12}=-\e^{12}=-\e_{\dot 1 \dot 2}=\e^{\dot 1 \dot 2}=1\, ,
\end{align}
which implies
\begin{align}
\e_{\a\b}\e^{\b \g}=\d_{\a}^{\g}\, , \qquad
\e_{\dot\a\dot\b}\e^{\dot\b \dot\g}=\d_{\dot\a}^{\dot\g}\, .
\end{align}
For example, in these conventions
\begin{equation}
    (\s^{\mu})^{ \a \dot \a} = \e^{\a \b}\s^{\mu}_{\b\dot \b}\e^{\dot \b\dot \a}=-(\bar \s^{\mu})^{  \dot \a \a}\,.
\end{equation}
Defining
\begin{align}
 \s^{\mu\nu}&=\frac14( \s^\mu \bar\s^\nu- \s^\nu \bar\s^\mu)\, , &  \bar \s^{\mu\nu}=\frac14(\bar \s^\mu  \s^\nu -\bar \s^\nu  \s^\mu )\,,
\end{align}
we obtain
\begin{align}
\s^{a \, b}& =\frac14[\s^a, \s^b]=\frac{i}{2} \e^{a b c}\s^ c =\bar \s^{a \, b}\, , &  \s^{4 \, a}=\frac{i}{2}\s^a=-\bar \s^{4\, a}  \, , \qquad a,b,c=1,2,3\, .
 \end{align}
 A four-dimensional vector $x^{\mu}$ can be represented as a bispinor according to
\begin{align}
 x_{\a\dot \a}&=x^{\mu} \s^{\mu}_{\a\dot \a}=\begin{pmatrix}
                                             x^3+ix^4 &x^1-ix^2 \\
                                             x^1+i x^2 & -x^3+i x^4
                                            \end{pmatrix}\, ,		\\
 x^{\dot \a \a}&=x^{\mu} (\bar\s^{\mu})^{\dot \a \a}=\begin{pmatrix}
                                             x^3-ix^4 &x^1-ix^2 \\
                                             x^1+i x^2 & -x^3-i x^4
                                            \end{pmatrix} \, .                                           
\end{align}
These definitions imply
\begin{align}
 x_{\a \dot \a} x^{\dot \a \b}&=x^2 \d_{\a}^{\b}\, ,  &  x^2\colonequals x^\mu x^\mu=-\det x_{\a \dot \a}\, .
\end{align}

\subsection*{Bulk superconformal algebra}

The symmetry algebra of $\mathcal{N}=4$ Super Yang-Mills theory is $\mathfrak{psu}(2,2|4)$. Its bosonic subalgebra consists of the four-dimensional conformal algebra $\mathfrak{so}(2,4)$ and the $\mathfrak{su}(4)_R$ R-symmetry. In what follows, $P_\mu$ and $K_\mu$ denote the generators of translations and special conformal transformations, respectively. $M_{\mu\nu}$ generates Lorentz transformations, and $D$ generates dilatations. The Poincar\'e and conformal supercharges are denoted by $Q$ and $S$, while $R^i_{\phantom{i}j}$ represent the $\mathfrak{su}(4)_R$ generators.

The commutation relations for the bosonic subalgebra can be written in spinor notation as
\begingroup
\allowdisplaybreaks[1]
\begin{equation}
\begin{alignedat}{4}
&[M_{\alpha}^{~\beta},M_{\gamma}^{\phantom{\gamma}\delta}]	&~=~&	\delta_{\gamma}^{~\beta}M_{\alpha}^{~\delta}-\delta_{\alpha}^{~\delta}M_{\gamma}^{~\beta}~,\\
&[M^{\dot \alpha}_{~\dot \beta},M^{\dot \gamma}_{~\dot \delta }]			&~=~&	\delta^{\dot \alpha}_{~\dot \delta }M^{\dot \gamma}_{~\dot \beta}-\delta^{\dot \gamma}_{~\dot \beta}M^{\dot \alpha}_{~\dot \delta }~,\\
&[M_{\alpha}^{~\beta},P_{\gamma\dot \gamma}]				&~=~&	\delta_{\gamma}^{~\beta}P_{\alpha\dot \gamma}-\tfrac12\delta_{\alpha}^{\phantom{\alpha}\beta}P_{\gamma\dot \gamma}~,\\
&[M^{\dot \alpha}_{~\dot \beta},P_{\gamma\dot \gamma}]				&~=~&	\delta^{\dot \alpha}_{~\dot \gamma}P_{\gamma\dot \beta}-\tfrac12\delta^{\dot \alpha}_{\phantom{\dot \alpha}\dot \beta}P_{\gamma\dot \gamma}~,\\
&[M_{\alpha}^{~\beta},K^{\dot \gamma\gamma}]				&~=~&	-\delta_{\alpha}^{~\gamma}K^{\dot \gamma\beta}+\tfrac12\delta_{\alpha}^{\phantom{\alpha}\beta}K^{\dot \gamma\gamma}~,\\
&[M^{\dot \alpha}_{~\dot \beta},K^{\dot \gamma\gamma}]				&~=~&	-\delta^{\dot \gamma}_{~\dot \beta}K^{\dot \alpha\gamma}+\tfrac12\delta^{\dot \alpha}_{\phantom{\dot \alpha}\dot \beta}K^{\dot \gamma\gamma}~,\\
&[D,P_{\alpha\dot \alpha}]							&~=~&	P_{\alpha\dot \alpha}~,\\
&[D,K^{\dot \alpha\alpha}]							&~=~&	- K^{\dot \alpha\alpha}~,\\
&[K^{\dot \alpha\alpha},P_{\beta\dot \beta}]					&~=~&	4 \delta_{\beta}^{\phantom{\beta}\alpha}\delta^{\dot \alpha}_{\phantom{\dot \alpha}\dot \beta}D+4\delta_{\beta}^{\phantom{\beta}\alpha}M^{\dot \alpha}_{\phantom{\dot \alpha}\dot \beta}+4\delta^{\dot \alpha}_{\phantom{\dot \alpha}\dot \beta}M_{\beta}^{\phantom{\beta}\alpha}~,
\end{alignedat}
\label{eq:confalg}
\end{equation}
\endgroup
where we have defined
\begin{eqnarray}
    P_{\alpha \dot \alpha} &=& (\sigma^\mu)_{\alpha \dot \alpha} P_\mu\,, \qquad M_{\beta}^{\phantom{\beta} \alpha}=-\tfrac14 (\bar{\sigma}^{\mu})^{ \dot \alpha \a} (\sigma_{\nu})_{ \beta \dot \alpha} M_{\mu \nu}\,, \nonumber \\
     \qquad
 K^{ \dot \alpha \alpha} &=& (\bar{\sigma}_\mu)^{\dot \alpha \alpha} K_\mu\,, \qquad \!M^{\dot \alpha}_{\phantom{\dot \alpha} \dot \beta}=- \tfrac14  (\bar{\sigma}^{\mu})^{ \dot \alpha \a} (\sigma_{\nu})_{ \alpha \dot \beta} M_{\mu \nu}\,.
\end{eqnarray}
The $\mathfrak{su}(4)_R$ generators $R^i_{\phantom{i}j}$ are traceless ($R^i_{\phantom{i}i} = 0$) and satisfy the standard $\mathfrak{su}(4)$ commutation relations:
\begin{equation}
[R^i_{\phantom{i}j}, R^k_{\phantom{k}l}] = \delta^k_{\phantom{k}j} R^i_{\phantom{i}l} - \delta^i_{\phantom{i}l} R^k_{\phantom{k}j}~.
\label{eq:Rsymmetry}
\end{equation}
The Poincar\'e and conformal supercharges satisfy the following anticommutation relations:
\begin{equation}
\begin{alignedat}{2}
&\{Q_{\alpha}^i,\,\tilde{Q}_{j\dot {\alpha}}\}  			&~=~~&\tfrac12	\delta^i_{\phantom{i}j} P_{\alpha\dot {\alpha}}~,\\
&\{\tilde{S}^{i\dot {\alpha}},\,S_{j}^{\phantom{\alpha}\alpha}\} &~=~~&	\tfrac12 \delta^i_{\phantom{i}j} K^{\dot {\alpha}\alpha}~,\\
&\{Q_{\alpha}^i,\,S^{\phantom{\alpha}\beta}_j\}     	&~=~~&	\tfrac12 \delta^i_{\phantom{i}j}\delta_{\alpha}^{\phantom{\alpha}\beta}D   + \delta^i_{\phantom{i}j} M_{\alpha}^{\phantom{\alpha}\beta}-\delta_\alpha^{\phantom{\alpha}\beta} \, R^i_{\phantom{i}j} ~,\\
&\{\tilde{S}^{i\dot {\alpha}},\,\tilde{Q}_{j\dot \beta}\}		&~=~~&	\tfrac12 \delta^i_{\phantom{i}j}\delta^{\dot {\alpha}}_{\phantom{\dot {\alpha}}\dot \beta}D + \delta^i_{\phantom{i}j} M^{\dot {\alpha}}_{\phantom{\dot {\alpha}}\dot \beta}+\delta^{\dot {\alpha}}_{\phantom{\dot {\alpha}}\dot \beta} \, R^i_{\phantom{i}j}  ~.
\end{alignedat}
\label{eq:Neq4Qcomm}
\end{equation}
Finally, the commutators of the supercharges with the bosonic generators are
\begingroup
\allowdisplaybreaks[1]
\begin{equation}
\label{eq:Neq4bosQcomm}
\begin{alignedat}{2}
&[M_{\alpha}^{~\beta},Q_{\gamma}^i]	=	\delta_{\gamma}^{~\beta} Q_{\alpha}^i -\tfrac12\delta_{\alpha}^{\phantom{\alpha}\beta} Q_{\gamma}^i~,	\qquad
&[&M^{\dot {\alpha}}_{~\dot \beta},\tilde{Q}_{i \dot \delta}]			= \delta^{\dot {\alpha}}_{~\dot \delta}\tilde{Q}_{i \dot \beta} -\tfrac12\delta^{\dot {\alpha}}_{\phantom{\dot {\alpha}}\dot \beta}\tilde{Q}_{i \dot \delta}~,\\
&[M_{\alpha}^{~\beta},S_{i}^{\phantom{\alpha}\gamma}]				=	-\delta_{\alpha}^{~\gamma}S_{i}^{\phantom{\alpha}\beta}+\tfrac12\delta_{\alpha}^{\phantom{\alpha}\beta} S_{i}^{\phantom{\alpha}\gamma}~,
\qquad &[&M^{\dot {\alpha}}_{~\dot \beta},\tilde{S}^{i\dot \gamma}]				=	-\delta^{\dot \gamma}_{~\dot \beta}\tilde{S}^{i\dot {\alpha}}+\tfrac12\delta^{\dot {\alpha}}_{\phantom{\dot {\alpha}}\dot \beta}\tilde{S}^{i\dot \gamma}~,\\
&[D,Q_{\alpha}^i]							=	\tfrac12 Q_{\alpha}^i~,\qquad
&[&D,\tilde{Q}_{i \dot {\alpha}}]							= 	\tfrac12 \tilde{Q}_{i \dot {\alpha}}~,\\
&[D, S_{i}^{\phantom{\alpha}\alpha}]							=	-\tfrac12  S_{i}^{\phantom{\alpha}\alpha}~,
&[&D, \tilde{S}^{i\dot {\alpha}} ]							=	-\tfrac12 \tilde{S}^{i\dot {\alpha}} ~,\\
&[R^i_{\phantom{I}j},Q_{\alpha}^k]=	\delta_{j}^{~k} Q_{\alpha}^i -\frac{1}{4} \delta_{j}^{i} Q_{\alpha}^k~,
&[&R^i_{\phantom{i}j},\tilde{Q}_{k \dot {\alpha}}]	=	-\delta_{k}^{~i} \tilde{Q}_{j \dot {\alpha}} +\frac{1}{4} \delta_{j}^{i} \tilde{Q}_{k \dot {\alpha}}~,\\
&[K^{\dot {\alpha}\alpha},Q_{\beta}^i]					=	2 \delta_{\beta}^{\phantom{\beta}\alpha}\tilde{S}^{i\dot {\alpha}}~,
&[&K^{\dot {\alpha}\alpha},\tilde{Q}_{i \dot \beta}]					=	2 \delta_{\dot \beta}^{\phantom{\dot \beta}\dot {\alpha}} S_{i}^{\phantom{\alpha}\alpha}~,\\
&[P_{\alpha\dot {\alpha}},S_{i}^{\phantom{\alpha}\beta}]					=	- 2 \delta_{\alpha}^{\phantom{\alpha}\beta}\tilde{Q}_{i \dot {\alpha}}~,
&[&P_{\alpha\dot {\alpha}},\tilde{S}^{i\dot \beta} ]					=	-2 \delta_{\dot {\alpha}}^{\phantom{\dot {\alpha}}\dot \beta} Q_{\alpha}^i~\, .
\end{alignedat}
\end{equation}
\endgroup
The commutators of $R^i_{\phantom{i}j}$ with $S$ and $\tilde{S}$ are omitted, as they follow from those of $Q$ and $\tilde{Q}$ with the same index structure. We have also omitted vanishing commutators for brevity.

\subsection*{Defect superconformal algebra}
\label{sec:Defect superconformal algebra}
The $1/2$ BPS surface defect considered in this work breaks the bulk superconformal symmetry algebra as
\begin{equation}
\mathfrak{psu}(2,2|4)  \rightarrow \mathfrak{psu}(1,1|2) \oplus \mathfrak{psu}(1,1|2) \oplus \mathfrak{u}(1)_- \, .
\end{equation}
The preserved bosonic symmetries include a two-dimensional conformal algebra $\mathfrak{so}(2,2)$ acting along the defect, an unbroken $\mathfrak{so}(4)_{\mathcal{R}}$ subalgebra of the R-symmetry, and an additional $\mathfrak{u}(1)_-$, which corresponds to a diagonal combination of orthogonal rotations and R-symmetry transformations. These are supplemented by half of the supercharges of the bulk theory.

We choose the following generators for the preserved conformal subalgebra:
\begin{equation}
\begin{alignedat}{2}
    \mathcal{P} &= \frac{1}{2}P_{1, \dot 1}\, , \quad &\bar{\mathcal{P}} &= \frac{1}{2}P_{2, \dot 2} \, , \\ 
    \mathcal{K} & = \frac{1}{2}K^{\dot 1 1}\, , \quad & \bar{\mathcal{K}} & = \frac{1}{2}K^{\dot 2 2}\, , \\ 
 \mathcal{D} &= \frac{1}{2}\left( D + M_1^{\phantom{1} 1} + M_{\phantom{\dot 1}\dot 1}^{ \dot 1}\right)\, , \quad &\bar{\mathcal{D}} &= \frac{1}{2}\left( D - M_1^{\phantom{1} 1} - M_{\phantom{\dot 1} \dot 1}^{\dot 1}\right)\, .
\end{alignedat}
\end{equation}
 Each column generates an $\mathfrak{so}(1,2)$ subalgebra. The non-vanishing commutation relations are
\begin{equation}
\begin{alignedat}{2}
    [\mathcal{K} ,\mathcal{P}] &= 2 \mathcal{D}  \, ,\quad &[\bar{\mathcal{K}} ,\bar{\mathcal{P}}] &= 2 \bar{\mathcal{D}}\, , \\ 
     [\mathcal{D} ,\mathcal{P}] &= \mathcal{P}\, , \quad &[\bar{\mathcal{D}} ,\bar{\mathcal{P}}] &= \bar{\mathcal{P}}\, , \\
       [\mathcal{D} ,\mathcal{K}] &= -\mathcal{K}\, , \quad  &[\bar{\mathcal{D}} ,\bar{\mathcal{K}}] &= -\bar{\mathcal{K}}\, .
\end{alignedat}
\end{equation}
The fermionic and $R$-symmetry generators of the preserved subalgebra are
\begin{align}\label{preservedgens}
\mathcal{R}^a_{\phantom{a}b} &= R^a_{\phantom{a}b} - \frac12 \delta^a_{\phantom{a}b} (R^1_{\phantom{1}1} + R^2_{\phantom{2}2})\, , \quad &\bar{\mathcal{R}}^a_{\phantom{a}b} &= R^{a+2}_{\phantom{a+2}b+2} - \frac12 \delta^a_{\phantom{a}b} (R^3_{\phantom{3}3} + R^4_{\phantom{4}4})\, ,\nonumber  \\
\mathcal{Q}_+^a &= Q^a_{1}\, , \quad &\bar{\mathcal{Q}}_+^a &= Q^{a+2}_{2}\, ,\nn  \\
\mathcal{Q}^{-}_a &= \tilde{Q}_{a,\dot 1}\, , \quad &\bar{\mathcal{Q}}^{-}_a &= \tilde{Q}_{a+2, \dot 2}\, , \\
\mathcal{S}_+^a &= \tilde{S}^{a, \dot 1}\, , \quad &\bar{\mathcal{S}}_+^a &= \tilde{S}^{a+2, \dot 2}\, ,\nonumber \\
\mathcal{S}^{-}_a &= S_a^{\phantom{a} 1}\, , \quad &\bar{\mathcal{S}}^{-}_a &= S_{a+2}^{\phantom{a+2}2}\, .\nonumber 
\end{align}
Together with the $\mathfrak{so}(1,2)$ subalgebra, each column corresponds to a copy of $\mathfrak{psu}(1,1|2)$, with $R$-symmetry indices $a,b = 1,2$.

In addition, there is a central generator,
\begin{align}\label{rminus}
    \mathcal{R}_- &=  \mathcal{R}_\perp - \mathcal{M}_\perp\, , 
\end{align}
where
\begin{equation}
   \mathcal{R}_\perp \colonequals R^1_{\phantom{1}1} + R^2_{\phantom{2}2}\, , \qquad \mathcal{M}_\perp \colonequals M_1^{\phantom{1} 1} - M_{\phantom{\dot 1}\dot 1}^{ \dot 1} \, .
\end{equation}
The non-vanishing commutation relations among the fermionic generators are
\begin{align}\label{QS}
\{\mathcal{Q}_{+}^a,\,\mathcal{Q}^{-}_b\}  			&~=~~\delta^a_{\phantom{a}b} \mathcal{P}~\, , \quad &\{\bar{\mathcal{Q}}_{+}^a,\,\bar{\mathcal{Q}}^{-}_b\}  			&~=~~\delta^a_{\phantom{a}b} \bar{\mathcal{P}}~\, , \nn \\
\{\mathcal{S}_{+}^a,\,\mathcal{S}^{-}_b\}  			&~=~~\delta^a_{\phantom{a}b} \mathcal{K}~\, , \quad &\{\bar{\mathcal{S}}_{+}^a,\,\bar{\mathcal{S}}^{-}_b\}  			&~=~~\delta^a_{\phantom{a}b} \bar{\mathcal{K}}~\, , \\
\{\mathcal{Q}_{+}^a,\,\mathcal{S}^{-}_b\}  			&~=~~\delta^a_{\phantom{a}b}\, \mathcal{D} - \mathcal{R}^a_{\phantom{a}b}-\frac12 \mathcal{R}_-\delta^a_{\phantom{a}b} ~\, , \quad &\{\bar{\mathcal{Q}}_{+}^a,\,\bar{\mathcal{S}}^{-}_b\}  			&~=~~\delta^a_{\phantom{a}b} \, \bar{\mathcal{D}} - \bar{\mathcal{R}}^a_{\phantom{a}b} +\frac12 \mathcal{R}_-\delta^a_{\phantom{a}b}~\, ,\nn \\
\{\mathcal{Q}^{-}_a,\,\mathcal{S}_{+}^b\}  			&~=~~\delta_a^{\phantom{a}b} \, \mathcal{D} + \mathcal{R}^{\phantom{a}b}_a +\frac12\mathcal{R}_-\delta_a^{\phantom{a}b} ~\, , \quad &\{\bar{\mathcal{Q}}^{-}_a,\,\bar{\mathcal{S}}_{+}^b\} &~=~~\delta_a^{\phantom{a}b} \, \bar{\mathcal{D}} + \bar{\mathcal{R}}^{\phantom{a}b}_a  -\frac12 \mathcal{R}_-\delta_a^{\phantom{a}b} ~\, ,\nn
\end{align}
while the $R$-symmetry generators satisfy
\begin{align}
    [\mathcal{R}^a_{\phantom{a}b},\mathcal{R}^c_{\phantom{c}d}] = \mathcal{R}^a_{\phantom{a}d } \delta^c_{\phantom{c}b} - \mathcal{R}^c_{\phantom{c}b} \delta^a_{\phantom{a}d}\, , \qquad
      [\bar{\mathcal{R}}^a_{\phantom{a}b},\bar{\mathcal{R}}^c_{\phantom{c}d}] = \bar{\mathcal{R}}^a_{\phantom{a}d } \delta^c_{\phantom{c}b} - \bar{\mathcal{R}}^c_{\phantom{c}b} \delta^a_{\phantom{a}d}\, .
\end{align}
The commutators of the supercharges with the bosonic symmetry generators are
\begin{align}
[\mathcal{P}, \mathcal{S}_{+}^a] &= - \mathcal{Q}_{+}^a\, ,& [\bar{\mathcal{P}}, \bar{\mathcal{S}}_{+}^a] &= - \bar{\mathcal{Q}}_{+}^a\, , \nn \\
[\mathcal{P}, \mathcal{S}^{-}_a] &= - \mathcal{Q}^{-}_a\, ,  & [\bar{\mathcal{P}}, \bar{\mathcal{S}}^{-}_a] &= - \bar{\mathcal{Q}}^{-}_a\, , \nn \\
[\mathcal{K}, \mathcal{Q}_{+}^a] &=  \mathcal{S}_{+}^a\, ,  & [\bar{\mathcal{K}}, \bar{\mathcal{Q}}_{+}^a] &=  \bar{\mathcal{S}}_{+}^a\, , \nn \\
[\mathcal{K}, \mathcal{Q}^{-}_a] &=  \mathcal{S}^{-}_a\, ,  & [\bar{\mathcal{K}}, \bar{\mathcal{Q}}^{-}_a] &= \bar{\mathcal{S}}^{-}_a\, ,\nn  \\
[\mathcal{D}, \mathcal{S}_{+}^a] &= - \frac12 \mathcal{S}_{+}^a\, ,  & [\bar{\mathcal{D}}, \bar{\mathcal{S}}_{+}^a] &= - \frac12 \bar{\mathcal{S}}_{+}^a\, , \nn \\
[\mathcal{D}, \mathcal{S}^{-}_a] &= - \frac12 \mathcal{S}^{-}_a\, ,  & [\bar{\mathcal{D}}, \bar{\mathcal{S}}^{-}_a] &= - \frac12 \bar{\mathcal{S}}^{-}_a\, ,\nn  \\
[\mathcal{D}, \mathcal{Q}_{+}^a] &=  \frac12 \mathcal{Q}_{+}^a\, ,  & [\bar{\mathcal{D}}, \bar{\mathcal{Q}}_{+}^a] &= \frac12 \bar{\mathcal{Q}}_{+}^a\, , \\
[\mathcal{D}, \mathcal{Q}^{-}_a] &=  \frac12 \mathcal{Q}^{-}_a\, , & [\bar{\mathcal{D}}, \bar{\mathcal{Q}}^{-}_a] &= \frac12 \bar{\mathcal{Q}}^{-}_a\, , \nn \\
[\mathcal{R}^a_{\phantom{a}b}, \mathcal{Q}_{+}^c] &=  \mathcal{Q}_{+}^a \delta^c_{\phantom{c}b} - \frac12  \mathcal{Q}_{+}^c \delta^a_{\phantom{a}b}\, ,  &[\bar{\mathcal{R}}^a_{\phantom{a}b}, \bar{\mathcal{Q}}_{+}^c] &=  \bar{\mathcal{Q}}_{+}^a \delta^c_{\phantom{c}b} - \frac12  \bar{\mathcal{Q}}_{+}^c \delta^a_{\phantom{a}b}\, , \nn \\ 
[\mathcal{R}^a_{\phantom{a}b}, \mathcal{Q}^{-}_c] &=  - \mathcal{Q}^{-}_b \delta^a_{\phantom{a}c } + \frac12  \mathcal{Q}^{-}_c \delta^a_{\phantom{a}b}\, ,  &[\bar{\mathcal{R}}^a_{\phantom{a}b}, \bar{\mathcal{Q}}^{-}_c] &=  - \bar{\mathcal{Q}}^{-}_b \delta^a_{\phantom{a}c } + \frac12  \bar{\mathcal{Q}}^{-}_c \delta^a_{\phantom{a}b}\, , \nn \\ 
[\mathcal{R}^a_{\phantom{a}b}, \mathcal{S}_{+}^c] &=  \mathcal{S}_{+}^a \delta^c_{\phantom{c}b} - \frac12  \mathcal{S}_{+}^c \delta^a_{\phantom{a}b}\, ,  &[\bar{\mathcal{R}}^a_{\phantom{a}b}, \bar{\mathcal{S}}_{+}^c] &=  \bar{\mathcal{S}}_{+}^a \delta^c_{\phantom{c}b} - \frac12  \bar{\mathcal{S}}_{+}^c \delta^a_{\phantom{a}b}\, , \nn \\
[\mathcal{R}^a_{\phantom{a}b}, \mathcal{S}^{-}_c] &=  - \mathcal{S}^{-}_b \delta^a_{\phantom{a}c } + \frac12  \mathcal{S}^{-}_c \delta^a_{\phantom{a}b}\, ,  &[\bar{\mathcal{R}}^a_{\phantom{a}b}, \bar{\mathcal{S}}^{-}_c] &=  - \bar{\mathcal{S}}^{-}_b \delta^a_{\phantom{a}c } + \frac12  \bar{\mathcal{S}}^{-}_c \delta^a_{\phantom{a}b}\, . \nn
\end{align}
The set is completed by a $\mathfrak{u}(1)$ generator,
\begin{equation}
    \mathcal{R}_+ = \mathcal{R}_\perp + \mathcal{M}_\perp\, ,
\end{equation}
which acts as an outer automorphism on the fermionic generators of the two $\mathfrak{psu}(1,1|2)$ factors:
\begingroup
\allowdisplaybreaks[1]
\begin{equation} \label{automorph}
\begin{alignedat}{2}
[  \mathcal{R}_+, \mathcal{S}_{+}^a] &=  \mathcal{S}_{+}^a\, , \quad & [  \mathcal{R}_+, \bar{\mathcal{S}}_{+}^a] &= - \bar{\mathcal{S}}_{+}^a\, , \\
[  \mathcal{R}_+, \mathcal{S}^{-}_a] &= -  \mathcal{S}^{-}_a\, , \quad & [  \mathcal{R}_+, \bar{\mathcal{S}}^{-}_a] &= \bar{\mathcal{S}}^{-}_a\, , \\
[  \mathcal{R}_+, \mathcal{Q}_{+}^a] &=   \mathcal{Q}_{+}^a\, , \quad & [  \mathcal{R}_+, \bar{\mathcal{Q}}_{+}^a] &= - \bar{\mathcal{Q}}_{+}^a\, , \\
[  \mathcal{R}_+, \mathcal{Q}^{-}_a] &= -  \mathcal{Q}^{-}_a\, , \quad & [  \mathcal{R}_+, \bar{\mathcal{Q}}^{-}_a] &=  \bar{\mathcal{Q}}^{-}_a\, .
\end{alignedat}
\end{equation}
\endgroup

\subsection*{The stress-tensor superprimary in different $\mathbf{SU(4)_R}$ notations}

In this subsection we review the explicit construction of the stress-tensor multiplet superconformal primary with manifest R-symmetry indices, detailing its relation to the index-free operator $\mathcal{O}_2$ utilized in the main text.

Due to the isomorphism $SU(4)_R \cong SO(6)_R$, operators are equivalently represented using either $SU(4)$ or $SO(6)$ indices. Specifically, $SO(6)_R$ vector indices (denoted by uppercase Latin letters) naturally map to antisymmetric pairs of $SU(4)_R$ fundamental indices (denoted by lowercase Latin letters). This correspondence is explicitly realized via the six-dimensional $\Sigma$-matrices:
\begin{equation}
 B_{kl}=-\frac12 \S_{kl}^M B^M\, , \qquad B^M=\frac12  (\bar \S^M)^{kl} B_{kl}\, .
\end{equation}
The $\Sigma$-matrices satisfy the properties
\begin{align}
 \S^M_{ij}&=-\S^M_{ji}\, , & (\bar \S^M)^{ij} \S^N_{jk}+(\bar\S^N)^{ij} \S^M_{jk}&=2 \d^{MN}\d^i_k\, , & (\bar\S^M)^{ij}=-\frac12 \e^{ijkl}\S^M_{kl}\, ,
\end{align}
and we define
\begin{align}
{(\S^{MN})^i}_j=\frac12\left[(\bar \S^M)^{ik} \S^N_{kj}-(\bar \S^N)^{ik} \S^M_{kj}\right]\, .
\end{align}
The superconformal primary of the stress-tensor multiplet is constructed from the six scalars $\Phi^M$ of $\mathcal{N}=4$ SYM as
\begin{equation}\label{tensorprimary}
\mathcal{O}^{MN}=\Tr\!\left(\Phi^M \Phi^N-\frac16 \delta^{MN} \Phi^L \Phi^L\right)\, ,
\end{equation}
where the trace is taken over the gauge group $U(N)$. The scalars $\Phi^M$ transform in the adjoint representation of $U(N)$ and the vector of $SO(6)_R$ (or $[0,1,0]$ Dynkin label). Translating their R-symmetry dependence to the $SU(4)_R$ fundamental notation via the $\Sigma$-matrices yields
\begin{equation}
 \Phi^M=\frac12 \Sigma^M_{ij} \Phi^{ij}=\frac12 (\bar{\Sigma}^M)^{ij} \Phi_{ij}\, ,
\end{equation}
with the reality condition
\begin{equation}
 \Phi^{ij}=-\frac12 \epsilon^{ijkl} \Phi_{kl}\, .
\end{equation}
The relations above do not fix the entries of the $\Sigma$-matrices. Throughout the paper we use the explicit realization
\begin{equation}\label{Sigmaexplicit}
\begin{aligned}
2\,\Phi^{34}&=\Phi^1+i\Phi^2\,, &\qquad 2\,\Phi^{12}&=\Phi^1-i\Phi^2\,,\\2\,\Phi^{14}&=-\Phi^3-i\Phi^4\,, &\qquad 2\,\Phi^{23}&=-\Phi^3+i\Phi^4\,,\\2\,\Phi^{13}&=\Phi^5+i\Phi^6\,, &\qquad 2\,\Phi^{24}&=-\Phi^5+i\Phi^6\,,
\end{aligned}
\end{equation}
that is $2\,\Phi^{ij}=C^{M,ij}\Phi^M$, with antisymmetric matrices $C^M$, together with $\Sigma^M=(C^M)^*$, $\bar\Sigma^M=-C^M$ and $\epsilon^{1234}=1$. This choice satisfies all the properties listed above, and the lower tensor is minus the Hermitian conjugate of the upper one, $\Phi_{ij}=-(\Phi^{ij})^\dagger$.

 The operator $\mathcal{O}^{MN}$ can also be expressed in terms of $SU(4)_R$ indices using the conversion rules introduced above,
\begin{equation}
 {\mathcal{O}^{ij}}_{kl}=\frac14  (\bar\Sigma^M)^{ij} \Sigma_{kl}^N \, \mathcal{O}^{MN}\, .
\end{equation}
$\mathcal{O}^{MN}$ is related to the index-free operator $\mathcal{O}_2$ defined in the main text \eqref{CPO} by contraction with the $SO(6)_R$ polarization vectors $Y^M$,
\begin{equation}
    \mathcal{O}_2 = \frac{8\pi^2}{\lambda\sqrt{2}} Y^M Y^N \mathcal{O}^{MN}\, .
\end{equation}
Here we reinstated the normalization used in \eqref{CPO}.
Given the index-free operator $\mathcal{O}_2$, we can recover $\mathcal{O}^{MN}$ by acting twice with the Todorov operator $D_M$,
\begin{equation}
  \frac{8\pi^2}{\lambda\sqrt{2}}  \mathcal{O}^{MN} = \frac{1}{12} D_M D_N \, \mathcal{O}_2\, ,
\end{equation}
where
\begin{align}
    D_M = \left(2 + Y \cdot \frac{\partial}{\partial Y} \right) \frac{\partial}{\partial Y^M} - \frac{Y^M}{2} \frac{\partial^2}{\partial Y \cdot \partial Y}\, .
\end{align}
This relation provides a straightforward way to extract the manifest-index correlators, such as $\langle \mathcal{O}^{MN}\rangle_\Sigma$ or $\langle \mathcal{O}^{MN} \mathcal{O}^{PQ}\rangle_\Sigma$, directly from the index-free expressions derived in the main text.

\section{Superconformal Ward identities for the one-point function}
\label{sec:WIs}
 We study superconformal Ward identities for one-point functions of chiral primary operators in $\mathcal{N}=4$ SYM in the presence of a surface defect that breaks the symmetries as in \eqref{symbreak}. Our aim is to show that these identities fix the one-point function to take the form \eqref{1ptbroken}. The operators in this appendix are not canonically normalized. However, all normalization factors can be absorbed into the one-point function coefficients, so the final results do not depend on them. The defect we are considering is bosonic, therefore only bosonic operators can acquire a non-trivial one-point function. We refer the reader to appendix \ref{app:conv} for the conventions used throughout this section.

To derive the Ward identities, we use the property that
\begin{equation}
 \langle\d \mathcal{O} \rangle_{\Sigma}=0\, ,
\end{equation}
for any insertion $\mathcal{O}$ if the variation is taken with respect to a preserved supersymmetry, i.e. $\delta \mathcal{O}_\Sigma = 0$.
We define the supersymmetry transformations as\footnote{The supersymmetry parameters, carrying indices, should not be confused with the cross-ratio \eqref{crossratio}.}
\begin{equation}
 \delta \mathcal{O} = \left[\delta, \mathcal{O} \right]= \left[\xi_{\alpha i}\, Q^{\alpha i} + \bar{\xi}^{i\dot \alpha}\, \tilde {Q}_{i\dot \alpha} , \mathcal{O} \right]\,,
 \qquad Q^{\alpha i}=\epsilon^{\alpha\beta}Q^i_\beta\,.
 \label{eq:delta}
\end{equation}
The parameters $\xi_{\alpha i}$ and $\bar\xi^{i\dot\alpha}$ transform in the representations of $SU(4)_R$ dual to those of $Q^i_\alpha$ and $\tilde Q_{i\dot\alpha}$ in \eqref{eq:Neq4Qcomm} and \eqref{eq:Neq4bosQcomm}. Here $\mathcal{O}_{ij}{}^{kl}\equiv{\mathcal{O}^{kl}}_{ij}$ denotes the superprimary of appendix \ref{app:conv}. In the following, symmetrization and antisymmetrization of indices are unweighted, e.g. $A_{[i}B_{j]}=A_iB_j-A_jB_i$, repeated Lorentz indices are summed, and we use the projectors
 \begin{equation}\label{traceproj}
 \begin{aligned}
 (\Pi X)^{ij}{}_{k}& =X^{ij}{}_{k}-\frac13\left(\delta^i_k X^{mj}{}_{m}-\delta^j_k X^{mi}{}_{m}\right)\,,\\
 (\bar\Pi X)^{k}{}_{ij}& =X^{k}{}_{ij}-\frac13\left(\delta^k_i X^{m}{}_{mj}-\delta^k_j X^{m}{}_{mi}\right)\,.
 \end{aligned}
 \end{equation}
 The transformations of the stress tensor multiplet in the non-interacting abelian case read\footnote{These standard transformations can be found, up to conventions, in e.g. \cite{Bianchi:1998nk}. Consistency of the supersymmetry transformations alone leaves some coefficients undetermined, and different choices of normalization are possible.}
\begin{align}
\label{susy}
 \delta \mathcal{O}_{ij}{}^{kl} = &  \, \bar{\xi}^{\dot{\alpha}[k}  \bar{\chi}_{\dot{\alpha}}{}^{l]}{}_{ij} - \frac{1}{2}\delta^{[k}_{[i}\, \bar{\xi}^{m\dot{\alpha}}  \bar{\chi}_{\dot{\alpha}}{}^{l]}{}_{mj]} + \chi^{kl\alpha}{}_{[i}\,  \xi_{\alpha j]} +  \frac{1}{2}\delta^{[k}_{[i}\, \chi^{ml]\alpha}{}_{j]}\,  \xi_{\alpha m}\, ,  \nn \\
 \delta \chi^{ij\alpha}{}_{k} = &  \, \Pi\Big[\epsilon^{i j m n}\epsilon^{\alpha \gamma}(\sigma^{\mu \nu})_{\gamma}^{\ \beta} B_{\mu \nu\, km}\,  \xi_{\beta n} + \epsilon^{i j m n} \epsilon^{\alpha \beta} \mathcal{E}_{m k}\,  \xi_{\beta n} \nn \\ & \quad - \epsilon^{\alpha \gamma}(\sigma^{\mu})_{\gamma \dot{\beta}} R_{\mu k}{}^{[i}\, \bar{\xi}^{j]\dot{\beta}} -\frac12 \epsilon^{\alpha \gamma} (\sigma^{\mu})_{\gamma \dot{\beta}} \partial_{\mu} \mathcal{O}_{kl}{}^{ij}\,  \bar{\xi}^{l\dot{\beta}}\Big] \, , \nn \\
 \delta \bar{\chi}_{\dot{\alpha}}{}^{k}{}_{ij} = &  \, \bar\Pi\Big[ - \epsilon_{i j m n} \bar{\xi}^{n \dot{\beta}}  \bar{B}_{\mu \nu}^{k m}  (\bar{\sigma}^{\mu \nu})^{\dot{\gamma}}_{ \ \dot{\beta}} \epsilon_{\dot{\gamma} \dot{\alpha}} + \epsilon_{i j m n}\bar{\xi}^{n \dot{\gamma}}   \bar{\mathcal{E}}^{m k} \epsilon_{\dot{\gamma} \dot{\alpha}} \nn \\&\quad + \xi_{\beta [i}  R_{\mu j]}{}^{k} (\bar{\sigma}^{\mu})^{\dot{\gamma} \beta}\epsilon_{\dot{\gamma} \dot{\alpha}} + \frac12  \xi_{\beta l}\,  \partial_{\mu} \mathcal{O}_{ij}{}^{kl} (\bar{\sigma}^{\mu})^{\dot{\gamma} \beta}\epsilon_{\dot{\gamma} \dot{\alpha}}\Big] \, , \nn \\
 \delta B_{\mu  \nu\, i j} = &  \, \frac{1}{2} \epsilon_{i j k l}\, \bar{\xi}^{k \dot{\gamma}}  \epsilon_{\dot{\gamma} \dot{\alpha}}(\bar{\sigma}^{\rho})^{\dot{\alpha} \beta} (\sigma_{\mu \nu})_{\beta}^{\ \gamma} J_{\rho}{}^{l}{}_{\gamma}  - \epsilon^{\alpha \gamma} \xi_{\gamma [i}
  (\sigma_{\mu \nu})_{\alpha}^{\ \beta}\Lambda_{\beta j]} \nn \\& + \frac{1}{8} \epsilon_{i j k l}\, \bar{\xi}^{n \dot{\gamma}} \epsilon_{\dot{\gamma} \dot{\alpha}}   (\bar{\sigma}^{\rho})^{\dot{\alpha} \beta}(\sigma_{\mu \nu})_{\beta}^{\ \gamma} \epsilon_{\gamma \lambda} \partial_{\rho} \chi^{kl\lambda}{}_{n} \, , \nn    \\
  \delta \bar{B}_{\mu  \nu}^{i j} = &  \, \frac{1}{2} \epsilon^{i j k l}\, \bar{J}_{\rho l}{}^{\dot{\gamma}}  ( \bar{\sigma}_{\mu \nu})^{\dot{\beta}}_{\ \dot{\gamma}} (\sigma^{\rho})_{\alpha \dot{\beta}}  \epsilon^{\alpha \gamma}    \xi_{\gamma k}  -  \bar{\Lambda}^{\dot{\beta} [j } (\bar{\sigma}_{\mu \nu})^{\dot{\alpha}}_{\ \dot{\beta}}\,  \bar{\xi}^{i]\dot{\gamma}} \epsilon_{\dot{\gamma} \dot{\alpha}} \nn \\ &  -\frac{1}{8} \epsilon^{i j k l}\, \partial_{\rho} \bar{\chi}_{\dot{\gamma}}{}^{n}{}_{kl}\,\epsilon^{\dot{\gamma} \dot{\alpha}} (\bar{\sigma}_{\mu \nu})^{\dot{\beta}}_{\ \dot{\alpha}} (\sigma^{\rho})_{ \gamma \dot{\beta}} \epsilon^{ \gamma \delta}     \xi_{\delta n}\, ,\nn \\
  \delta \mathcal{E}_{i j} = &  \, \epsilon^{\alpha \lambda} \xi_{\lambda (i}  \Lambda_{\alpha j)} - \frac{1}{8} \epsilon_{m n k ( i}\, \bar{\xi}^{k\dot{\gamma}}     \epsilon_{\dot{\gamma} \dot{\alpha}} (\bar{\sigma}^{\mu})^{\dot{\alpha} \beta} \epsilon_{\beta \gamma} \partial_{\mu} \chi^{mn\gamma}{}_{j)} \, , \nn \\
  \delta  \bar{\mathcal{E}}^{i j} = &  - \bar{\xi}^{\dot{\alpha}( i }   \bar{\Lambda}^{j)\dot{\beta}}  \epsilon_{\dot{\beta} \dot{\alpha} } - \frac{1}{8} \epsilon^{m n k ( i}\, \partial_{\mu} \bar{\chi}_{\dot{\beta}}{}^{j)}{}_{mn}\, \epsilon^{\alpha \gamma} (\sigma^{\mu})_{\gamma \dot{\lambda}} \epsilon^{\dot{\lambda} \dot{\beta}}  \xi_{\alpha k}\, , \\
 \delta R_{\mu i}{}^{j} = &  \, \epsilon^{\alpha \beta} \xi_{\beta i}  J_{\mu}{}^{j}{}_{\alpha} - \frac{1}{4} \epsilon^{\alpha \beta}\delta^{j}_{i}\, \xi_{\beta k}  J_{\mu}{}^{k}{}_{\alpha}  -\frac12 \epsilon^{\alpha \gamma} \xi_{\gamma k}\,   (\sigma_{\mu}^{\ \nu})_{\alpha}^{\ \beta} \epsilon_{\beta \lambda} \partial_{\nu} \chi^{jk\lambda}{}_{i} \nn \\& +  \bar{J}_{\mu i}{}^{\dot{\beta}}\epsilon_{\dot{\beta} \dot{\alpha}}  \bar{\xi}^{j \dot{\alpha}}   -  \frac{1}{4}\delta^{j}_{i}\, \bar{J}_{\mu k}{}^{\dot{\beta}}\epsilon_{\dot{\beta} \dot{\alpha}}  \bar{\xi}^{k \dot{\alpha}} +\frac12 \bar{\xi}^{k \dot{\alpha}}\,   (\bar{\sigma}_{\mu}^{\ \nu})^{\dot{\gamma}}_{\ \dot{\lambda}} \epsilon_{\dot{\gamma} \dot{\alpha}} \epsilon^{\dot{\lambda} \dot{\beta}} \partial_{\nu} \bar{\chi}_{\dot{\beta}}{}^{j}{}_{ik}   \, , \nn \\
 \delta J_{\mu}{}^{i}{}_{\alpha} = &  \, (\sigma^{\nu})_{\alpha \dot{\beta}} T_{\mu \nu}\,  \bar{\xi}^{i\dot{\beta}}  -\frac12  \left( (\sigma^{\rho})_{\alpha \dot{\beta}} (\bar{\sigma}_{\mu}^{\ \nu})^{\dot{\beta}}_{\ \dot{\gamma}} - \frac{1}{3} (\sigma_{\mu}^{\ \nu})_{\alpha}^{\ \beta} (\sigma^{\rho})_{\beta \dot{\gamma}} \right) \partial_{\nu} R_{\rho k}{}^{i}\,  \bar{\xi}^{k\dot{\gamma}} \nn \\& +\frac14 \left((\sigma^{\rho \sigma})_{\alpha}^{\ \beta} ( \sigma_{\mu \nu})_{\beta}^{\ \gamma} + \frac{1}{3}(\sigma_{\mu \nu})_{\alpha}^{\ \beta} ( \sigma^{\rho \sigma})_{\beta}^{\ \gamma}\right) \epsilon^{i k l m} \partial^{\nu} B_{\rho \sigma\, kl}\,  \xi_{\gamma m} \, , \nn \\
 \delta \bar{J}_{\mu i}{}^{\dot{\alpha}} = &   \, \xi_{\beta i}\,   T_{\mu \nu} (\bar{\sigma}^{\nu})^{\dot{\alpha} \beta} +\frac12  \xi_{\gamma k}\,  \partial^{\nu} R_{\rho i}{}^{k} \left( (\sigma_{\mu \nu})_{\beta}^{\ \gamma}(\bar{\sigma}^{\rho})^{\dot{\alpha} \beta}  -  \frac{1}{3}(\bar{\sigma}^{\rho})^{\dot{\beta} \gamma} (\bar{\sigma}_{\mu \nu})^{\dot{\alpha}}_{\ \dot{\beta}}  \right)    \nn \\& +\frac14 \bar{\xi}^{m \dot{\gamma}} \epsilon_{i k l m} \partial^{\nu} \bar{B}_{\rho \sigma}^{k l} \left( (\bar{\sigma}_{\mu \nu})^{\dot{\beta}}_{\ \dot{\gamma}}  (\bar{\sigma}^{\rho \sigma})^{\dot{\alpha}}_{\ \dot{\beta}} + \frac{1}{3} (\bar{\sigma}^{\rho \sigma})^{\dot{\beta}}_{\ \dot{\gamma}}  (\bar{\sigma}_{\mu \nu})^{\dot{\alpha}}_{\  \dot{\beta}} \right)\, , \nn \\
 \delta \Lambda_{\alpha i} = &  \, \xi_{\alpha i}\,  \mathcal{L} -\frac14 (\sigma^{\mu})_{\alpha \dot{\beta}} \partial_{\mu} \mathcal{E}_{i k}\,  \bar{\xi}^{k\dot{\beta}} +\frac14 (\sigma^{\mu \nu})_{\alpha}^{\ \beta}  (\sigma^{\rho})_{\beta \dot{\gamma}} \partial_{\rho} B_{\mu \nu\, i j}\,  \bar{\xi}^{j\dot{\gamma}}\, , \nn    \\
 \delta \bar{\Lambda}^{i\dot{\alpha}} = &  \, \bar{\mathcal{L}}\,  \bar{\xi}^{i\dot{\alpha}} +\frac14 \xi_{\beta k}\,  \partial_{\mu} \bar{\mathcal{E}}^{i k} (\bar{\sigma}^{\mu})^{\dot{\alpha} \beta} +\frac14 \xi_{\gamma j}\,   \partial_{\rho} \bar{B}_{\mu \nu}^{i j} (\bar{\sigma}^{\rho})^{\dot{\beta} \gamma}(\bar{\sigma}^{\mu \nu})^{\dot{\alpha}}_{\ \dot{\beta}}\, , \nn \\
 \delta T_{\mu \nu} = &  \, \frac14 \epsilon^{\alpha \gamma} \xi_{\gamma i}\,  \partial_{\rho} J_{(\nu}{}^{i}{}_{\beta}\,  (\sigma_{\mu) }^{\ \ \rho})_{\alpha}^{\ \beta}  +\frac14  (\bar{\sigma}_{(\mu}^{\ \ \rho})^{\dot{\alpha}}_{\ \dot{\beta}}\, \partial_{\rho} \bar{J}_{\nu) i}{}^{\dot{\beta}}\,
 \bar{\xi}^{i \dot{\gamma}}\epsilon_{\dot{\gamma} \dot{\alpha}}\, , \nn \\
 \delta \mathcal{L} = &  \,  \frac12 \bar{\xi}^{i\dot{\gamma}}   \epsilon_{\dot{\gamma} \dot{\alpha}} (\bar{\sigma}^{\mu})^{\dot{\alpha}\beta} \partial_{\mu} \Lambda_{\beta i}  +\epsilon^{\alpha \lambda} \xi_{\lambda i}  \left(  \partial^{\mu} J_{\mu}{}^{i}{}_{\alpha} - 2 (\sigma^{\mu})_{\alpha  \dot{\beta}} (\bar{\sigma}^{\nu})^{\dot{\beta} \gamma} \partial_{\mu} J_{\nu}{}^{i}{}_{\gamma} \right)\, , \nn    \\
 \delta \bar{\mathcal{L}} =  &  \, \frac12  \partial_{\mu} \bar{\Lambda}^{i \dot{\beta}} (\sigma^{\mu})_{\alpha \dot{\beta}} \epsilon^{\alpha \gamma}   \xi_{\gamma i} -  \partial^{\mu} \bar{J}_{\mu i}{}^{\dot{\alpha}}\,   \bar{\xi}^{i\dot{\gamma}} \epsilon_{ \dot{\gamma} \dot{\alpha}} + 2\,  \partial_{\mu} \bar{J}_{\nu i}{}^{\dot{\gamma}} (\sigma^{\nu})_{\beta \dot{\gamma}} (\bar{\sigma}^{\mu})^{\dot{\alpha} \beta}\,    \bar{\xi}^{i\dot{\lambda}}\epsilon_{\dot{\lambda} \dot{\alpha}}\, , \nn
 \end{align}
 with the operators in the supermultiplet satisfying
 \begin{align}\label{closure}
 (\delta_1 \delta_2 - \delta_2 \delta_1 )\mathcal{O} &  = \left[\left[\delta_1, \delta_2\right],\mathcal{O}\right] = - \left(\xi_{1, \alpha i}\, \bar{\xi}_{2}^{j\dot \alpha} - \xi_{2, \alpha i}\,\bar{\xi}_{1}^{j\dot \alpha} \right)  \left[\{ Q^{\alpha i} , \tilde{Q}_{j\dot \alpha} \} , \mathcal{O} \right]\nn \\
 & = - \frac{1}{2} \left(\xi_{1, \alpha i}\, \bar{\xi}_{2}^{i\dot \alpha} - \xi_{2, \alpha i}\,\bar{\xi}_{1}^{i\dot \alpha} \right)  \epsilon^{\alpha\beta}(\sigma^\mu)_{\beta \dot \alpha}\, \partial_\mu \mathcal{O} \,,
 \end{align}
 in agreement with \eqref{eq:Neq4Qcomm}.

 To break the symmetry from $\mathfrak{psu}(2,2|4)$ into two copies of $\mathfrak{psu}(1,1|2)$, we choose the conventions of appendix \ref{app:conv}. With this choice, $x_1$ and $x_2$ are spacetime coordinates orthogonal to the defect, while $x_3$ and $x_4$ are parallel to it.

The Ward identities for one-point functions require that the action of the preserved supercharges on fermionic operators vanishes, namely 
\begin{equation}
  \langle \mathcal{Q}^a_{+} \mathcal{O} \rangle_{\Sigma} =   \langle \mathcal{Q}^-_{a} \mathcal{O} \rangle_{\Sigma} =  \langle \mathcal{ \bar Q}^a_{+}  \mathcal{O} \rangle_{\Sigma} = \langle \mathcal{ \bar Q}^-_{a}  \mathcal{O} \rangle_{\Sigma} = 0\, , 
\end{equation}
whenever $\mathcal{O}$ is a fermionic operator in the multiplet.

As expected, one-point functions are independent of the coordinates parallel to the defect:
\begin{align}
    \partial_{\textsc{x}} \langle \mathcal{O} \rangle_{\Sigma} &= \partial_{\bar {\textsc{x}}} \langle \mathcal{O} \rangle_{\Sigma} = 0\, ,
\end{align}
where $\textsc{x} = x_3 + i \, x_4$ and $\bar{\textsc{x}} = \textsc{x}^*$. Moreover, some operators cannot acquire a one-point function:
\begin{align}
    \langle B^M_{\mu \nu} \rangle_{\Sigma} &=  \langle \mathcal{E}^{M N P} \rangle_{\Sigma} = \langle \mathcal{L} \rangle_{\Sigma} = 0\, .
\end{align}
In index-free notation, the one-point function of the $\mathbf{20}'$ operator which satisfies the Ward identities reads
\begin{equation}
\label{1ptindexfree}
 \langle \mathcal{O}_2(\zeta,\bar \zeta)\rangle_{\Sigma} =  \frac{8\pi^2
 }{\lambda \sqrt{2}} Y^M Y^N \langle \mathcal{O}^{M N}
 (\zeta,\bar \zeta)\rangle_{\Sigma} = a_{\mathcal{O}_2}^{(-2)}\frac{\bar y^2}{\zeta^2}+a_{\mathcal{O}_2}^{(0)}\frac{y \bar y}{\zeta \bar \zeta}+ a_{\mathcal{O}_2}^{(2)} \frac{y^2}{\bar \zeta^2}  \, .
\end{equation}
Here $\zeta = x_1 + i \, x_2$ and $\bar \zeta = \zeta^*$ denote the spacetime coordinates orthogonal to the defect and $y=Y^1+iY^2$, $\bar y=Y^1-iY^2$. To compare it with the one-point function of equation \eqref{1ptbroken}, we parametrize the embedding coordinates as
\begin{align}
\label{perp_embedding}
    P_{\perp}&=\left(\frac{\zeta+\bar \zeta}{2},\frac{\zeta-\bar \zeta}{2i}\right) \, , \nn \\ Y_{\perp}&=\left(\frac{y+\bar y}{2},\frac{y-\bar y}{2i}\right)\, 
\end{align}
and see that the one-point function \eqref{1ptindexfree} agrees with the one in equation \eqref{1ptbroken}.

We have also checked that the one-point functions of the stress tensor and of the R-symmetry current are uniquely determined by \eqref{1ptindexfree} and do not depend on $a^{(\pm 2)}_{\mathcal{O}_2}$. They are therefore the same whether or not $\mathcal{R}_+$ is broken.

\section{Cross-ratios}\label{app:cross}

When the defect preserves both transverse and R-symmetry rotations, the two-point function of bulk operators depends on four independent conformal invariants. Following the conventions of \cite{Billo:2016cpy}, we define the cross-ratios
\begin{align}\label{crossratio}
    \xi&=-\frac{2P_1 \cdot P_2}{\left(P_1 \circ P_1\right)^{1 / 2}\left(P_2 \circ P_2\right)^{1 / 2}}\, ,
    & 
    \xi_R&=-\frac{2Y_1 \cdot Y_2}{\left(Y_1 \circ Y_1\right)^{1 / 2}\left(Y_2 \circ Y_2\right)^{1 / 2}}\,,\\
    \cos \phi&=\frac{P_1 \circ P_2}{\left(P_1 \circ P_1\right)^{1 / 2}\left(P_2 \circ P_2\right)^{1 / 2}}\, ,
    &
    \cos \phi_R&=\frac{Y_1 \circ Y_2}{\left(Y_1 \circ Y_1\right)^{1 / 2}\left(Y_2 \circ Y_2\right)^{1 / 2}} \,.
    \nonumber
\end{align}
Here $P_i^I$ denote the embedding-space coordinates of the operator insertions, while $Y_i^I$ encode the R-symmetry polarizations. For codimension-two defects, one can additionally construct the parity-odd structures
\begin{align}\label{sinphi}
    \sin \phi&=\frac{\epsilon_{IJ} P_1^I P_2^J}{\left(P_1 \circ P_1\right)^{1 / 2}\left(P_2 \circ P_2\right)^{1 / 2}}\, ,
    &
    \sin \phi_R&=\frac{\epsilon_{IJ} Y_1^I Y_2^J}{\left(Y_1 \circ Y_1\right)^{1 / 2}\left(Y_2 \circ Y_2\right)^{1 / 2}}  \,.
\end{align}
Here $\epsilon_{IJ}$ is the antisymmetric tensor of the transverse plane, with orientation $\epsilon_{21}=-\epsilon_{12}=1$, not to be confused with the spinor tensors of appendix \ref{app:conv}. With the parametrization \eqref{ptheta}, this choice gives $\phi=\theta_1-\theta_2$ and $\phi_R=\theta_{1R}-\theta_{2R}$, as used below.
 It is often convenient to trade the cross-ratios \eqref{crossratio} for lightcone variables, defined by
\begin{align}\label{lightcone}
    \cos \phi&=\frac{z+\bar{z}}{2 \sqrt{z \bar{z}}}\, ,
    &
    \cos \phi_R&=\frac{\alpha+\bar{\alpha}}{2 \sqrt{\alpha \bar{\alpha}}}\,,
    \nonumber \\
    \xi&=\frac{(1-z)(1-\bar z)}{\sqrt{z \bar z}} \,,
    &
    \xi_R &= \frac{(1-\a)(1-\bar \a)}{\sqrt{\a \bar \a}}\,.
\end{align}
In the Euclidean regime the variables are complex conjugates, $\bar z = z^*$, whereas in Lorentzian signature they are independent real variables \cite{Lemos:2017vnx}.

We also introduce radial coordinates,
\begin{equation}\label{radialcoord}
    r=\sqrt{z \bar{z}} \,, \qquad
    w=\sqrt{\frac{z}{\bar{z}}}=e^{i\phi}\,, \qquad
    r_R=\sqrt{\alpha \bar \alpha}\,, \qquad
    w_R=\sqrt{\frac{\bar \alpha}{\alpha}}=e^{i\phi_R}\, .
\end{equation}
If the defect preserves only the diagonal subgroup $U(1)_{-}$ of transverse and R-symmetry rotations, the two-point function depends on the additional cross-ratio \eqref{theta}. Its origin can be understood by choosing a convenient conformal frame. Using the conformal and R-symmetry transformations that preserve the defect, excluding transverse rotations, the two bulk insertion points can be brought to a plane orthogonal to the defect both in spacetime and in R-symmetry space. One may furthermore fix one operator to lie at unit distance from the defect. The configuration is then described by two radial variables, identified with the cross-ratios $r$ and $r_R$, together with four angular variables,
\begin{equation}
    \theta_1,\qquad \theta_2,\qquad \theta_{1R},\qquad \theta_{2R}\,,
\end{equation}
which specify the orientations of the operators in spacetime and in the internal R-symmetry space.

When transverse rotations are preserved, the generators $\mathcal{M}_{\perp}$ and $\mathcal{R}_{\perp}$ can be used to eliminate two angular variables, leaving only the standard combinations
\begin{equation}
    \phi=\theta_1-\theta_2\,, \qquad
    \phi_R=\theta_{1R}-\theta_{2R}\,.
\end{equation}
In the present case, however, only the diagonal combination $\mathcal{R}_\perp - \mathcal{M}_\perp$ remains unbroken. Consequently, one additional independent angular variable survives, which we parametrize as
\begin{equation}
    \vartheta = \theta_1+ \theta_2+ \theta_{1R}+ \theta_{2R}\,.
\end{equation}
With the definition of $\eta$ given in \eqref{etatheta}, one finds
\begin{equation}\label{udef}
    \eta_1+\eta_2=\vartheta\,, \qquad
    \eta_1-\eta_2=\phi+\phi_R\,, \qquad
    \omega=e^{i(\eta_1-\eta_2)}
    =e^{i(\phi+\phi_R)}
    =\sqrt{\frac{z \bar \alpha}{\bar z \alpha}}\,,
\end{equation}
where we have defined the mixed phase $\omega$ for convenience.

\section{Twist and twist again} \label{twist}
We briefly review the chiral algebra construction of  \cite{Beem:2013sza} and we make the explicit connection with the topological sector captured by the localization construction of \cite{Pestun:2009nn}. 

\subsection*{The first twist}
Consider $\mathcal{N}=4$ SYM on $\mathbb{R}^4$ and restrict to a plane $(x_1, x_2)$ parameterized by complex coordinates
\begin{equation}
	\zeta= x_1 + i x_2 \, ,  \qquad \bar \zeta = x_1 - i x_2 \, , 
\end{equation}
with corresponding translation generators $P_\z$ and $P_{\bar \z}$. We view four-dimensional $\mathcal{N}=4$ SYM as an $\mathcal{N}=2$ vector multiplet plus an adjoint hypermultiplet.  The $SU(4)_R$ symmetry is decomposed as
\begin{equation}\label{breakingN2}
SU(4)_R \to SU(2)_R\times SU(2)_F\times U(1)_ \mathfrak{r} \, .
\end{equation}
The group $SU(2)_F$ is a flavor symmetry from the $\mathcal{N}=2$ point of view. In particular, we use the following identification for the generators:
\begin{equation}
\begin{split}
    R^+ &= R^3_{\phantom{3} 1} \, , \quad  R^- = R^1_{\phantom{1} 3} \, , \quad R =\frac{1}{2} \left( R^3_{\phantom{3} 3}- R^1_{\phantom{1} 1}\right) \, , \\
     F^+ &= R^4_{\phantom{4} 2} \, , \quad  F^- = R^2_{\phantom{2} 4}\, , \quad F = \frac{1}{2} \left(R^4_{\phantom{4} 4} - R^2_{\phantom{2} 2}\right) \, ,
     \end{split}
\end{equation}
and
\begin{equation}
     \mathfrak{r} = R^1_{\phantom{1} 1} + R^3_{\phantom{3} 3} \, .
\end{equation}
The Lorentz spins are denoted by $(j_1,j_2)$, and the scaling dimension by $\Delta$.  The $SU(2)_R$ and $SU(2)_F$ Cartan generators are denoted by $R$ and $F$ while the $U(1)_ \mathfrak{r} $ charge is called $\mathfrak{r}$. 

The chiral algebra emerges for a subsector of local operators that belong to the cohomology of a nilpotent supercharge $\mathbbmtt{Q}$, a linear combination of the fermionic supercharges $Q$ and $S$. These are called Schur operators and they satisfy
\begin{equation}
	\label{Schur}
	\Delta - (j_1 + j_2) - 2R = 0\, ,
	\qquad
	 \mathfrak{r}  + j_1 - j_2 = 0\, .
\end{equation}
Following the standard construction, the twisted translation is defined as
\begin{equation}\label{twistedtransl}
	\widehat P_{\bar \z} \colonequals P_{\bar \z} + R^{-}\, ,
\end{equation}
which is $\mathbbmtt{Q}$-exact with respect to a specific fermionic supercharge $\mathcal G$,
\begin{equation}
	\widehat P_{\bar \z} = \{\mathbbmtt{Q}, \mathcal G\} \, , \quad 	[\mathbbmtt{Q}, 	\widehat P_{\bar \z}] = 0\, .
\end{equation}
Twisted-translated operators are then obtained by using $\widehat P_{\bar \z} $ to translate a Schur operator away from the origin
\begin{equation}
	\mathcal O(\z, \bar \z) \colonequals e^{\z P_\z + \bar \z	\widehat P_{\bar \z}} \mathcal{O}(0,0) e^{-\z P_\z - \bar \z	\widehat P_{\bar \z}} \, .
\end{equation}
The cohomology class of twisted-translated operators
\begin{equation}
	\mathcal O(\z) \colonequals [\mathcal{O}(\z, \bar \z)]_{\mathbbmtt{Q}}
\end{equation}
is purely holomorphic with a meromorphic OPE, characteristic of a two-dimensional chiral algebra,
\begin{equation}
	\mathcal O_i(\z)\,\mathcal O_j(0)
	\sim \sum_k \frac{c_{ij}{}^k\,\mathcal O_k(0)}{\z^{h_i + h_j - h_k}}\, ,
\end{equation}
where the two-dimensional conformal weights are identified as
\begin{equation}
	h_i \colonequals \frac{1}{2}\left( \Delta_i + (j_1 +j_2)_i \right) = \Delta_i - R_i\, .
\end{equation}
Equivalently, one can think of the twisted translation as a spacetime-dependent polarization vector $Y^M$. Let us see how this works for dimension-two chiral primary operators \eqref{CPO}. For a scalar operator, the conditions \eqref{Schur} reduce to
\begin{equation}
    \Delta=2R\, , \qquad  \mathfrak{r} =0\,, 
\end{equation}
which tell us that scalar Schur operators must be charged under $SU(2)_R$.
The six scalars in $\mathcal{N}=4$ SYM decompose under \eqref{breakingN2} as
\begin{equation}
\mathbf 6=[0,1,0]\longrightarrow
(\mathbf 2,\mathbf 2)_0
\oplus (\mathbf 1,\mathbf 1)_{+1}
\oplus (\mathbf 1,\mathbf 1)_{-1} 
\end{equation}
and we need to select the $(\mathbf 2,\mathbf 2)_0$ component. This can be done by setting $Y^5 = Y^6 = 0$ and writing
\begin{equation}
	Y^\mu = (Y^1, Y^2, Y^3, Y^4) \, .
\end{equation}
These components can be organized in terms of indices $a=1,2$ and $\dot a=1,2$, associated with $SU(2)_R$ and $SU(2)_F$, respectively, as 
\begin{equation}
	\label{Ybroken}
 	Y_{a \dot a} = Y^\mu ( \sigma^{\mu})_{a \dot a}
 	= \begin{pmatrix}
 		Y^3 + i Y^4 &  Y^1 - i Y^2 \\
 		Y^1 + i Y^2 &  -Y^3 + i Y^4 
 	\end{pmatrix} \, .
\end{equation}
This matrix has vanishing determinant thanks to the null condition on the vector $Y$, therefore we can introduce auxiliary vectors  $u_a$ and $\tilde u_{\dot a}$ associated with $SU(2)_R$ and $SU(2)_F$, respectively, such that
\begin{equation}
	Y_{a \dot a} = u_a \tilde u_{\dot a} \, .
\end{equation}
 The twist is implemented by the choice $u_a = (1, \bar \z)$. The scalar product then becomes
\begin{equation}
\label{Yprodchiral}
 	Y_1 \cdot Y_2 = \frac{1}{2} (\bar \z_1 - \bar \z_2)\, \tilde u^{12} \, ,
 \end{equation}
where we defined
\begin{equation}
\label{uijtilde}
 	\tilde u^{ij} \colonequals \tilde u^i_{\dot 1} \tilde u^j_{\dot 2} - \tilde u^i_{\dot 2} \tilde u^j_{\dot 1} \, .
 \end{equation}

 Considering, for example, the dimension-two chiral primary operators \eqref{CPO}, in the $\mathcal{N}=2$ picture they decompose as
\begin{equation}
    \mathbf{20'}=[0,2,0]\to (\mathbf{3},\mathbf{3})_0\oplus(\mathbf{2,2})_{\pm 1}\oplus(\mathbf{1,1})_{\pm2}\oplus (\mathbf{1,1})_{0} \, .
\end{equation}
The Schur operator is $(\mathbf{3},\mathbf{3})_0$ and it corresponds to the moment map operator, i.e. the highest-weight operator of the $\mathcal{N}=2$ flavor current multiplet \cite{Beem:2016wfs}.  In the chiral algebra, flavor symmetries give rise to affine Kac-Moody currents ($h = \Delta - R = 1$), obtained via twisted translations of the moment map $\mu^{A \,  a \, b}$
\begin{equation}\label{chiralcurrent}
	J^A(\z) \colonequals [\kappa_J \, u_a(\bar \z) u_b(\bar \z) \mu^{A \,  a \, b}(\z , \bar \z) ]_{\mathbbmtt{Q}}\, ,
\end{equation}
where $A$ labels the adjoint representation of the flavor algebra and $\kappa_J$ is a normalization constant.  

Let us see how this works at the level of correlation functions. The two-point function of $\mathcal{O}_2$ in embedding space is
\begin{equation}
\label{free2pt}
	\langle \mathcal{O}_2(P_1, Y_1)\mathcal{O}_2(P_2, Y_2) \rangle 
	= \left( \frac{Y_1 \cdot Y_2}{-2 \, P_1 \cdot P_2} \right)^2 \, .
\end{equation}
Substituting \eqref{Yprodchiral} into \eqref{free2pt}, we find
\begin{equation}
 	\tilde u^1_{\dot a} \tilde u^1_{\dot b}\tilde u^2_{\dot c} \tilde u^2_{\dot d} \, \langle  J^{\dot a \dot b}(\z_1) J^{\dot c \dot d}(\z_2) \rangle 
 	= \kappa_J^2\, \frac{(\tilde u^{12})^2}{4 (\z_1 - \z_2)^2} \, .
 \end{equation}
 To rewrite this with adjoint indices for the currents, we introduce the Pauli matrices
\begin{equation}
 	(\sigma^A)^{\dot a}_{\ \dot b} = (\sigma^1, \sigma^2, \sigma^3) \, ,
 \end{equation}
 or equivalently
\begin{equation}
	(\sigma^A)_{\dot a \dot b} = (\sigma^A)^{\dot c}_{\ \dot a}\, \epsilon_{\dot c \dot b} \, .
\end{equation}
Defining 
\begin{equation}
\label{defJA}
 	J^A \colonequals (\sigma^A)_{\dot a \dot b}\, J^{\dot a \dot b} \, ,
\end{equation}
 and opening the indices with $\partial_{\tilde u^1_{\dot a}} \partial_{\tilde u^1_{\dot b}} \partial_{\tilde u^2_{\dot c}} \partial_{\tilde u^2_{\dot d}}$, we obtain
\begin{equation}
\label{twopointchiralnodef}
	\langle J^A(\z_1) J^B(\z_2) \rangle = - \frac{\kappa_J^2 \delta^{AB}}{2 (\z_1 - \z_2)^2} \, .
\end{equation}
We can match this result by working directly in the chiral algebra with affine Lie algebra $\mathfrak{su}(2)_{2k}$. The OPE between two Kac-Moody currents reads 
\begin{equation}
\label{OPEchiral}
    J^A(\z_1) J^B(\z_2) \sim k\,  \frac{\delta^{A B}}{(\z_1 - \z_2)^2} + \frac{i \varepsilon^{A B C} J^C(\z_2)}{\z_1 - \z_2}\, .
\end{equation}
Comparing with \eqref{twopointchiralnodef}, we immediately find that the level is $2 k =- \kappa_J^2$. 

Similarly, the three-point function of $\mathfrak{su}(2)_{2k}$ currents is
\begin{equation}
    \langle J^A(\z_1) J^B(\z_2) J^C(\z_3) \rangle = k \, \frac{ i \, \varepsilon^{A B C}}{\z_{12} \, \z_{23} \, \z_{13}}\, ,
\end{equation}
with $\z_{ij} \colonequals \z_i - \z_j$. On the other hand, the three-point function of $\mathbf{20'}$ operators is
\begin{equation}
	\langle \mathcal{O}_2(P_1, Y_1)\mathcal{O}_2(P_2, Y_2)\mathcal{O}_2(P_3, Y_3) \rangle 
	= \lambda_{\mathcal{B}_{[0,2,0]}} \frac{Y_1 \cdot Y_2}{-2 P_1 \cdot P_2}  \frac{Y_2 \cdot Y_3}{-2 P_2 \cdot P_3}   \frac{Y_1 \cdot Y_3}{-2 P_1 \cdot P_3}  \, .
\end{equation}
Reducing this expression to the chiral algebra as above gives
\begin{equation}
     \langle J^A(\z_1) J^B(\z_2) J^C(\z_3) \rangle = \lambda_{\mathcal{B}_{[0,2,0]}}\, \kappa_J^3  \frac{i \, \varepsilon^{A B C}}{4 \, \z_{12} \, \z_{23}\,  \z_{13}}\, .
\end{equation}
For $k = - \kappa_J^2/2$, this fixes the normalization to
\begin{equation}\label{k1sulambda}
\kappa_J = - \frac{2}{\lambda_{\mathcal{B}_{[0,2,0]}}}\, .
\end{equation}

\subsubsection*{Defect operators}
We now consider the defect algebra in the $\mathcal{N}=2$ setting. We show the explicit map between our notation and the one used in \cite{Bianchi:2019sxz}. For the supercharges we have
\begin{align}
    G^-_{-1/2} = \mathcal{Q}_1^{-}\, , \quad \bar G^-_{-1/2} =  \mathcal{\bar Q}_1^{-}\, , \quad   G^+_{1/2} = \mathcal{S}^1_{+}\, , \quad \bar G^+_{1/2} =  \mathcal{\bar S}^1_{+}\, ,
\end{align}
with the algebra
\begin{equation}
\begin{split}
    \{ G^+_r, G^-_s\} & = L_{r+s} + \frac{r-s}2 \mathcal{J}\, , \\
    \{ \bar G^+_r, \bar G^-_s\} & = \bar L_{r+s}+  \frac{r-s}2 \bar {\mathcal{J}}\, .
\end{split}
\end{equation}
The identification of the 2d conformal generators is the usual one
\begin{align}
    L_{-1}&=\mathcal{P}\, ,  & L_{0}&=\mathcal{D}\, , & L_1=\mathcal{K}\, , \nn \\
    \bar L_{-1}&=\mathcal{\bar P}\,,  & \bar L_{0}&=\mathcal{\bar D} \, ,& \bar L_1=\mathcal{\bar K} \, ,
\end{align}
while the global symmetry generators introduced in \cite{Bianchi:2019sxz} are expressed in terms of the $\mathcal{N}=2$ R-symmetry generators as
\begin{align}
    \mathcal{J} &= - 2 R - \mathcal{M}_\perp +  \mathfrak{r}  \, , & \mathcal{\bar J} &= 2 R + \mathcal{M}_\perp +  \mathfrak{r}\,, & \mathcal{C}=R+\mathcal{M}_\perp\, .
\end{align}
These can be expressed in terms of the $\mathfrak{psu}(1,1|2) \oplus \mathfrak{psu}(1,1|2) \oplus \mathfrak{u}(1)_{-}$ generators as
\begin{equation}
\mathcal{J} = 2 \mathcal{R} + \mathcal{R}_{-}\, , \qquad \mathcal{\bar J} = 2 \mathcal{\bar R} - \mathcal{R}_{-}\,  ,\qquad \mathcal{C}=\frac12(\mathcal{\bar R}-\mathcal{R}-\mathcal{R}_-+\mathcal{M}_\perp)\, ,
\end{equation}
where we used the map
\begin{align}\label{mapchiral}
    \mathcal{R} & = \frac12 \left(  \mathfrak{r}  - R + F \right) \, , \quad \mathcal{\bar R} = \frac12 \left(  \mathfrak{r} + R - F \right) \, , \quad \mathcal{R}_{-} = - R - F - \mathcal{M}_\perp\, .
\end{align}
Notice that the central charge $\mathcal{C}$ in the $\mathcal{N}=2$ setting is not identified with the central generator $\mathcal{R}_-$ in the $\mathcal{N}=4$ picture. This is important when identifying which defect operators are in cohomology. 

A defect operator in cohomology at the origin must satisfy \cite{Bianchi:2019sxz}
\begin{equation}
 L_0 = - \frac12 \mathcal{J} \, , \quad \bar L_0 = \frac12 \mathcal{\bar J}\, .
\end{equation}
For a scalar operator this implies
\begin{equation}
  \hat \Delta = - \mathcal{R}  + \bar {\mathcal{R}} - \mathcal{R}_{-}\, , \qquad \mathcal{R}+\bar{\mathcal{R}}=0\, .
\end{equation}
As explained in \cite{Bianchi:2019sxz}, the condition for an operator at the origin to be degenerate with the defect identity is $\mathcal{C} = 0$. For a scalar operator this translates to the constraint $\hat \Delta = -s$, where $s$ is the transverse spin.
Let us take as an example the $[0,2,0]$ operator. In the $\mathcal{N} = 2$ picture it has $R = 1$. The previous condition therefore becomes $\hat \Delta = -s = 1$ and the eigenvalues of the three Cartan generators in \eqref{mapchiral} now read
\begin{align}\label{mapchiral_defect}
    \mathcal{R} & = \frac12 \left(  F -1 \right) \, , \quad \bar{\mathcal{R}} = \frac12 \left( 1 - F \right) \, , \quad \mathcal{R}_{-} = - F \, .
\end{align}

\subsection*{Second twist: the extra $\mathcal{N}=4$ topological twist}

For a generic $\mathcal{N}=2$ theory the story stops with a holomorphic chiral algebra.  In $\mathcal{N}=4$ SYM the chiral algebra has small $\mathcal{N}=4$ supersymmetry.  In particular the $SU(2)_F$ current algebra is actually an $R$-symmetry algebra for the extended two-dimensional superconformal algebra.  One can therefore perform a further two-dimensional topological twist using the $SU(2)_F$ Cartan. This further twist selects operators with two-dimensional conformal weights given by $h=-F$ leading to the conditions
\begin{equation}
    -F-R=j_1+j_2\, , \qquad \mathfrak{r}=j_2-j_1 \, ,\qquad \Delta=R-F\, .
\end{equation}
For a scalar operator this reduces to 
\begin{equation}
    R=-F\, , \qquad \mathfrak{r}=0\, , \qquad \Delta=2R=-2F\, .
\end{equation}
Translating to $SU(4)$ quantum numbers this selects operators with labels $[0,p,0]$ and $\Delta=p$, i.e. precisely the chiral primary operators \eqref{CPO}.

At the level of polarization the second twist is simply implemented by selecting 
\begin{equation}
    u= (1,\bar \z)\, , \qquad \tilde u=(1,\z)\, .
\end{equation}
With this choice the $Y$ vector reads
\begin{equation}
Y(\z,\bar \z)=
\frac12\bigl(
 \z+\bar \z,
 i(\z-\bar \z),
 1-\z\bar \z,
 -i(1+\z\bar \z),
 0,
 0
 \bigr) \, .
 \end{equation} 
 We can see that correlation functions become topological by the fact that 
 \begin{equation}
     Y_1\cdot Y_2 = -\frac12\, \zeta_{12}\,  \bar \zeta_{12}=P_1\cdot P_2
 \end{equation}
 in the restricted planar kinematics. 
To map the problem on the $S^2$ sphere, we introduce stereographic coordinates 
\begin{equation}
 n^1={\frac{\z+\bar \z}{1+\z\bar \z} }\, ,
 \qquad
 n^2={\frac{i(\z-\bar \z)}{ 1+\z\bar \z}}\, ,
 \qquad
 n^3={\frac{1-\z\bar \z}{1+\z\bar \z}}\, ,
 \qquad
 n^i n^i=1 \, .
\end{equation}
Then
\begin{equation}
Y(\z,\bar \z)
=\frac 12\, (1+\z\bar \z)
\bigl(n^1,n^2,n^3,-i,0,0\bigr).
\end{equation}
The overall factor is the expected Weyl factor for a scalar primary of dimension one.  Therefore, after mapping to an $S^2$ of radius $\rho$, with $n^i=x^i/\rho$, the natural sphere representative is
\begin{equation}\label{Olocdef}
 \mathcal{O}^{\mathrm{loc}}_p(x)=
 \Tr\left(
 {\frac{x^i}{\rho}}\Phi_i(x)-i\Phi_4(x)
 \right)^p\, ,
 \qquad x\in S^2\,  .
\end{equation}
Remarkably, this is precisely the local-operator polarization that appears in the $S^2$ localization picture \cite{Pestun:2009nn} (up to a convention-dependent minus sign). Therefore, the localization result on the two-sphere can be compared with the result of the correlator after the double twist.

\section{Superconformal blocks}\label{app:blocks}
\subsection{Bulk channel}
\subsection*{Bulk bosonic blocks}
In four dimensions, in the presence of a surface defect preserving transverse rotations, the bulk
spacetime conformal blocks $f^{(\Delta_{12})}_{\Delta,\ell}(\xi,\phi)$ are eigenfunctions of the quadratic Casimir\footnote{Its symbol should not be confused with the generator $\mathcal{J}$ introduced in the previous appendix.} of $SO(2,4)$
\begin{equation}\label{Cas}
    \mathcal{J}^2=-\frac{1}{2}\left(\mathcal{J}_1^{MN}+\mathcal{J}_2^{MN}\right)^2\,,
\end{equation}
where, in embedding space,
\begin{equation}\label{spacecas}
    \mathcal{J}_i^{MN} = P_i^{M} \frac{\partial}{\partial P_i^{N}}-P_i^{N} \frac{\partial}{\partial P_i^{M}}\,.
\end{equation}
The Casimir equation takes the form
\begin{equation}\label{bulkcasimireq}
    \mathcal{J}^2 \left[\frac{\xi^{-\frac{\Delta_1+\Delta_2}{2}} f^{(\Delta_{12})}_{\Delta, \ell}(\xi,\phi)}{(P_1 \circ P_1)^{\frac{\Delta_1}{2}}(P_2 \circ P_2)^{\frac{\Delta_2}{2}} }\right]= (\Delta \left(\Delta-4\right)+\ell(\ell+2)) \, \frac{\xi^{-\frac{\Delta_1+\Delta_2}{2}} f^{(\Delta_{12})}_{\Delta, \ell}(\xi,\phi)}{(P_1 \circ P_1)^{\frac{\Delta_1}{2}}(P_2 \circ P_2)^{\frac{\Delta_2}{2}} }\,.
\end{equation}
Here $\Delta$ and $\ell$ are the scaling dimension and spin of the exchanged operator, while $\Delta_{12} \colonequals \Delta_1-\Delta_2$ denotes the difference between the scaling dimensions of the external operators. For identical operators $\Delta_{12} = 0$ and equation \eqref{bulkcasimireq} reduces to
\begin{eqnarray}\label{boscas}
&&\mathcal{C}^{(0)}_{SO(2,4)} \, f^{(0)}_{\Delta, \ell}(z, \bar{z})=0, \\
&& \mathcal{C}^{(0)}_{SO(2,4)}\colonequals D_z+D_{\bar{z}}+2 \frac{(1-z)(1-\bar{z})}{1-z \bar{z}}\left(z \partial_z+\bar{z} \partial_{\bar{z}}\right)-\frac{1}{2}(\Delta \left(\Delta-4\right)+\ell(\ell+2)), \nonumber\\
&& D_z=(1-z)^2 z \partial_z^2+(1-z)^2 \partial_z\,, \nonumber
\end{eqnarray}
where we have expressed the equation in terms of the lightcone cross-ratios introduced in \eqref{lightcone}. This equation can be mapped to the Casimir equation for four-point functions without defects \cite{Billo:2016cpy}, whose solution is \cite{Dolan:2000ut}
\begin{equation}\label{bulkblock1}
f^{(0)}_{\Delta, \ell}(z, \bar{z}) = \frac{(1-z)(1-\bar{z})}{1-z \bar{z}}\left(k^{0,0}_{\Delta-\ell-2}(1-z) k^{0,0}_{\Delta+\ell}(1-\bar{z})+(-1)^{\ell}(z \leftrightarrow \bar{z})\right)\, ,
\end{equation}
with
\begin{equation}
    k^{r,s}_\beta (z) = z^{\beta / 2}{ }_2 F_1\left(\frac{\beta+r}{2}, \frac{\beta-s}{2}, \beta, z\right)\,.
\end{equation}

The R-symmetry blocks $h^{(\Delta_{12})}_R(\alpha,\bar\alpha)$ are eigenfunctions of the Casimir of $SU(4)_R\cong SO(6)_R$,
\begin{equation}
\mathcal{L}^2=-\frac{1}{2}\left(\mathcal{L}_1^{MN}+\mathcal{L}_2^{MN}\right)^2\,,
\end{equation}
where
\begin{equation}\label{rsymcas}
    \mathcal{L}_i^{MN}=Y_i^{M} \frac{\partial}{\partial Y_i^{N}}-Y_i^{N} \frac{\partial}{\partial Y_i^{M}}\,.
\end{equation}
The structure of $\mathcal{L}^2$ is identical to that of the spacetime Casimir after the formal replacement $P_i\to Y_i$. The associated eigenvalue equation is
\begin{equation}
    \mathcal{L}^2 \left[\frac{\xi^{\frac{\Delta_1+\Delta_2}{2}}_R \,h^{(\Delta_{12})}_{R} (\xi_R,\phi_R)}{(Y_1 \!\circ\! Y_1)^{-\frac{\Delta_1}{2}}(Y_2 \circ Y_2)^{-\frac{\Delta_2}{2}}}\right]= 2(n(n+3)+m(m+1)) \,\frac{\xi^{\frac{\Delta_1+\Delta_2}{2}}_R \,h^{(\Delta_{12})}_{R} (\xi_R,\phi_R)} {(Y_1 \!\circ\! Y_1)^{-\frac{\Delta_1}{2}}(Y_2 \circ Y_2)^{-\frac{\Delta_2}{2}}}\,,
\end{equation}
for representations with Dynkin labels $R=[n-m,\,2m,\,n-m]$. For $\Delta_{12}=0$, comparison with the spacetime case yields
\begin{equation}\label{rblock}
  h^{(0)}_{[n-m, \, 2m, \, n-m]}(\alpha,\bar{\alpha}) =  f^{(0)}_{-m-n, n-m}\, (\alpha,\bar{\alpha})\,,
\end{equation}
where the blocks are expressed in the $(\alpha,\bar \alpha)$ coordinates defined in \eqref{lightcone}. Since $m,n\in\mathbb{Z}_{\ge 0}$, the hypergeometric functions truncate, and the R-symmetry blocks reduce to polynomials.

When the defect does not preserve transverse rotations and R-symmetry separately, as in the setup considered in the main text, the bosonic conformal blocks for $\Delta_{1}=\Delta_2=p$ and fixed Fourier charge $q$ in the decomposition \eqref{Fourier} satisfy
\begin{eqnarray}
    \mathcal{J}^2 \left[\frac{\xi^{-p} e^{i q \frac{\vartheta}{2}} f^{(q)}_{\Delta, \ell}(\xi,\phi)}{(P_1 \circ P_1)^{p/2}(P_2 \circ P_2)^{p/2} }\right]&=& (\Delta \left(\Delta-4\right)+\ell(\ell+2)) \,\frac{\xi^{-p} e^{i q \frac{\vartheta}{2}} f^{(q)}_{\Delta, \ell}(\xi,\phi)}{(P_1 \circ P_1)^{p/2}(P_2 \circ P_2)^{p/2} } \\
    \mathcal{L}^2 \left[\frac{\xi^p_R  \,e^{i q \frac{\vartheta}{2}}\,h^{(q)}_{R} (\xi_R,\phi_R)}{(Y_1 \!\circ\! Y_1)^{-p/2}(Y_2 \circ Y_2)^{-p/2}}\right]&=& 2(n(n+3)+m(m+1)) \,\frac{\xi^p_R \, e^{i q \frac{\vartheta}{2}}\,h^{(q)}_{R} (\xi_R,\phi_R)}{(Y_1 \!\circ\! Y_1)^{-p/2}(Y_2 \circ Y_2)^{-p/2}}\,.\nonumber
    \end{eqnarray}
Using \eqref{Cas} and \eqref{rsymcas}, one finds that these equations reduce to those satisfied by rotation-preserving conformal blocks with $\Delta_{12}=\pm q$. For the spacetime sector, this gives
\begin{equation}\label{Cas2}
     \mathcal{C}^{(q)}_{SO(2,4)} f^{(q)}_{\Delta, \ell}(z,\bar z)=0\,, \qquad
    \mathcal{C}^{(q)}_{SO(2,4)} =\mathcal{C}^{(0)}_{SO(2,4)}-\frac{q^2}{16}\left(z+\bar z+\frac{1}{z}+\frac{1}{\bar z}-4\right)\,.
\end{equation}
The differential operator of the R-symmetry sector is obtained through the replacement $(z,\bar z)\to(\alpha,\bar\alpha)$ together with $\Delta(\Delta-4)+\ell(\ell+2)\to 2(n(n+3)+m(m+1))$. Imposing invariance under $q \leftrightarrow -q$, the solution to \eqref{Cas2} is \cite{Isachenkov:2018pef}
\begin{eqnarray}\label{bulkblockbosq}
    f^{(q)}_{\Delta, \ell}(z,\bar z)=\frac{1}{2}\left[ \tilde f^{(q)}_{\Delta, \ell}(z,\bar z)+\tilde f^{(-q)}_{\Delta, \ell}(z,\bar z)\right]\,,
\end{eqnarray}
where
\begin{equation}
    \tilde f^{(q)}_{\Delta, \ell}(z,\bar z) = \frac{(1-z)(1-\bar{z})}{1-z \bar{z}}(z \bar z)^{q/4}\left(k^{q,0}_{\Delta-\ell-2}(1-z) k^{q,0}_{\Delta+\ell}(1-\bar{z})+(-1)^{\ell}(z \leftrightarrow \bar{z})\right)\,.
\end{equation}
Similarly, the R-symmetry blocks are
\begin{equation}
    h^{(q)}_{[n-m, \, 2m, \, n-m]}(\alpha,\bar{\alpha}) = f^{(q)}_{-m-n, n-m}\, (\alpha,\bar{\alpha})\,.
\end{equation}

\subsection*{Bulk superconformal blocks}

The two-point function of $\mathcal{O}_p$ admits an expansion in bulk superconformal blocks,
\begin{equation}
\mathcal{F}(z,\bar z,\alpha,\bar\alpha,\vartheta) = \sum_{q}\sum_{\mathcal{O}}(-2)^{n_\mathcal{O} -p}
\lambda_{\mathcal{O}_p\mathcal{O}_p\mathcal{O}}\, a^{(q)}_{\mathcal{O}}\,
e^{i q \frac{\vartheta}{2}}\, \left(\frac{\xi}{\xi_R}\right)^{-p} \mathcal{G}^{(q)}_{\mathcal{M}_\mathcal{O}}(z,\bar z,\alpha,\bar\alpha)\,,
\end{equation}
where $\mathcal{O} \in [a,b,a]$ and $n_\mathcal{O} = a + b/2$.
Each block $\mathcal{G}^{(q)}_{\mathcal{M}}$ can be expanded as a linear combination of products of bosonic spacetime and R-symmetry blocks,
\begin{equation}\label{ansatzblockb}
\mathcal{G}^{(q)}_{\mathcal{M}}(z,\bar z,\alpha,\bar\alpha) = \sum_{\Delta,\ell,R} c^{(q)}_{\Delta,\ell,R}\, f^{(q)}_{\Delta,\ell}(z,\bar z)\, h^{(q)}_{R}(\alpha,\bar\alpha)\,,
\end{equation}
where the coefficients $c^{(q)}_{\Delta,\ell,R}$ are fixed by imposing the Ward identities \eqref{WI} on the ansatz \eqref{ansatzblockb}. The sum runs over the conformal primaries contained in the supermultiplet $\mathcal{M}$, labeled by their scaling dimension $\Delta$, spin $\ell$, and $SU(4)_R$ representation $R$.

For the two-point function of $\mathcal{O}_2$, the supermultiplets $\mathcal{M}$ appearing in the bulk OPE are listed in \eqref{O2bulkope}, while the operator content of each multiplet was determined in \cite{Dolan:2002zh}. The corresponding bulk superconformal blocks for the allowed charge sectors $q=0,\pm2,\pm4$ are collected in tables \ref{tab:blocks} and \ref{tab:blocksq}. Each superblock $\mathcal{G}_{\mathcal{M}}^{(q)}(z,\bar z,\alpha,\bar\alpha)$ admits a decomposition into holomorphic and non-holomorphic contributions, reproducing the general solution to the Ward identities in \eqref{wardsolution},
\begin{eqnarray}
    &&\mathcal{G}_{\mathcal{M}}^{(q)}(z,\bar z,\alpha,\bar\alpha) = \mathcal{C}^{(q)}_{\mathcal{M}} + \frac{(z-\alpha)(\bar\alpha z-1)(\alpha\bar z-1)(\bar z-\bar\alpha)}{\alpha \bar \alpha \, z \bar z} G^{(q)}_{\mathcal{M}}(z,\bar z,\alpha,\bar\alpha)\\
&&+\frac{ (\bar\alpha z-1)(\alpha\bar z-1)
\left[g^{(q)}_{\mathcal{M}}(z,\alpha)\!+\!g^{(q)}_{\mathcal{M}}(\bar z,\bar\alpha)\right] }{ (\alpha\bar\alpha-1)(z\bar z-1) }-\frac{ (z-\alpha)(\bar z-\bar\alpha) \left[g^{(q)}_{\mathcal{M}}\!\left(z,\frac{1}{\bar\alpha}\right) \!+\!g^{(q)}_{\mathcal{M}}\!\left(\bar z,\frac{1}{\alpha}\right)\right] }{ (\alpha\bar\alpha-1)(z\bar z-1)}\,,
\nonumber
\end{eqnarray}
where $g^{(q)}_{\mathcal{M}}(z,\alpha) \colonequals \mathcal{G}_{\mathcal{M}}^{(q)}(z,\bar z,\alpha,\bar z)$ and $\mathcal{C}^{(q)}_{\mathcal{M}}\colonequals -g^{(q)}_{\mathcal{M}}(z,z)$.
Table \ref{tab:blocks} presents this decomposition instead of the full superblocks, whose explicit expressions are more cumbersome.
We note that long multiplets $\mathcal{A}^{\Delta}_{[0,0,0],\ell}$ contribute neither to the holomorphic sector nor to charge sectors with $q\neq0$.

\newpage
\begin{table}[htbp!]
\begin{center}
\renewcommand{\arraystretch}{1.5}
\begin{tabular}{|c|c|c|c|}
\hline
$\mathcal{M}$ & $G^{(0)}_{\mathcal{M}}(z,\bar z,\alpha,\bar \alpha)$ & $g^{(0)}_{\mathcal{M}}(z,\alpha)$ & $\mathcal{C}^{(0)}_{\mathcal{M}}$ \\
\hline\hline

$\id$  
& $0$  
& $\begin{aligned}\\
&1\\
&
\end{aligned}$  
& $-1$ \\
\hline

$\mathcal{B}_{[0,2,0]}$ 
& $0$ 
& $\begin{aligned}
\\
    &\tfrac{\alpha\!+\!1}{2(1\!-\!\alpha)}k_2(1\!-\!z)-\tfrac{1}{12}k_4(1\!-\!z)\\
    &
\end{aligned}$ 
& $-1$ \\
\hline

$\mathcal{B}_{[2,0,2]}$ 
& $-\tfrac{\alpha \bar \alpha\, z \bar z\, f^{(0)}_{6,0}(z,\bar z)}{(1-\alpha)^2(1-\bar \alpha)^2(1-z)^2(1-\bar z)^2}$ 
& $\begin{aligned}
\\
&k_4(1\!-\!z)\!-\!\tfrac{\alpha\!+\!1}{2(1\!-\!\alpha)}k_6(1\!-\!z)
\!+\!\tfrac{9}{140}k_8(1\!-\!z)\\
&
\end{aligned}$ 
& $0$ \\
\hline

$\mathcal{B}_{[0,4,0]}$ 
& $\frac{\alpha \bar \alpha\, z \bar z\, f^{(0)}_{4,0}(z,\bar z)}{(1-\alpha)^2(1-\bar \alpha)^2(1-z)^2(1-\bar z)^2}$ 
& $\begin{aligned}
\\
&\tfrac{\alpha(\alpha\!+\!4)\!+\!1}{6(1\!-\!\alpha)^2}k_4(1\!-\!z)\!-\!\tfrac{\alpha\!+\!1}{30(1\!-\!\alpha)}k_6(1\!-\!z)\\
&
\end{aligned}$ 
& $-1$ \\
\hline

$\mathcal{C}_{[0,2,0],\ell}$ 
& $-\frac{\alpha \bar \alpha\, z \bar z\, f^{(0)}_{\ell+6,\ell+2}(z,\bar z)}{(1-\alpha)^2(1-\bar \alpha)^2(1-z)^2(1-\bar z)^2}$ 
& $\begin{aligned}
\\
&\tfrac{\alpha\!+\!1}{2(1\!-\!\alpha)}k_{2\ell+6}(1\!-\!z)\!-\!\tfrac{(\alpha\!+\!1)^2}{4(1\!-\!\alpha)^2}k_{2\ell+8}(1\!-\!z)\\
&+\tfrac{(\alpha\!+\!1)(\ell\!+\!4)^2}{8(1\!-\!\alpha)(2\ell\!+\!7)(2\ell\!+\!9)}k_{2\ell+10}(1\!-\!z)\\
&

\end{aligned}$ 
& $0$ \\
\hline

$\mathcal{C}_{[1,0,1],\ell}$ 
& $-\frac{\alpha \bar \alpha\, z \bar z\, f^{(0)}_{\ell+7,\ell+1}(z,\bar z)}{(1-\alpha)^2(1-\bar \alpha)^2(1-z)^2(1-\bar z)^2}$ 
& $\begin{aligned}
\\
&k_{2\ell+6}(1\!-\!z)\!-\!\tfrac{\alpha\!+\!1}{2(1\!-\!\alpha)}k_{2\ell+8}(1\!-\!z)\\
&+\tfrac{(\ell\!+\!4)^2}{4(2\ell\!+\!7)(2\ell\!+\!9)}k_{2\ell+10}(1\!-\!z)\\ 
&
\end{aligned}$ 
& $0$ \\
\hline

$\mathcal{A}^{\Delta}_{[0,0,0],\ell}$ 
& $\frac{\alpha \bar \alpha\, z \bar z\, f^{(0)}_{\Delta+4,\ell}(z,\bar z)}{(1-\alpha)^2(1-\bar \alpha)^2(1-z)^2(1-\bar z)^2}$  
& $\begin{aligned}
    \\
    &0\\
    &
\end{aligned}$ 
& $0$ \\
\hline

\end{tabular}
\renewcommand{\arraystretch}{1}
\caption{Bulk superconformal blocks for the $q=0$ sector. We use the shorthand notation $k_{\beta}(1-z)\equiv k_{\beta}^{0,0}(1-z)$.}
\label{tab:blocks}
\end{center}
\end{table}
\newpage
\begin{table}[h!]
\begin{center}
\renewcommand{\arraystretch}{1.5}
\begin{tabular}{|Sc|Sc|Sc|}
\hline
$\mathcal{M} $ & $\mathcal{G}_{\mathcal{M}}^{(\pm2)}(z,\bar z,\alpha,\bar\alpha)$ &$\mathcal{G}_{\mathcal{M}}^{(\pm4)}(z,\bar z,\alpha,\bar\alpha)$ \\
\hline\hline
$\id$  
& $0$  
& $0$
\\
\hline

$\mathcal{B}_{[0,2,0]}$ 
& $f^{(\pm2)}_{2,0}(z,\bar z)\,h_{[0,2,0]}^{(\pm2)}(\alpha , \bar \alpha)$
& $0$
\\
\hline

$\mathcal{B}_{[2,0,2]}$ 
& $0$ 
& $0$ 
\\
\hline

$\mathcal{B}_{[0,4,0]}$ 
& $\begin{aligned}
&f^{(\pm2)}_{4,0}(z,\bar z)\,h_{[0,4,0]}^{(\pm2)}(\alpha , \bar \alpha)-\tfrac{1}{20} f^{(\pm2)}_{5,1}(z,\bar z)\,h_{[1,2,1]}^{(\pm2)}(\alpha , \bar \alpha)\\
&-\tfrac{1}{350} f^{(\pm2)}_{6,2}(z,\bar z)\,h_{[0,2,0]}^{(\pm2)}(\alpha , \bar \alpha)
\end{aligned}$ 
&\scalebox{0.8}{$\displaystyle
f^{(\pm4)}_{4,0}(z, \!\bar z)h_{[0,4,0]}^{(\pm4)}(\alpha,\! \bar \alpha)
$}
\\
\hline

$\mathcal{C}_{[0,2,0],\ell}$ 
& $\begin{aligned}
   &\Big[f^{(\pm2)}_{\ell+4,\ell}(z,\bar z)+\tfrac{3 (\ell+2) (\ell+5)}{10 (2 \ell+5) (2 \ell+9)}f^{(\pm2)}_{\ell+6,\ell+2}(z,\bar z)\\
   &+\tfrac{(\ell+3) (\ell+4) (\ell+5) (\ell+6)}{16 (2 \ell+7) (2 \ell+9)^2 (2 \ell+11)}f^{(\pm2)}_{\ell+8,\ell+4}(z,\bar z)\Big]h_{[0,2,0]}^{(\pm2)}(\alpha , \bar \alpha)\\
    &+ \Big[f^{(\pm2)}_{\ell+5,\ell+1}(z,\bar z)+\tfrac{(\ell+3) (\ell+5)}{4 (2 \ell+7) (2 \ell+9)}f^{(\pm2)}_{\ell+7,\ell+3}(z,\bar z)\Big]h_{[1,2,1]}^{(\pm2)}(\alpha , \bar \alpha) \\
    &-f^{(\pm2)}_{\ell+6,\ell+2}(z,\bar z) h_{[0,4,0]}^{(\pm2)}(\alpha , \bar \alpha)
\end{aligned}$ 
& $0$
 \\
\hline
$\mathcal{C}_{[1,0,1],\ell}$ 
& $0$ 
& $0$ \\
\hline

$\mathcal{A}^{\Delta}_{[0,0,0],\ell}$ 
& $0$  
& $0$ \\
\hline

\end{tabular}
\renewcommand{\arraystretch}{1}
\caption{Bulk superconformal blocks for the $q=\pm2,\pm4$ sectors.}
\label{tab:blocksq}
\end{center}
\end{table}

\subsection{Defect channel}
\label{app:defect-superblocks}
In this section we review the representation theory of the preserved superconformal algebra, we derive the bosonic defect blocks appropriate to a surface
defect that preserves only the diagonal generator
$\mathcal{R}_{-}$ and describe the construction of the corresponding superconformal blocks.
\subsubsection*{Representation theory of $\mathfrak{psu}(1,1|2)\oplus\mathfrak{psu}(1,1|2)\oplus \mathfrak{u}(1)_{-}$} \label{app:irreppsu}
We summarize the representations of the superalgebra $\mathfrak{psu}(1,1|2)\oplus\mathfrak{psu}(1,1|2)\oplus \mathfrak{u}(1)_-$ that are relevant for the defect spectrum. For a single copy of $\mathfrak{psu}(1,1|2)$, the representation theory can be found for instance in \cite{Agmon:2020pde}. Here we simply extend their discussion to the case of a surface defect. The algebra is given in appendix \ref{app:conv}. Highest-weight representations are labeled by
\begin{equation}
 [h,\bar h,r,\bar r]_{r_-}\, .
\end{equation}
Since $\mathfrak{u}(1)_-$ is a central extension, each supermultiplet will be labeled by a particular value of $r_-$. 
The long supermultiplet $\mathcal{\hat A}_{[h,\bar h,r,\bar r]_{r_-}}$ contains $2^8$ states, obtained by acting with the eight Poincar\'e supercharges on a generic superprimary. Positivity of the norms of the level-one descendants, using the anticommutators in \eqref{QS}, gives
\begin{equation}
 h-r\geq \frac{|r_-|}{2}\, ,
 \qquad
 \bar h-\bar r\geq \frac{|r_-|}{2}\, .
\end{equation}
A representation is long when both inequalities are strict. When one or both bounds are saturated, the possible shortening conditions depend on the sign of the central charge $r_-$. For $r_->0$ they are
\begin{center}
\begin{tabular}{|c|c|c|}
\hline
\textbf{Name} & \textbf{Superprimary} & BPS  \\
\hline
$ \mathcal{\hat B}^{1/4,0}$ & $[r+\frac{r_-}{2},\bar h,r,\bar r]$ & $1/8$ \\
\hline
$ \mathcal{\hat B}^{0,1/4}$ & $[h,\bar r+\frac{r_-}{2},r,\bar r]$ &$1/8$ \\
\hline
$ \mathcal{\hat B}^{1/2,0}$ & $[\frac{r_-}{2},\bar h,0,\bar r]$ &$1/4$  \\
\hline
$ \mathcal{\hat B}^{0,1/2}$ & $[h,\frac{r_-}{2},r,0]$ &$1/4$ \\
\hline
$ \mathcal{\hat B}^{1/4,1/4}$ & $[r+\frac{r_-}{2},\bar r+\frac{r_-}{2},r,\bar r]$ & $1/4$ \\
\hline
$ \mathcal{\hat B}^{1/2,1/4}$ & $[\frac{r_-}{2},\bar r+\frac{r_-}{2},0,\bar r]$ & $3/8$ \\
\hline
$ \mathcal{\hat B}^{1/4,1/2}$ & $[r+\frac{r_-}{2},\frac{r_-}{2},r,0]$ & $3/8$ \\
\hline
$ \mathcal{\hat B}^{1/2,1/2}$ & $[\frac{r_-}{2},\frac{r_-}{2},0,0]$ & $1/2$  \\
\hline
\end{tabular}
\end{center}
For $r_-<0$ we have the conjugated multiplets

\begin{center}
\begin{tabular}{|c|c|c|}
\hline
\textbf{Name} & \textbf{Superprimary} & BPS \\
\hline
$ \hat{\bar{\mathcal{B}}}^{1/4,0}$ & $[r-\frac{r_-}{2},\bar h,r,\bar r]$ & $1/8$ \\
\hline
$ \hat{\bar{\mathcal{B}}}^{0,1/4}$ & $[h,\bar r-\frac{r_-}{2},r,\bar r]$ &$1/8$ \\
\hline
$ \hat{\bar{\mathcal{B}}}^{1/2,0}$ & $[-\frac{r_-}{2},\bar h,0,\bar r]$ &$1/4$ \\
\hline
$ \hat{\bar{\mathcal{B}}}^{0,1/2}$ & $[h,-\frac{r_-}{2},r,0]$ &$1/4$  \\
\hline
$ \hat{\bar{\mathcal{B}}}^{1/4,1/4}$ & $[r-\frac{r_-}{2},\bar r-\frac{r_-}{2},r,\bar r]$ & $1/4$ \\
\hline
$ \hat{\bar{\mathcal{B}}}^{1/2,1/4}$ & $[-\frac{r_-}{2},\bar r-\frac{r_-}{2},0,\bar r]$ & $3/8$ \\
\hline
$ \hat{\bar{\mathcal{B}}}^{1/4,1/2}$ & $[r-\frac{r_-}{2},-\frac{r_-}{2},r,0]$ & $3/8$ \\
\hline
$ \hat{\bar{\mathcal{B}}}^{1/2,1/2}$ & $[-\frac{r_-}{2},-\frac{r_-}{2},0,0]$ & $1/2$  \\
\hline
\end{tabular}
\end{center}

Finally, for $r_-=0$ we have the self-conjugate multiplets
\begin{center}
\begin{tabular}{|c|c|c|}
\hline
\textbf{Name} & \textbf{Superprimary} & BPS \\
\hline
$ \mathcal{\hat B}_0^{1/2,0}$ & $[r,\bar h,r,\bar r]$ & $1/4$ \\
\hline
$ \mathcal{\hat B}_0^{0,1/2}$ & $[h,\bar r,r,\bar r]$ & $1/4$ \\
\hline
$ \mathcal{\hat B}_0^{1/2,1/2}$ & $[r,\bar r,r,\bar r]$ & $1/2$ \\
\hline
\end{tabular}
\end{center}

For the defect OPE of a chiral primary $\mathcal{O}_p$ in the $[0,p,0]$ representation of $SO(6)_R$, the relevant representations have $h=\bar h=\hat\Delta/2$ and $r=\bar r$. These are $\mathcal{\hat B}^{1/4,1/4}$, $\mathcal{\hat B}^{1/2,1/2}$, $\hat{\bar{\mathcal{B}}}^{1/4,1/4}$, $\hat{\bar{\mathcal{B}}}^{1/2,1/2}$ and $\hat{\mathcal{B}}_0^{1/2,1/2}$. In the main text, we suppress the superscript indicating the BPS conditions and use the notation
\begin{equation}\label{notationdefectmult}
    \mathcal{\hat B}_{r,r_-}\colonequals \left\{\begin{array}{ll}
    \mathcal{\hat B}^{1/4,1/4}_{[r+\frac{r_-}{2},r+\frac{r_-}{2},r,r]} & r_->0\\
    \mathcal{\hat B}_{0,[r,r,r,r]}^{1/2,1/2} & r_-=0\\
    \hat{\bar{\mathcal B}}^{1/4,1/4}_{[r-\frac{r_-}{2},r-\frac{r_-}{2},r,r]} & r_-<0
    \end{array}\right. \, , \qquad 
    \mathcal{\hat B}_{0,r_-}\colonequals \left\{\begin{array}{ll}
    \mathcal{\hat B}^{1/2,1/2}_{[\frac{r_-}{2},\frac{r_-}{2},0,0]} & r_->0\\
    \hat{\bar{\mathcal{B}}}^{1/2,1/2}_{[-\frac{r_-}{2},-\frac{r_-}{2},0,0]} & r_-<0
    \end{array}\right.  \, .
\end{equation}
For the long supermultiplet we use the notation $\mathcal{\hat A}^{\hat{\Delta}}_{r,r_-}\equiv\mathcal{\hat A}_{[\hat{\Delta}/2,\hat{\Delta}/2,r, r]_{r_-}}$.

\subsubsection*{Bulk-to-defect two-point function}\label{bulkdefect}

When $\mathcal{R}_+$ is broken, the transverse spin $s$ and $r_\perp$ are not good quantum numbers for a defect operator.
Defect operators are instead labeled by
\begin{equation}
  r_-=r_\perp-s\, .
  \label{eq:rminus-definition-blocks}
\end{equation}
The covariant bulk-to-defect two-point function can be written as \eqref{eq:bulk-defect-broken-rotations} and the coefficient   $B_{\hat{\Delta},n,r_-}(\eta)$ is a function of the parameter $\eta$ defined in \eqref{etatheta}
\begin{equation}\label{eq:B-eta-corrected}
 B_{\hat{\Delta},n,r_-}(\eta)=\sum_{m=0}^{p-n}
    b_{\hat{\Delta},n,r_-,p-n-2m}\,
    e^{i(p-n-2 m)\eta}\, ,
\end{equation}
with $n\leq p$. For $p=2$ we have 
\begin{align}
    B_{\hat{\Delta},2,r_-}(\eta)&=
    b_{\hat{\Delta},2,r_-,0} \qquad B_{\hat{\Delta},1,r_-}(\eta)=
    b_{\hat{\Delta},1,r_-,1} e^{i\eta}+b_{\hat{\Delta},1,r_-,-1} e^{-i\eta}  \nonumber\\
    B_{\hat{\Delta},0,r_-}(\eta)&=
    b_{\hat{\Delta},0,r_-,2} e^{2i\eta}+b_{\hat{\Delta},0,r_-,0} + b_{\hat{\Delta},0,r_-,-2} e^{-2i\eta}\, .
\end{align}
When the rotational symmetry is restored, each mode in the function $B(\eta)$ would correspond to a specific $r_+$ charge of the defect operator. The choice of the prefactor in \eqref{eq:bulk-defect-broken-rotations} is such that the coefficient of $\eta$ in the exponent of \eqref{eq:B-eta-corrected} is precisely the $r_{\perp}$ charge  of the operator appearing in the defect OPE when the symmetry is restored. Indeed, a single Fourier mode in \eqref{eq:B-eta-corrected}, together with the phase $e^{-ir_-\theta}$ that accompanies $B_{\hat{\Delta},n,r_-}(\eta)$ in the defect OPE, gives
\begin{equation}
  e^{-ir_-\theta}e^{i r_{\perp}\eta}
  =
  e^{i(r_\perp-r_-)\theta}e^{ir_\perp\theta_R}=  e^{is\theta}e^{ir_\perp\theta_R}.
  \label{eq:single-fourier-mode}
\end{equation}
Alternative choices in the prefactor would induce a dependence on $r_-$ in \eqref{eq:B-eta-corrected}.

\subsubsection*{Bosonic defect blocks}\label{app:bosonicdefectblocks}

Consider the exchange of a defect primary of charge $r_-$ between the two
bulk operators. The two defect OPEs contain the phases
\begin{equation}
  e^{-ir_-\theta_1},
  \qquad
  e^{+ir_-\theta_2},
\end{equation}
because the second OPE involves the conjugate defect operator.
Their product is\footnote{Notice that the factor $w^{-r_-}$ in \eqref{eq:corrected-defect-bosonic-block} has the correct
rotation-preserving limit. Indeed, a contribution with definite
$R_\perp$ charge $r_\perp$ satisfies
\begin{equation}
  w^{-r_-}(ww_R)^{r_\perp}
  =
  w^{r_\perp-r_-}w_R^{r_\perp}
  =
  w^s w_R^{r_\perp}\, ,
  \qquad
  s=r_\perp-r_-\, ,\nn 
\end{equation}
reproducing the ordinary product of the transverse spacetime and
R-symmetry blocks.}
\begin{equation}
  e^{-ir_-(\theta_1-\theta_2)}
  =w^{-r_-}.
  \label{eq:w-rminus-origin}
\end{equation}
It follows that the bosonic block for a defect primary of dimension $\hat{\Delta}$, $SO(4)_\mathcal{R}$ representation $[n/2,n/2]$ and charge $r_-$ is
\begin{equation}
  \widehat g_{\widehat\Delta,n,r_-}(r,r_R,w)
  =
  f_{\widehat\Delta}(r)\,
  f_{-n}(r_R)\,
  w^{-r_-}.
  \label{eq:corrected-defect-bosonic-block}
\end{equation}
where
\begin{equation}
  f_a(x)
  =
  x^a\,{}_2F_1(a,1;a;x^2)\, ,
  \label{eq:defect-radial-function}
\end{equation}
with the R-symmetry solution $f_{-n}$ understood through the appropriate
polynomial continuation of the Casimir equation. In particular, the
simplification
\begin{equation}
  {}_2F_1(a,1;a;x^2)=\frac{1}{1-x^2}
\end{equation}
should not be applied naively when $a$ is a non-positive integer. In section \ref{sec:holo-witten} we also use the R-symmetry block
\begin{equation}
    \hat h^{(0)}_{[n/2,n/2]_m}(r_R, w_R)
    = f_{-n}(r_R)\, w_R^m\,.
\end{equation}
The defect block decomposition of the two-point function of $\mathcal{O}_p$ takes the form \cite{Holguin:2025bfe}
\begin{equation}\label{defblockOPE}
    \mathcal{F}(r,r_R,w,w_R,\vartheta)
    = \sum_{\hat{\mathcal{O}}}
  B_{\hat{\Delta},n,r_-}(\eta_1)B_{\hat{\Delta},n,-r_-}(\eta_2)\,  \hat{g}_{\hat{\Delta},n,r_-}(r,r_R,w)\,.
\end{equation}
The product $B_{\hat\Delta,n,r_-}(\eta_1)B_{\hat\Delta,n,-r_-}(\eta_2)$ is a finite Laurent polynomial in the variable $\omega=w w_R$, introduced in \eqref{udef}, and a finite Fourier polynomial in $\vartheta=\eta_1+\eta_2$. In particular, we can write it as
\begin{equation}\label{Beta1Beta2}
\begin{aligned}
B_{\hat\Delta,n,r_-}(\eta_1)\,B_{\hat\Delta,n,-r_-}(\eta_2)=
\sum_{q}
\sum_{m=0}^{p-n-\frac{|q|}{2}}
e^{iq\frac{\vartheta}{2}}
\omega^{2m+n-p+\frac{|q|}{2}}
\widetilde b^{\,2}_{\hat\Delta,n,r_-,q,m}\, ,
\end{aligned}
\end{equation}
where $\widetilde b^{\,2}_{\hat\Delta,n,r_-,q,m}$ can be
determined from the products of the coefficients in \eqref{eq:B-eta-corrected}. We define the shorthand notation
\begin{equation}\label{B2}
     \texttt{B}^{(q)}_{\hat{\Delta},n,r_-}\!(\omega)
     =
     \sum_{m=0}^{p-n-\frac{|q|}{2}}
     \omega^{\,2m+n-p+\frac{|q|}{2}}
     \widetilde b^{\,2}_{\hat\Delta,n,r_-,q,m}\,,
\end{equation}
which is used in the main text for compactness.
From \eqref{Beta1Beta2}, we see that operators transforming in the
$[\frac n2,\frac n2]$ representation of $SO(4)_\mathcal{R}$ contribute
to the even Fourier sectors $|q|\leq2(p-n)$.
Consequently, the contribution to the Fourier sector of charge $q$ is
\begin{equation}
\begin{aligned}
\mathcal F^{(q)}
=
\!\!\!\sum_{\substack{\hat\Delta,n,r_-\\
                0\leq n\leq p-\frac{|q|}{2}}}
\!\!\!\!\!\!\sum_{m=0}^{p-n-\frac{|q|}{2}}
\,b_{\hat\Delta,n,r_-,\,2m+n-p+\frac{q+|q|}{2}}
b_{\hat\Delta,n,-r_-,\,-2m-n+p+\frac{q-|q|}{2}}
\omega^{2m+n-p+\frac{|q|}{2}}
\widehat g_{\hat\Delta,n,r_-}.
\end{aligned}\label{eq:general-bosonic-defect-expansion}
\end{equation}
For $p=2$, the neutral sector is
\begin{align}\label{eq:defect-expansion-q0}
  \mathcal F^{(0)}
  ={}&
  \sum_{\widehat\Delta,r_-}
  \bigg[
    b_{\widehat\Delta,0,r_-,0}\,
    b_{\widehat\Delta,0,-r_-,0}
    +
    b_{\widehat\Delta,0,r_-,2}\,
    b_{\widehat\Delta,0,-r_-,-2}\,\omega^2
   \nonumber\\
   &+
    b_{\widehat\Delta,0,r_-,-2}\,
    b_{\widehat\Delta,0,-r_-,2}\,\omega^{-2}
  \bigg]
  \widehat g_{\widehat\Delta,0,r_-}
  \\
  &+
  \sum_{\widehat\Delta,r_-}
  \bigg[
    b_{\widehat\Delta,1,r_-,1}\,
    b_{\widehat\Delta,1,-r_-,-1}\,\omega
    +
    b_{\widehat\Delta,1,r_-,-1}\,
    b_{\widehat\Delta,1,-r_-,1}\,\omega^{-1}
  \bigg]
  \widehat g_{\widehat\Delta,1,r_-}
  \nonumber\\
  &+
  \sum_{\widehat\Delta,r_-}
  b_{\widehat\Delta,2,r_-,0}\,
  b_{\widehat\Delta,2,-r_-,0}\,
  \widehat g_{\widehat\Delta,2,r_-}\, ,\nonumber
\end{align}
where we have suppressed the dependence on the cross-ratios to avoid cluttering.
The charge-two sectors are
\begin{eqnarray}
    \mathcal F^{(+2)}
  &=&
  \!\sum_{\widehat\Delta,r_-}
  \bigg[
    b_{\widehat\Delta,0,r_-,2}\,
    b_{\widehat\Delta,0,-r_-,0}\,\omega
    +
    b_{\widehat\Delta,0,r_-,0}\,
    b_{\widehat\Delta,0,-r_-,2}\,\omega^{-1}
  \bigg]
  \widehat g_{\widehat\Delta,0,r_-}
  \nonumber\\
  &+&
  \sum_{\widehat\Delta,r_-}
  b_{\widehat\Delta,1,r_-,1}\,
  b_{\widehat\Delta,1,-r_-,1}\,
  \widehat g_{\widehat\Delta,1,r_-},
  \label{eq:defect-expansion-qplus2}
  \\[2mm]
  \mathcal F^{(-2)}
  &=&
  \sum_{\widehat\Delta,r_-}
  \bigg[
    b_{\widehat\Delta,0,r_-,0}\,
    b_{\widehat\Delta,0,-r_-,-2}\,\omega
    +
    b_{\widehat\Delta,0,r_-,-2}\,
    b_{\widehat\Delta,0,-r_-,0}\,\omega^{-1}
  \bigg]
  \widehat g_{\widehat\Delta,0,r_-}
  \nonumber\\
  &+&
  \sum_{\widehat\Delta,r_-}
  b_{\widehat\Delta,1,r_-,-1}\,
  b_{\widehat\Delta,1,-r_-,-1}\,
  \widehat g_{\widehat\Delta,1,r_-}.
  \label{eq:defect-expansion-qminus2}
\end{eqnarray}

Finally, the maximally charged sectors are
\begin{align}
  \mathcal F^{(+4)}
  &=
  \sum_{\widehat\Delta,r_-}
  b_{\widehat\Delta,0,r_-,2}\,
  b_{\widehat\Delta,0,-r_-,2}\,
  \widehat g_{\widehat\Delta,0,r_-},
  \nonumber\\
  \mathcal F^{(-4)}
  &=
  \sum_{\widehat\Delta,r_-}
  b_{\widehat\Delta,0,r_-,-2}\,
  b_{\widehat\Delta,0,-r_-,-2}\,
  \widehat g_{\widehat\Delta,0,r_-}.
  \label{eq:defect-expansion-q4}
\end{align}
The sum over $r_-$ includes both signs. For each operator with charge $r_-$ a conjugate operator with charge $-r_-$ is exchanged with the same OPE coefficient, but with a conjugate block $\widehat{g}_{\widehat{\Delta},n,-r_-}$.
When the terms with charges $r_-$ and $-r_-$ are grouped together, their
tensor structures must remain correlated with the sign of the charge.
They therefore appear in combinations of the form
\begin{equation}
  \omega^k\,
  \widehat g_{\widehat\Delta,n,r_-}
  +
  \omega^{-k}\,
  \widehat g_{\widehat\Delta,n,-r_-},
  \qquad r_->0,
  \label{eq:correlated-bosonic-pair}
\end{equation}
with the appropriate products of bulk-to-defect coefficients multiplying
the two terms.

\begin{table}[h!]
\centering
\renewcommand{\arraystretch}{1.8}
\begin{tabular}{|Sc|Sc|}
\hline
$\text{multiplet}$
&
$\widehat{\mathcal G}_{\widehat{\mathcal M}}
(z,\bar z,\alpha,\bar\alpha)$
\\
\hline\hline
$\hat{\id}$
&
$1$
\\
\hline\hline

\multicolumn{2}{|c|}{$r_-=0$}
\\
\hline

$\widehat{\mathcal B}_{1/2,0}$
&
$\displaystyle
\frac{
  (z\bar z)^{\frac12}
  \left(
    1-\alpha\bar z-z\bar\alpha+\alpha\bar\alpha
  \right)
}{
  (\alpha\bar\alpha)^{\frac12}(1-z\bar z)
}$
\\
\hline

$\widehat{\mathcal B}_{1,0}$
&
$\displaystyle
\frac{
  z\bar z
}{
  \alpha\bar\alpha(1-z\bar z)
}
\left[
  1-z\bar\alpha
  -\alpha(\bar z-\bar\alpha)
  \left(
    1-z\bar\alpha+\alpha\bar\alpha
  \right)
\right]$
\\
\hline\hline

\multicolumn{2}{|c|}{$r_->0$}
\\
\hline

$\widehat{\mathcal B}_{0,r_-}$
&
$\displaystyle
\frac{
  \bar z^{r_-}
  (1-\alpha\bar z)
  (\bar\alpha-\bar z)
}{
  \bar\alpha(1-z\bar z)
}$
\\
\hline

$\widehat{\mathcal B}_{1/2,r_-}$
&
$\displaystyle
\frac{
  z^{\frac12}\bar z^{r_-+\frac12}
  (1-\alpha\bar z)
  (\bar\alpha-\bar z)
  (1-z\bar\alpha+\alpha\bar\alpha)
}{
  \alpha^{\frac12}\bar\alpha^{\frac32}
  (1-z\bar z)
}$
\\
\hline
$\widehat{\mathcal A}^{\widehat\Delta}_{0,r_-}$
&
$\displaystyle
\frac{
  z^{\frac{\widehat\Delta-r_-}{2}}
  \bar z^{\frac{\widehat\Delta+r_-}{2}}
  (\alpha-z)
  (1-\alpha\bar z)
  (\bar\alpha-\bar z)
  (1-z\bar\alpha)
}{
  \alpha\bar\alpha(1-z\bar z)
}$
\\
\hline

\end{tabular}
\renewcommand{\arraystretch}{1}
\caption{Defect superconformal blocks for $r_-\geq0$;
the blocks with negative $r_-$ are obtained by charge conjugation. The
$r_-=0$ entries are self-conjugate.}
\label{tab:defblocks}
\end{table}

\subsubsection*{Defect superconformal blocks}

\label{sec:defect-superblocks}

The two-point function admits an expansion in defect superconformal blocks,
\begin{equation}
\mathcal F(z,\bar z,\alpha, \bar \alpha,\vartheta)
= \sum_q \sum_{\hat{\mathcal{O}}}\texttt{B}_{\hat{\mathcal{O}}}^{(q)}(\omega) e^{i q \frac{\vartheta}{2}}\widehat{\mathcal G}_{\widehat{\mathcal{M}}}(z,\bar z,\alpha, \bar \alpha)\,,
\end{equation}
where $\hat{\mathcal{O}}$ is the superprimary of the multiplet $\widehat{\mathcal{M}}$ and the corresponding coefficient $\texttt{B}_{\hat{\mathcal{O}}}^{(q)}(\omega)$ is defined in \eqref{B2}.

A defect supermultiplet contains several bosonic conformal primaries with
different dimensions and $SO(4)_\mathcal{R}$ representations. Since $\mathcal{R}_-$ is
central in the preserved superalgebra, all conformal primaries in a given
supermultiplet carry the same value of $r_-$. A defect superconformal
block with fixed charge $r_-$ can therefore be written as
\begin{equation}
  \widehat{\mathcal G}_{\widehat{\mathcal M}}
  =
  \sum_I
  c_I(\omega)\,
  \widehat g_{\widehat\Delta_I,n_I,r_-}(r,r_R,w)\, ,
  \label{eq:defect-superblock-ansatz}
\end{equation}
where $I$ runs over the bosonic conformal primaries contained in the
supermultiplet $\widehat{\mathcal M}$. The functions
$\widehat g_{\widehat\Delta_I,n_I,r_-}$ are the bosonic blocks in
\eqref{eq:corrected-defect-bosonic-block}, while the coefficients $c_I(\omega)$ are
finite Laurent polynomials. Their powers of $\omega$ account for the different
$r_+$ charges of the conformal primaries inside the supermultiplet. Even when $\mathcal{R}_+$ is not preserved, it is still an outer automorphism of the preserved superalgebra, therefore the $r_+$ charge of the superdescendants relative to the superprimary is well defined.

Notice that
\begin{equation}
  \left.
  (\partial_z+\partial_\alpha)\,\omega
  \right|_{\alpha=z}
  =0\, ,
  \qquad
  \left.
  (\partial_{\bar z}+\partial_{\bar\alpha})\,\omega
  \right|_{\bar\alpha=\bar z}
  =0\, .
  \label{eq:u-Ward-identities}
\end{equation}
The Ward identities therefore do not determine an arbitrary overall
function of $\omega$. For a single power $\omega^{k}$, the exponent $k$ would be fixed by the $r_+$ charge of the superprimary if $\mathcal{R}_+$ were preserved. We fix this overall power by requiring that the superprimary of the multiplet appears with its bosonic block, i.e. $c_1(\omega)=1$.

Rewriting the ansatz as a function of $(z,\bar z,\alpha,\bar\alpha)$ we determine the unknown coefficients by imposing
\begin{equation}
  \left.
  (\partial_z+\partial_\alpha)
  \widehat{\mathcal G}_{\widehat{\mathcal M}}
  \right|_{\alpha=z}
  =0\, ,
  \qquad
  \left.
  (\partial_{\bar z}+\partial_{\bar\alpha})
  \widehat{\mathcal G}_{\widehat{\mathcal M}}
  \right|_{\bar\alpha=\bar z}
  =0\, .
  \label{eq:notebook-defect-Ward-identities}
\end{equation}
The result of this analysis leads to the blocks in table \ref{tab:defblocks} for $r_-\geq0$.\footnote{The $\widehat{\mathcal B}_{1,r_-}$ superblock with $r_->0$ is absent from the table, since the corresponding multiplet contains a conformal primary transforming in the $[\frac{3}{2},\frac{3}{2}]$ representation of $SO(4)_{\mathcal R}$, which is excluded by group theory.} The conjugate blocks with $r_-<0$ are obtained from those with $r_->0$ through
\begin{equation}
  (z,\bar z,\alpha,\bar\alpha,r_-)
  \longleftrightarrow
  (\bar z,z,\bar\alpha,\alpha,-r_-)\, .
  \label{eq:defect-block-charge-conjugation}
\end{equation}

\section{Weak coupling result}\label{app:weak}
In this appendix we collect the weak-coupling results for the specific case of a Gukov-Witten surface defect. These expressions were originally derived in \cite{Holguin:2025bfe}. At leading order in the small $g_{\mathrm{YM}}$ expansion, the two-point function of $\mathcal{O}_p$ in the presence of a Gukov-Witten defect is given by\footnote{The expression reported here differs from \cite{Holguin:2025bfe} by an overall normalization factor due to different conventions, see \eqref{CPO} and equation $(7.1)$ of \cite{Holguin:2025bfe}. Moreover, we have expressed the result in terms of the cross-ratio  $\xi=\chi_{\text{there}}-2\cos{\phi}$, and included the disconnected contribution $(-2)^{-p}\tfrac{\xi_R^p}{\xi^p}+A_{\mathcal{O}_p}(\eta_1)A_{\mathcal{O}_p}(\eta_2)$.}
\begin{eqnarray}\label{Weakcoupling}
\mathcal{F}(\xi,\phi,\xi_R,\phi_R,\vartheta)
&=&
-\frac{p \, (4\pi^2)^{p-1}}{2 \, N^p} g_{\mathrm{YM}}^{2-2p}  \Bigg[\, \sum_{k_1,k_2=0}^{p-1}\!\!\! C(k_1,k_2)\,
e^{i k_1' \eta_1} e^{i k_2' \eta_2}\, \xi_R
\nonumber \\
& \times &
\!\!\sum_{s=-\infty}^{\infty}\!\!
e^{i \, s\,\phi}\,
(\xi+2\cos{\phi})^{-(1+|s|)}\,
{}_2F_1\!\left(
\tfrac{|s|+1}{2},
\tfrac{|s|}{2}+1;
|s|+1;
\tfrac{4}{(\xi+2\cos{\phi})^2}
\right) \nonumber\\
&+&\mathcal{O} (g^2_{\mathrm{YM}}) \Bigg]+(-2)^{-p}\frac{\xi_R^p}{\xi^p}+A_{\mathcal{O}_p}(\eta_1)A_{\mathcal{O}_p}(\eta_2)\, ,
\end{eqnarray}
where $k'_i=p-2k_i-1$, $\eta_1=\tfrac{\vartheta+\phi+\phi_R}{2}$, $\eta_2=\tfrac{\vartheta-\phi-\phi_R}{2}$ and\footnote{The factors of $\sqrt{2}$ originate from the different normalization of the defect profile in \eqref{classprof} relative to equation $(2.5)$ of \cite{Holguin:2025bfe}.} 
\begin{equation}
C(k_1,k_2) = \binom{p-1}{k_1} \binom{p-1}{k_2} \sum_{l=1}^M N_l \left(\frac{\beta_l-i\gamma_l}{\sqrt{2}}\right)^{2p-k_1-k_2-2} \left(\frac{\beta_l+i\gamma_l}{\sqrt{2}}\right)^{k_1+k_2}\,.
\end{equation}
Here $\beta_l$ and $\gamma_l$ are the parameters entering the
definition of the defect \eqref{classprof}.
The correlator \eqref{Weakcoupling} can be further simplified by applying the identity
\begin{equation}
    \sum_{s=-\infty}^{\infty}
e^{i \, s\,\phi}\,
(\xi+2\cos{\phi})^{-(1+|s|)}\,
{}_2F_1\!\left(
\frac{|s|+1}{2},
\frac{|s|}{2}+1;
|s|+1;
\frac{4}{(\xi+2\cos{\phi})^2}
\right) = \xi^{-1}\,,
\end{equation}
which can be derived using an integral representation of the hypergeometric function and exchanging the order of summation and integration.

Specializing to the $p = 2$ case, in lightcone coordinates the two-point function takes the form 
\begin{eqnarray}\label{O2weak}
\mathcal{F}(z,\bar z,\alpha,\bar \alpha,\vartheta)&=& \sum_q e^{iq\frac{\vartheta}{2}} \, \mathcal{F}^{(q)}(z,\bar z,\alpha,\bar \alpha)\,, \quad q=0,\pm2,\pm4\,, \nonumber\\
    \mathcal{F}^{(0)}(z,\bar z,\alpha,\bar \alpha)&=& -\frac{2 \pi^2}{g^2_{\mathrm{YM}} N^2} \left[\sum_{l=1}^M N_l(\beta^2_l+\gamma^2_l)\right]\frac{\bar \alpha z+\alpha  \bar z}{\alpha \bar \alpha} \frac{(1-\alpha ) (1-\bar \alpha)}{(1-z) (1-\bar z)}\nonumber\\
    &+&\left[a^{(0)}_{\mathcal{O}_2}\right]^2+a^{(+2)}_{\mathcal{O}_2} a^{(-2)}_{\mathcal{O}_2} \left(\frac{\alpha \bar z}{\bar{\alpha} z}\!+\!\frac{\bar{\alpha} z}{\alpha \bar z}\right)+\mathcal{O} (g^0_{\mathrm{YM}})\,, \\
    \mathcal{F}^{(\pm2)}(z,\bar z,\alpha,\bar \alpha)&=&-\frac{2 \pi^2}{g^2_{\mathrm{YM}} N^2} \left[\sum_{l=1}^M N_l(\beta_l\mp i\gamma_l)^2\right] \frac{(1-\alpha ) (1-\bar \alpha) \sqrt{z \bar z}}{(1-z) (1-\bar z) \sqrt{\alpha  \bar \alpha}}\nonumber\\
    &+&a^{(0)}_{\mathcal{O}_2} a^{(\pm 2)}_{\mathcal{O}_2} \left(\sqrt{\frac{\bar{\alpha} z}{\alpha \bar z}}\!+\!\sqrt{\frac{\alpha \bar z}{\bar{\alpha} z}}\ \right)+\mathcal{O} (g^0_{\mathrm{YM}})\,, \nonumber \\
    \mathcal{F}^{(\pm4)}(z,\bar z,\alpha,\bar \alpha)&=&\left[a^{(\pm 2)}_{\mathcal{O}_2}\right]^2+\mathcal{O} (g^0_{\mathrm{YM}})\,,\nonumber 
\end{eqnarray}
where the one-point function coefficients are given explicitly in \eqref{a2broken}.

The results for the $q=\pm 2, \pm 4$ sectors of \eqref{O2weak} match the bootstrap predictions \eqref{chargedres} upon substituting the explicit expression for the coefficient $a\lambda_{\mathcal{B}_{[0,2,0]}}^{(q)}$, provided in \eqref{lambdaO2} and \eqref{a2broken}. The holomorphic functions $f^{(q)}(z,\alpha)$ can be extracted by comparing \eqref{O2weak} with the general solution to the Ward identities \eqref{wardsolution}. For the $q = 0$ sector, this gives
\begin{eqnarray}\label{smallfweak}
    f^{(0)}(z,\alpha)&=& \left[a^{(0)}_{\mathcal{O}_2}\right]^2+a^{(+2)}_{\mathcal{O}_2} a^{(-2)}_{\mathcal{O}_2} \left(\frac{\alpha}{z}+\frac{ z}{\alpha}\right)\nonumber\\
    &-&\frac{2 \pi^2}{g^2_{\mathrm{YM}} N^2}\left[\sum_{l=1}^M N_l\!\left(\beta^2_l+\gamma^2_l\right)\!\right]\frac{(\alpha-1)(\alpha+z)}
    {\alpha(z-1)}+\mathcal{O}(g_{\mathrm{YM}}^0)\,.
\end{eqnarray}
This expression exactly reproduces the weak-coupling expansion of the analytic bootstrap result \eqref{smallf0}.

\section{Witten diagrams}\label{app:witten}
In this appendix we collect the explicit expressions for the Witten diagrams entering the holographic computation of section \ref{sec:holo-witten}, see figure \ref{fig:diagrams}. Most of these expressions were originally derived in \cite{Gimenez-Grau:2023fcy} for holographic defects of arbitrary dimension $p_\text{\tiny{def}}$, dual to probe branes wrapping $AdS_{p_\text{\tiny{def}}+1}\subset AdS_{d+1}$. We summarize them here and specialize to the setup considered in this work ($p_\text{\tiny{def}}=2$, identical external operators), in order to establish our conventions and keep the presentation self-contained.

We define the bulk-boundary propagator as
\begin{equation}
 K_{\Delta}\left(X,Z\right)
 = \left( \frac{Z^0}{\left(Z-X\right)^2} \right)^{\Delta}\,,
\end{equation}
where $X$ denotes the coordinates of a point on the conformal boundary of $AdS_5$, while $Z=\left(Z^0,Z^a,Z^i\right)$ denotes a point in the bulk of $AdS_5$. The indices $a$ and $i$ label directions parallel and orthogonal to the probe brane, respectively. Throughout this appendix, the terms ``bulk'' and ``boundary'' always refer to the holographic $AdS$ geometry, rather than to the position of CFT operators with respect to the defect on the boundary. The bulk-bulk propagator is given by \cite{DHoker:1999mqo}
\begin{align}
  G_{\Delta}\left(Z_1,Z_2\right)
& = \frac{\mathcal{C}_\Delta \left(Z^0_1 Z^0_2\right)^{\Delta}}
         {(Z_1 - Z_2)^{2\Delta}} \,
    {}_2F_1 \left(
    \Delta, \Delta-\frac d2+\frac12, 2\Delta - d + 1,
    -\frac{4 Z^0_1 Z^0_2}{(Z_1 - Z_2)^2}
    \right) \, , \\
  \mathcal{C}_\Delta
& = \frac{\Gamma(\Delta)}{2 \pi^{d/2} \Gamma(\Delta-d/2+1)} \, .
\end{align}

The simplest diagram appearing in figure \ref{fig:diagrams} is the contact diagram
\begin{align}
  \begin{tikzpicture}[valign]
    \tikzstyle{every node}=[font=\small]
    \pgfmathsetmacro{\x}{sqrt(2)/2}
    \pgfmathsetmacro{\y}{sqrt(2)/2}
    \pgfmathsetmacro{\z}{0.5}
    \pgfmathsetmacro{\r}{-0.75}
    \draw [thick] (0,0) circle [radius=1];
    \draw [thick, blue] (+\x,+\y) to[out=240,in=120] (+\x,-\y);
    \draw [dashed]      (-\x,+\y) -- (\z,0);
    \draw [dashed]      (-\x,-\y) -- (\z,0);
    \node at (-0.95, 0.85) {$X_1$};
    \node at (-0.95,-0.95) {$X_2$};
    \node at (-\r,0.05) {$\hat{Z}$};
    \node at ( 0.3, 0.8) [anchor=north,rotate=-30] {$\Delta_1$};
    \node at (-0.4, 0.15)[anchor=north,rotate= 30] {$\Delta_2$};
  \end{tikzpicture}
 \;\; = \;\;
 \int_{\hat{Z}} 
 K_{\Delta_1}\left(X_1, \hat{Z}\right) K_{\Delta_2}\left(X_2, \hat{Z}\right)
 \; = \;
 \frac{C_{\Delta_1\Delta_2}(z,\bar z)}
      {\left|X^i_1\right|^{\Delta_1} \left|X_2^i\right|^{\Delta_2}} \, ,
\end{align}
where the integration is performed over the brane worldvolume, parametrized by the coordinate $\hat{Z}$. The function $C_{\Delta_1\Delta_2}(z,\bar z)$ can be computed in closed form as \cite{Gimenez-Grau:2023fcy}
\begin{align}
 C_{\Delta_1\Delta_2}(z,\bar z)
&= \frac{\pi ^{\frac{p_\text{\tiny{def}}+1}{2}}}{2^{\Delta_1+\Delta_2}}
   \frac{\Gamma \! \left(\frac{\Delta_1+\Delta_2-p_\text{\tiny{def}}}{2} \right)}
        {\Gamma \! \left(\frac{\Delta_1+\Delta_2+1}{2} \right)} \,
   {}_2F_1 \left( \Delta_1,\Delta_2,
            \frac{\Delta_1+\Delta_2+1}{2};-\frac{(1- \sqrt{z \bar z})^2}{4 \sqrt{z \bar z}} \right) \, ,
\end{align}
where $z$ and $\bar z$ are the lightcone cross-ratios defined in \eqref{lightcone}. For the case of interest in this paper, namely $p_\text{\tiny{def}}=2$ and $\Delta_1=\Delta_2=2$, the contact diagram evaluates to
\begin{equation}
  C_{2,2}(z,\bar z)  =\frac{\pi  z \bar z \left(-z \bar z+\tfrac{1}{2}\left(z \bar z+1\right) \log (z \bar z)+1\right)}{\left(z \bar z-1\right)^3}\,.
\end{equation}
The bulk scalar-exchange Witten diagram reads
\begin{align}
   \begin{tikzpicture}[valign]
    \tikzstyle{every node}=[font=\small]
    \pgfmathsetmacro{\x}{-sqrt(2)/2}
    \pgfmathsetmacro{\y}{sqrt(2)/2}
    \pgfmathsetmacro{\z}{-0.5}
    \pgfmathsetmacro{\r}{0.4}
    \draw [thick] (0,0) circle [radius=1];
    \draw [thick, blue] (-\x,+\y) to[out=240,in=120] (-\x,-\y);
    \draw [dashed]      (+\x,+\y) -- (-\r,0);
    \draw [dashed]      (+\x,-\y) -- (-\r,0);
    \draw [dashed]      (-\r,  0) -- (-\z,0);
    \node at (0.02, 0.2) {{\footnotesize $(\Delta,\!0)$}};
    \node at (-0.95, 0.85) {$X_1$};
    \node at (-0.95,-0.95) {$X_2$};
    \node at (-0.7,0.05) {$Z$};
    \node at (0.75,0.0) {$\hat{Z}$};
    \end{tikzpicture}
 \;\; = \;\;
 \int_{Z,\hat{Z}}
    K_{\Delta_1}\left(X_1,Z\right)
    K_{\Delta_2}\left(X_2,Z\right)
    G_{\Delta}\left(Z,\hat{Z}\right)
  \; = \;
  \frac{E^{\Delta,0}_{\Delta_1,\Delta_2}(z,\bar z)}
       {\left|X^i_1\right|^{\Delta_1} \left|X_2^i\right|^{\Delta_2}} \, ,
\end{align}
where the integration is performed over both the bulk point $Z$ and the defect worldvolume coordinate $\hat{Z}$. The integral admits a closed-form representation in Mellin space, namely~\cite{Gimenez-Grau:2023fcy}
\begin{align}
 \mathcal{M}_{\Delta_1,\Delta_2}^{\Delta,0}(\delta,\rho)
&= \frac{\pi ^{p_\text{\tiny{def}}/2}
         \Gamma\!\left(\frac{\Delta -p_\text{\tiny{def}}}{2}\right)
         \Gamma\!\left(\frac{\Delta_1+\Delta_2+\Delta-d}{2}\right)}
        {16 \Gamma (\Delta_1) \Gamma (\Delta_2)
         \Gamma\!\left(\frac{2 \Delta-d+2}{2}\right) \!
         \left(\delta - \frac{\Delta_1+\Delta_2-\Delta}{2}\right)} \notag \\
& \qquad \qquad \times \,
  _3F_2 \! \left( \!\! \begin{array}{*{20}{c}}
    {1+\frac{\Delta-d+p_\text{\tiny{def}}}{2},
     1+\frac{\Delta-\Delta_1-\Delta_2}{2},
    \delta +\frac{\Delta-\Delta_1-\Delta_2}{2}} \\
    {\Delta+1-\frac{d}{2},
     \delta+1+\frac{\Delta-\Delta_1-\Delta_2}{2}}
    \end{array}; 1\right) \, ,
\end{align}
where the Mellin amplitude is defined as
\begin{equation}\label{mellindiag}
 E(\xi,\phi)
 = \int \frac{d\delta \, d\rho}{(2\pi i)^2} \,
   \frac{\Gamma (\delta ) \Gamma (\rho ) \Gamma \! \left(\frac{\Delta_1-\delta-\rho}{2} \right) \Gamma \! \left(\frac{\Delta_2-\delta -\rho}{2}\right)}{\xi^\delta \, (2\cos \phi)^\rho} \,
   \mathcal{M}(\delta,\rho)
   \,,
\end{equation}
and $\xi, \phi$ are the cross-ratios defined in \eqref{crossratio}.
The spinning exchange diagrams can be computed either by Mellin-space techniques or by means of differential recursion relations. Since this technology is rather involved, we refer the reader to \cite{Gimenez-Grau:2023fcy} for the details. As an example, the spin-$2$ exchange diagram can be expressed in terms of contact diagrams as
\begin{equation}
    E^{4,2}_{2,2}
 = \frac{4}{3 \xi } \big( \cos \phi\,  C_{2,2} - C_{1,3} \big)\,.
\end{equation}
As discussed in the main text, a more efficient way to organize the bulk exchange diagrams is to introduce Polyakov-Regge blocks, namely exchange diagrams supplemented by suitable contact diagrams so that the resulting object is Regge superbounded. As shown in \cite{Gimenez-Grau:2023fcy}, these blocks can be constructed directly by applying the conformal dispersion relation \eqref{dispersion} to a single bulk conformal block, up to normalization conventions.\footnote{This derivation is analogous to the one of the Polyakov-Regge blocks of a four-point function in absence of a defect, see \cite{Caron-Huot:2020adz}.} Explicitly,
\begin{eqnarray}
 P_{\Delta_1,\Delta_2}^{\Delta,\ell}(r,w)
 &\propto& \int_{0}^{r} \frac{dw'}{2\pi i}\, \frac{\text{Disc}_{w<r}\Big[
 \xi^{-\frac{\Delta_1+\Delta_2}{2}} f^{(\Delta_{12})}_{\Delta,\ell}(r,w') \Big]}{w'-w} \nonumber\\
 &+& \int_{1/r}^{\infty} \frac{dw'}{2\pi i}\, \frac{\text{Disc}_{w>1/r}\!\Big[
 \xi^{-\frac{\Delta_1+\Delta_2}{2}} f^{(\Delta_{12})}_{\Delta,\ell}(r,w') \Big]}{w'-w}\, .
\end{eqnarray}
Here $r$ and $w$ are the radial coordinates defined in \eqref{radialcoord}, and $f^{(\Delta_{12})}_{\Delta,\ell}(r,w)$ denotes the bulk conformal block \eqref{bulkblockbosq}. One can prove that each Polyakov-Regge block is indeed a specific linear combination of exchange and contact diagrams. In the normalization of \cite{Gimenez-Grau:2023fcy}, one finds
\begin{align}
 P^{\Delta,0}_{\Delta_1,\Delta_2}
 = k_1 E^{\Delta,0}_{\Delta_1,\Delta_2} \, , \qquad
 P^{\Delta,2}_{\Delta_1,\Delta_2}
 = k_1 k_2 \left( E^{\Delta,2}_{\Delta_1,\Delta_2}
 + k_3 C_{\Delta_1,\Delta_2} \right) \, ,
\end{align}
where
\begin{align}
 k_1
&= \frac{2^{\Delta+2} (-1)^{\frac{\Delta_1+\Delta_2-\Delta+2}{2}}
          \Gamma \left(\frac{\Delta +1}{2}\right)
          (1-\Delta_1)_{\frac{\Delta_1+\Delta_2-\Delta}{2}}
          (1-\Delta_2)_{\frac{\Delta_1+\Delta_2-\Delta}{2}}}
         {\pi^{\frac{p_\text{\tiny{def}}+1}{2}}
          \Gamma \! \left(\frac{\Delta -p_\text{\tiny{def}}}{2}\right)
          \left(\frac{\Delta_1+\Delta_2-\Delta-2}{2}\right)!
          \left(\frac{d-\Delta -\Delta_1-\Delta_2+2}{2}\right)
             _{\frac{\Delta_1+\Delta_2-\Delta -2}{2}}}\, , \\
 k_2
&= \frac{8 \left(\Delta ^2-1\right)}
        {(\Delta_1+\Delta_2-\Delta)
         (\Delta_1+\Delta_2+\Delta-d) (\Delta^2 -\Delta^2_{12})}
         \, , \\[0.4em]
 k_3
&= \frac{d+1}{2d}
  \!-\! \frac{\Delta_{12}^2 (\Delta_1+\Delta_2-d)^2}{2d\Delta  (\Delta -d)}
  + \frac{\left(\Delta_{12}^2-1\right) \left((\Delta_1+\Delta_2-d)^2-1\right)}{2d(\Delta -1) (\Delta-d+1)} \, ,
\end{align}
and $\Delta_{12}=\Delta_1-\Delta_2$. The Polyakov-Regge blocks relevant for the holographic computation of section \ref{sec:holo-witten} are
\begin{eqnarray}\label{polyakovb}
    P_{2,2}^{2,0}(z,\bar z)&\propto&\frac{z \bar z (\log (z \bar z))}{(z-1) (\bar z-1) (z \bar z-1)}\,,\nonumber\\
    P_{2,2}^{3,1}(z,\bar z)&\propto&-\frac{12 z \bar z ((z-2) \bar z+1)}{(z-1) (\bar z-1) (z \bar z-1)^2}+\frac{6 z \bar z (z \bar z+1) ((z-2) \bar z+1) \log (z \bar z)}{(z-1) (\bar z-1) (z \bar z-1)^3}\,,\\
    P_{2,2}^{4,2}(z,\bar z)&\propto&-\frac{90 z \bar z (z \bar z+1)}{(z-1) (\bar z-1) (z \bar z-1)^2}+\frac{30 z \bar z \left(z^2 \bar z^2+4 z \bar z+1\right) \log (z \bar z)}{(z-1) (\bar z-1) (z \bar z-1)^3}\nonumber\,,
\end{eqnarray}
where we omit possible overall normalization factors, as these play no role in the Witten diagram bootstrap discussed in the main text.
The last class of diagrams appearing in figure \ref{fig:diagrams} consists of defect-exchange diagrams. For the exchange of a scalar defect field one has
\begin{align}
\begin{tikzpicture}[valign]
    \tikzstyle{every node}=[font=\small]
    \pgfmathsetmacro{\x}{sqrt(1)/2}
    \pgfmathsetmacro{\y}{sqrt(3)/2}
    \pgfmathsetmacro{\xx}{sqrt(2)/2}
    \pgfmathsetmacro{\yy}{sqrt(2)/2}
    \pgfmathsetmacro{\z}{0.5}
    \pgfmathsetmacro{\rx}{0.32}
    \pgfmathsetmacro{\ry}{0.45}
    \draw [thick] (0,0) circle [radius=1];
    \draw [dashed]      (-\xx,+\yy) -- (\rx,\ry);
    \draw [dashed]      (-\xx,-\yy) -- (\rx,-\ry);
    \draw [dashed,blue] (\rx,\ry) to[out=210,in=150] (\rx,-\ry);
    \draw [thick, blue] (\x,+\y) to[out=240,in=120] (\x,-\y);
    \node at (-0.95, 0.85) {$X_1$};
    \node at (-0.95,-0.95) {$X_2$};
    \node at (0.6, 0.3) {$\hat{Z}_1$};
    \node at (0.6,-0.3) {$\hat{Z}_2$};
    \node at (-0.2,0) { $\hat{\Delta}$};
    \end{tikzpicture}
 \; & = \;
 \int_{\hat{Z}_1, \hat{Z}_2}
 K_{\Delta_1}\left(X_1, \hat{Z}_1\right)
 K_{\Delta_2}\left(X_2, \hat{Z}_2\right)
 G_{\hat{\Delta}}\left(\hat{Z}_1,\hat{Z}_2\right)
 \; \equiv \;
  \frac{\widehat{E}^{\hat{\Delta},0}_{\Delta_1,\Delta_2}(z,\bar z)}
       {\left|X^i_1\right|^{\Delta_1} \left|X_2^i\right|^{\Delta_2}} \,.
\end{align}
This diagram also admits a closed-form Mellin representation,
\begin{align}
 \widehat{\mathcal{M}}_{\Delta_1\Delta_2}^{\hat{\Delta},0}
&=
 -\frac{\pi^{p_\text{\tiny{def}}/2}
        \Gamma\!\left(\frac{\Delta_1+\hat{\Delta}-p_\text{\tiny{def}}}{2}\right)
        \Gamma\!\left(\frac{\Delta_2+\hat{\Delta}-p_\text{\tiny{def}}}{2}\right)}
       {8 \Gamma (\Delta_1) \Gamma (\Delta_2) \Gamma(\hat{\Delta}\!+\!1-\frac{p_\text{\tiny{def}}}{2})
        (\delta\!+\!\rho\!-\!\hat{\Delta}) }
  {}_3F_2 \! \left( \!\! \begin{array}{*{20}{c}}
  {1\!-\!\frac{\Delta_1-\hat{\Delta}}{2}, 1\!-\!\frac{\Delta_2-\hat{\Delta}}{2},
   \frac{\hat{\Delta}-\delta-\rho}{2}} \\
  {\hat{\Delta}\!+\!1\!-\!\frac{p_\text{\tiny{def}}}{2},1 \!+\! \frac{\hat{\Delta}-\delta-\rho}{2}} \\
  \end{array}; 1 \right)\! .
\end{align}
The corresponding position-space diagram is then obtained through the Mellin transform \eqref{mellindiag}. Similar expressions can be derived for defect-exchange diagrams involving fields with non-vanishing transverse spin \cite{Gimenez-Grau:2023fcy}. Here we simply report the result for the transverse spin $s=\pm1$ diagram,
\begin{align}
 \widehat{E}^{\hat{\Delta},\pm1}_{\Delta_1,\Delta_2}(z,\bar z)
 = 2 \Delta_1 \Delta_2 \,\left(\frac{z}{\bar z}\right)^{\pm \frac{1}{2}}
   \widehat{E}^{\hat{\Delta},0}_{\Delta_1+1,\Delta_2+1}(z,\bar z) \, .
\end{align}
The defect-exchange diagrams entering the holographic computation of section \ref{sec:holo-witten} explicitly read
\begin{eqnarray}
    \widehat{E}^{2,0}_{2,2}(z,\bar z) &=&-\frac{\pi ^3}{48}+\frac{\pi  z \bar z \log ^2(\sqrt{z \bar z})}{4-4 z\bar z}+\frac{\pi  \left(z \bar z+1\right) \text{Li}_2\left(1-z \bar z\right)}{8-8 z \bar z}\, ,\nonumber\\
    \widehat{E}^{3,\pm1}_{2,2}(z,\bar z) &=& -\left(\frac{z}{\bar z}\right)^{\pm \frac{1}{2}}\frac{\pi (z \bar z+1) \left(\pi ^2 (z \bar z-1)^2-6 z \bar z\right)}{24 \sqrt{z \bar z} (z \bar z-1)^2}\\
    &+&\left(\frac{z}{\bar z}\right)^{\pm \frac{1}{2}}\frac{\pi \sqrt{z \bar z} \left(z^2 \bar z^2-4 z \bar z+1\right) \log (z \bar z)}{4(z \bar z-1)^3}+\left(\frac{z}{\bar z}\right)^{\pm \frac{1}{2}}\frac{\pi  (z \bar z)^{\frac{3}{2}} \log ^2(z \bar z)}{8 (1- z \bar z)}\nonumber\\
    &-&\left(\frac{z}{\bar z}\right)^{\pm \frac{1}{2}}\frac{\pi  \left(z^2 \bar z^2+1\right) \text{Li}_2(1-z \bar z)}{4 \sqrt{z \bar z} (z \bar z-1)}\, ,\nonumber\\
    \widehat{E}^{1,\pm1}_{2,2}(z,\bar z) &=&8 \left(\frac{z}{\bar z}\right)^{\pm\frac{1}{2}}\left(\frac{\pi  \sqrt{z \bar z} \left(z \bar z+1\right)}{32 \left(z\bar z-1\right)^2} -\frac{\pi  (z \bar z)^{\frac{3}{2}}\log \left(z \bar z\right)}{16 \left(z \bar z-1\right)^3}\right)  \,.\nonumber
\end{eqnarray}

\newpage
\bibliographystyle{JHEP}
\bibliography{references.bib}

\end{document}